\documentclass[usenatbib]{mnras}

\usepackage{newtxtext,newtxmath}

\usepackage[T1]{fontenc}

\DeclareRobustCommand{\VAN}[3]{#2}
\let\VANthebibliography\thebibliography
\def\thebibliography{\DeclareRobustCommand{\VAN}[3]{##3}\VANthebibliography}

\usepackage{graphicx}	
\usepackage{amsmath}	
\usepackage{ulem}
\usepackage{orcidlink}
\usepackage{bm}
\usepackage{colortbl} 

\usepackage{amsmath}
\usepackage[nopatch]{microtype}
\usepackage{booktabs}
\usepackage{hyperref} 
\hypersetup{colorlinks,citecolor=blue,linkcolor=blue,urlcolor=blue}

\usepackage{caption} 
\usepackage{amsfonts}

\usepackage{amssymb} 
\usepackage{amsmath}
\usepackage{graphicx}  
\usepackage{float} 

\usepackage[dvipsnames]{xcolor}

\def\v2{V_2\left(\mathbf{U},\Delta \nu = 0\right)}

\def\V{\mathcal{V}}

\newcommand{\vcg}{\mathcal{V}_{cg}}

\newcommand{\HI}{\ion{H}{i}} 
\newcommand{\cl}{C_{\ell}}
\newcommand{\dnu}{\Delta\nu}

\def\n{\hat{\mathbf{n}}}

\def\pk{P(k_{\perp}, k_{\parallel})}

\newcommand{\kpar}[1][]{k_{\parallel{#1}}}

\newcommand{\U}{\mathbf{U}}

\newcommand{\kk}{{\bf k}}
\newcommand{\kpp}{k_{\perp}}

\defcitealias{Chatterjee2024}{Paper~I}
\defcitealias{Elahi2025}{Paper~II}

\title[TTGE III: Multiple pointings]{The Tracking Tapered Gridded Estimator for the 21-cm power spectrum from the Murchison Widefield Array (MWA) drift scan observations -- III. Improved upper limits at $z = 8.2$ from multiple pointings}

\author[S. Sarkar et al.]{Shouvik Sarkar$^{1}$\thanks{E-mail: shouvik.s@physics.iitm.ac.in},
Khandakar Md Asif Elahi$^{1}$\,\orcidlink{0000-0003-1206-8689},
Samir Choudhuri$^{1}$
\orcidlink{0000-0002-2338-935X},
Somnath Bharadwaj$^{2}$,
\newauthor
Suman Chatterjee$^{3}$,
Baijayanta Bhattacharyya$^{2}$,
Shiv Sethi$^{4}$, and
Akash Kumar Patwa$^{4}$\\
\\
$^{1}$ Centre for Strings, Gravitation and Cosmology, Department of Physics, Indian Institute of Technology Madras, Chennai 600036, India\\
$^{2}$ Department of Physics, Indian Institute of Technology Kharagpur, Kharagpur - 721302, India\\
$^{3}$ Department of Physics and Astronomy, University of the Western Cape, 7535 Bellvill, Cape Town, South Africa\\
$^{4}$ Raman Research Institute, C. V. Raman Avenue, Sadashivanagar, Bengaluru 560080, India\\
}

\date{Accepted XXX. Received YYY; in original form ZZZ}

\pubyear{\the\year{}}

\begin{document}
\label{firstpage}
\pagerange{\pageref{firstpage}--\pageref{lastpage}}
\maketitle

\begin{abstract} 
We analyze zenith-pointing $(\delta=-26.7^{\circ})$ Murchison Widefield Array (MWA) $\nu_c=154.2 \,{\rm MHz}$ drift scan observations covering $349.0^{\circ} \le \alpha \le 70.0^{\circ}$ with 163 pointing centers (PCs) spaced by $0.5^{\circ}$. We measure $D_{\ell}$, the mean-squared angular brightness temperature fluctuations, as a function of $\alpha$. A broad peak at $\alpha \approx 50.0^{\circ}$ corresponds to the bright extended source Fornax~A in the main lobe of the primary beam. A smaller peak at $\alpha \approx 5.0^{\circ}$ possibly corresponds to Fornax~A in the first sidelobe. For $\alpha \leq 22.0^{\circ}$ and $\ell \ge 200$, we find $D_{\ell} \propto \ell^2$, which we interpret as Poisson fluctuations from point sources. We present $\Delta^2(k)$, the mean-squared 21-cm brightness temperature fluctuations from the Epoch of Reionization, as a function of $\alpha$. Fornax~A causes strong contamination near $\alpha \approx 50.5^{\circ}$, elsewhere several PCs are consistent with noise. The range $358.5^{\circ} \leq \alpha \leq 11.5^{\circ}$ is relatively foreground-free and best suited for EoR science. The PC at $\alpha = 11.0^{\circ}$ yields the best $2\sigma$ upper limit $\Delta^{2}_{\rm UL}(k) = (173.13)^{2}\,{\rm mK^{2}}$ at $k = 0.161\,{\rm Mpc^{-1}}$. We incoherently combine $23$ PCs to obtain $\Delta_{\rm UL}^2(k)=(98.67)^{2}\,{\rm mK}^{2}$ at $k=0.156\,{\rm Mpc}^{-1}$. This is the tightest upper limit from the MWA, being $\approx3$ times lower than earlier MWA limits at $z = 8.2$, but $\approx2$ and $\approx21$ times higher than the LOFAR and HERA limits, respectively, and $\approx3$ orders of magnitude above theoretical predictions. 
\end{abstract}

\begin{keywords}
methods: statistical, data analysis -- techniques: interferometric -- cosmology: diffuse radiation, dark ages, reionization, first stars, large-scale structure of Universe
\end{keywords}



\section{Introduction}
\label{sec:intro}
The redshifted 21-cm line from the hyperfine transition of neutral hydrogen (\HI{}) is a promising probe to study the high redshift universe, especially the Epoch of Reionization (EoR), when the universe became ionized from a neutral state. It can also help us understand the nature of the first stars and galaxies formed during this stage. There are several efforts towards the statistical detection of the 21-cm emission in terms of the power spectrum (PS) using different radio interferometric observations, e.g., the Murchison Widefield Array (MWA; \citealt{Tingay2013, Nunhokee2025}), LOw Frequency ARray (LOFAR; \citealt{vanHarlem2013, Mertens2025}), Hydrogen Epoch of Reionization Array (HERA; \citealt{Deboer2017, HERA2023}),  Giant Metrewave Radio Telescope (GMRT; \citealt{Swarup1991, Gupta2017, Paciga2011}) and the upcoming SKA-low \citep{Mellema2013, Koopmans2015}. The current best $2\sigma$ upper limit on the EoR 21-cm PS is $\Delta^2(k) < (21.4)^2\,{\rm mK}^2$ at $k = 0.34\, h\,{\rm Mpc}^{-1}$ at $z = 7.9$ \citep{HERA2023}.

The amplitude of this 21-cm signal from such a high redshift is expected to be faint and of the order of tens of mK. In contrast, the extra-galactic (such as AGNs, star-forming galaxies, radio halos, and relics) and Galactic foregrounds (such as free-free emission, synchrotron emission) are 4 to 5 orders of magnitude stronger than the signal \citep{Ali2008, Bernardi2009, Ghosh2012, Paciga2013, Patil2017}. Factors such as radio frequency interference (RFI), ionospheric contamination, instrumental systematics, and other terrestrial effects further complicate the detection of this signal. There are two broadly different methods for dealing with spectrally smooth astrophysical foregrounds: `foreground removal' and `foreground avoidance'. The former technique aims to subtract out the spectrally smooth foregrounds using different approximation techniques (e.g., \citealt{Chapman2012, Mertens2018, Elahi2023b}). While the latter uses the cylindrical PS $\pk$ and avoids the wedge shaped region in the $(\kpp,\kpar)$ space \citep{Datta2010, Morales2012, Vedantham2012, Trott2012, Pober2016, Elahi2023}.

The Tapered Gridded Estimator (TGE) \citep{Choudhuri2014, Choudhuri2016b} is a visibility-based PS estimator that convolves the measured visibilities with a Gaussian window function to suppress the side-lobe contribution from bright extra-galactic point sources. 
The convolved visibilities are evaluated onto a regular rectangular grid.
The TGE correlates the gridded visibilities, making it computationally much faster than correlating each visibility individually. It also subtracts out the noise bias internally to provide an unbiased estimate of the PS. Another major impediment in measuring the 21-cm PS is the missing frequency channels in the visibility data. As a result of which, a Fourier transform along the frequency introduces a ripple-like structure in the delay space PS \citep{Morales2004, Parsons2009}. Several efforts to address this issue have been reported in the literature \citep{Parsons2009, Parsons2012b, Parsons2014, Trott2016a, Kolopanis2019, Ewall-Wice2021, Kennedy2023}. A detailed discussion of different methods is given in Section 1 of \citetalias{Elahi2025}.





To overcome the issue of the missing frequency channels, 
\citealt{Bharadwaj2018} modified the TGE first to estimate the multi-frequency angular power spectrum (MAPS; \citealt{Datta2007}, \citealt{Mondal2019}) $C_{\ell}\left(\Delta\nu\right)$, and then Fourier transform the estimated $C_\ell (\Delta\nu)$ along $\Delta\nu$ to obtain the $\pk$. The key idea is that even in the presence of missing frequency channels $\nu$, there are no missing separations $\Delta\nu$, and the PS remains free from artefacts. In this approach, the PS can be obtained by using the available channels only, without the need for compensating for the missing channels.
They showed that even if the data consists of 80\% random flagging of the frequency channels, TGE can still recover the PS. TGE has been extensively applied to GMRT data which had quite severe flagging, to put upper limits on the 21-cm PS from the EoR \citep{Pal2020} and also from the post-EoR Universe \citep{Pal2022, Elahi2024}. Furthermore, TGE has also been extensively used in Galactic astrophysics, including measurements of the angular power spectrum (APS) of the diffuse Galactic synchrotron emission (DGSE) \citep{Choudhuri2017, Choudhuri2020} and the magnetohydrodynamic turbulence from supernova remnant \citep{Saha2019, Saha2021}.

\citet{Chatterjee2022} have developed the Tracking Tapered Gridded Estimator (TTGE) for estimating the 21-cm PS from drift scan observations (e.g., MWA) while sharing the same advantages as the original TGE. They also validated the TTGE using simulated MWA drift scan observations at a single frequency. \cite{Chatterjee2024} (hereafter \citetalias{Chatterjee2024}), the first paper in this series, has upgraded the TTGE to measure $\pk$ from drift scan observations and validated it with a simulation using the same flagging as the actual MWA data. In \citetalias{Chatterjee2024}, we applied the TTGE to a single pointing drift scan data centered at $(\alpha, \delta) = (6.0^{\circ},-26.7^{\circ})$ where $\alpha$ is the Right Ascension and $\delta$ is the Declination. The measured $\pk$ from the real data shows a periodic pattern of spikes along $\kpar$, which exactly matches the period of the pattern of flagged channels $(1.28 \, {\rm MHz})$. Although the amplitude of these spikes is found to be much smaller than the amplitude at $k_\parallel = 0$, these spikes contaminated most of the $(\kpp,\kpar)$ plane. We used a small region ($0.05 \leq \kpp \leq 0.16 \, {\rm Mpc^{-1}}$, $0.9 \leq \kpar \leq 4.6 \, {\rm Mpc^{-1}}$) which is relatively uncontaminated and put a $2\sigma$ upper limit of $\Delta_{\rm UL}^2(k) = (1.85\times10^4)^2\, {\rm mK^2}$ at $k=1 \,{\rm Mpc}^{-1}$ on the mean squared 21-cm brightness temperature fluctuations. 

In \cite{Elahi2025} (hereafter \citetalias{Elahi2025}), we have investigated the above issue in detail and found that the spikes originate from a combination of the missing channels and the strong spectral dependence of the foregrounds. We demonstrated that the periodic spikes along the $\kpar$ direction can be significantly mitigated if we filter out the spectrally smooth component from the gridded visibility data before estimating $C_\ell (\Delta\nu)$ and $\pk$. We have applied this technique, termed as Smooth Component Filtering (SCF), on the same data as in \citetalias{Chatterjee2024}, and found a large region in $(\kpp,\kpar)$ plane to be free from those spikes. We found the $2\sigma$ upper limit of $\Delta_{\rm UL}^2(k)=(934.60)^2\,{\rm mK^2}$  at $k=0.418\,{\rm Mpc^{-1}}$ using 17 minutes of a single pointing centre (PC) of the drift scan data. 

In addition to the 21-cm PS, the EoR 21-cm bispectrum (BS) is also predicted to contain a wealth of information \citep{Majumdar2018, Gill2024}. Recently, \citet{Gill2025a} and \citet{Gill2025b} have developed an estimator for the 21-cm BS, and they have applied it to the same MWA data as in \citetalias{Chatterjee2024} and \citetalias{Elahi2025}. 
They have evaluated the binned  21-cm BS for triangles of all possible sizes and shapes. 
The best $2 \sigma$  upper limits they obtain for $\mid \Delta^3(k)\mid$, the mean cubed EoR 21-cm 
brightness temperature fluctuations, are $(1.81 \times 10^3)^3 \, {\rm mK}^3$ at 
$k = 0.008 \,  {\rm Mpc}^{-1}$  and  $(2.04 \times 10^3)^3 \, {\rm mK}^3$ at 
$k = 0.012 \, {\rm Mpc}^{-1}$  for equilateral and squeezed triangles, respectively \citep{Gill2025}. 


In this paper, we present a comprehensive analysis of all 163 PCs of the MWA drift scan observation. This zenith pointing observation, 55 hours in duration, spans the $\alpha$ range from $349.0^{\circ}$ to $70.0^{\circ}$ at an interval of  $0.5^{\circ}$  with a fixed $\delta$ at $-26.7^{\circ}$. The main motivation for this work is threefold. First, we aim to measure the variation of the total sky signal, predominantly due to foregrounds,  covering the entire sky region under observation. Following \citet{chatterjee2025},  we have used the observed visibility data to estimate the APS $(C_{\rm \ell})$ of the two-dimensional (2D) brightness temperature fluctuations on the sky as a function of $\alpha$.  We expect this study to provide important inputs to guide the choice of target fields for future observations towards detecting the EoR 21-cm signal.   Secondly, we aim to measure the variation of the PS of the three-dimensional (3D) 21-cm brightness temperature fluctuations as a function of $\alpha$. Following \citetalias{Elahi2025}, we have applied the TTGE and SCF to estimate the 21-cm PS. The observations here are not deep enough to detect the EoR 21-cm signal, and the measured 21-cm PS is either consistent with the expected system noise contribution or is foreground contaminated, depending on the PC.  We use these to obtain upper limits on $\Delta_{\rm UL}^2(k)$, the mean squared brightness temperature fluctuations of the 21-cm signal, for the entire $\alpha$ range probed by the observation.  We expect this to provide additional guidance for the choice of target fields for future EoR 21-cm observations, the fields where $\Delta_{\rm UL}^2(k)$ is not foreground contaminated being more favourable in comparison to those where this is foreground dominated. We expect the above studies to be particularly important in view of the fact that SKA-Low \citep{Koopmans2015} is being built at exactly the same location as MWA. The region of sky considered here will be overhead for SKA-Low, and these studies are expected to provide useful guidance for selecting target fields for future SKA-Low observations to detect the EoR 21-cm signal. We note in this context that \cite{Jong2025} recently investigated the suitability of several candidate fields for EoR science. Finally, we expect the system noise to go down if we combine observations from multiple PCs. The third aim here is to incoherently combine the  21-cm PS estimated for the different PCs and use this to constrain the EoR 21-cm signal.

A brief outline of the paper is as follows: Section \ref{sec:data} describes the MWA  drift scan data analyzed, and Section \ref{sec:methodology1} presents the methodology of our analysis. The results are presented in Section~\ref{sec:Results}. Finally, we present the summary and conclusions in Section \ref{sec:summary}. Throughout our analysis, we have used the cosmological parameters from \cite{Planck2020f}.

\section{Data}
\label{sec:data}

In this paper, we utilized a specific drift scan observation (Project ID G0031) carried out using the MWA Phase II compact configuration  \citep{Wayth2018}.  The observation and data processing 
have been carried out by \citealt{Patwa2021}, who presents the details. This is 
 also summarized in \citetalias{Chatterjee2024}, and \citetalias{Elahi2025}. This zenith pointing drift scan observation has $\delta=-26.7^{\circ}$ fixed, and it covers $\alpha=349.0^{\circ}$ to $70.0^{\circ}$, with a total span of $81.0^{\circ}$ (as shown in Figure 1 of \citetalias{Chatterjee2024}), and observing time 5 hr 24 mins per night.     The visibility data correspond to $163$ PCs at an interval of $\Delta \alpha = 0.5^ {\circ}$.  The same observation was made on 10 consecutive nights from October 3rd to 12th, 2016. 

 The observation has been performed at the central frequency of $\nu_c=154.2 \, {\rm MHz}$ with $N_c = 768$ channels of resolution of $\dnu_c=40 \, {\rm kHz}$  covering the total observing bandwidth of $B_{\rm bw} = 30.72$ MHz. This is further divided into  24 coarse bands, each containing  32 channels {\it i.e.,} $1.28 \, {\rm MHz}$.  Due to the  MWA design (\citealt{Prabu2015}), in each coarse band, 4 channels at both ends and the central one are not usable (for more details refer to Figure 2 of \citetalias{Chatterjee2024}).

The initial visibility data is recorded at a time resolution of 0.5 seconds. We have used COTTER \citep{Offringa2015} (which in turn uses AOFlagger \citep{Offringa2010}) to avoid  RFI, defunct frequency channels, and antennas, and to subsequently average to 10 s resolution. This is written out in \textit{Measurement Sets} (MS), which are calibrated utilizing  \texttt{CASA}\footnote{\url{https://casa.nrao.edu/}} (\citealt{casa07}). Bandpass and flux density calibration were carried out using the unresolved calibrator  Pictor A with $\left(\alpha,\delta\right) = (79.95^{\circ}, -45.78^{\circ})$, which has a flux density $ 381.88 \, \pm \, 5.36\, {\rm Jy}$  and a flux spectral index $ -0.76 \, \pm \, 0.01$ at 150 MHz (\citealt{Jacobs_2013}). Here, we have not incorporated polarization calibration, and we only consider the visibilities $\V^{\rm XX}$ and $\V^{\rm YY}$, which correspond to the two linear polarization states recorded in MWA. We observe Pictor A at the end of our observation every night, and we apply the same gain solution to all PCs.

 Data obtained from each night was flagged and calibrated individually, and the $10$ nights data were then averaged. We note that the first $2 \, {\rm hr}$ of data is missing on the $6$-th night.  As a consequence, the nights of observations are $N_{\rm nights}=10$  for PCs $\alpha>18.5^{\circ}$, whereas it is $N_{\rm nights}=9$ for other PCs.   Finally, the MS for each PC 
 contains visibility data with $11$ different time stamps each with $t_{\rm int}= N_{\rm nights} \times 10~{\rm s}$ effective integration time. Furthermore, the root mean square (r.m.s.) of system noise $(\sigma_{\rm N})$ for the real (and also imaginary) part of the measured visibilities is predicted to be  $\sigma_{\rm N} = {60 \,\rm Jy}/\sqrt{N_{\rm nights}}$  (eq. 1 of \citetalias{Chatterjee2024}).


As mentioned earlier, the subsequent analysis is in two parts. In the first part we quantify 
the 2D angular fluctuations of the sky signal using  $C_{\ell}$,   
considering a fixed frequency of $154\,{\rm MHz}$. Here, similarly to \citet{chatterjee2025}, we collapse  17 frequency channels,  a total of $0.68\,{\rm MHz}$ and analyze the resulting visibilities $\V_i$ with corresponding baselines $\U_i$. Note that the frequency width is sufficiently small to avoid bandwidth smearing \citep{taylor1999synthesis}. Furthermore, we do not explicitly show the frequency $\nu$ which is held fixed. The different polarizations are assumed to have the same signal, but independent noise, just like the different timestamps, and we do not explicitly show the polarization. In the second part, which is a 3D analysis, we quantify the angular and frequency dependence of the sky signal. Here we use the full visibility data $\V_i^{\rm XX}(\nu)$ and $\V_i^{\rm YY}(\nu)$ where the corresponding baselines $\U_i$ are at the central frequency $\nu_c$ and $\nu$ spans all $768$ frequency channels available in the data.

\section{Methodology}
\label{sec:methodology1}
We first consider the 2D analysis, where we have the visibilities $ \V_i$, which quantify  
brightness temperature fluctuations on the sky at a fixed frequency. We have used the TGE to estimate the APS $(C_{\ell})$ for each PC of the drift scan observation. The detailed mathematical formalism of the TGE has been discussed in several earlier works \citep{Choudhuri2016b, Choudhuri2017, Choudhuri2020}. Following the same procedure as in \citet{chatterjee2025}, we taper the sky response with a window function ${\cal W}(\theta)=\exp[-\theta^2/\theta_w^2]$, which is a Gaussian with a FWHM $\theta_{\rm FWHM}=\theta_w/0.6$ with $\theta_{\rm FWHM}= 15^{\circ}$. We implement the tapering by convolving the measured visibilities $ \V_i$ with $\tilde{w}(\mathbf{U})$, the Fourier transform of ${\cal W}(\theta)$. We evaluate the tapered visibilities on a rectangular grid using 
\begin{equation}
    \V_{cg} = \sum_{i}\tilde{w}(\mathbf{U}_g-\mathbf{U}_i) \, \V_i
    \label{eq:vcg}
\end{equation}
which gives the convolved gridded visibility $\V_{cg}$  at the grid point $g$ with corresponding baseline $\U_g$, and the angular multipole value $\ell_g= 2 \pi \left| \mathbf{U}_g \right|$.
In TGE, we evaluate  $C_{\ell}$ using, 
\begin{equation}
C_{\ell_g} = M_g^{-1} \, \mathcal{R}e \Big\langle \left( \left| \V_{cg} \right|^2 -\sum_i \left|
\tilde{w}(\mathbf{U}_g-\mathbf{U}_i) \right|^2 \left| \V_i \right|^2 \right) \Big \rangle \,,
\label{eq:a6}
\end{equation}
where $M_g$ is a normalising factor, which we calculate using simulated visibilities corresponding to a unit angular power spectrum (UAPS), $C_{\ell}=1$. To reduce the statistical fluctuations, we have used $50$ realisations of UAPS to estimate $M_g$ (details can be found in \citealt{Chatterjee2022}). We have further binned the $C_{\ell_g}$ to get the binned APS $C_{\ell}$ at an effective angular multipole $\ell$, which the average of all $\ell_g$'s in a particular bin.

We now consider the 3D analysis that also considers the frequency dependence of the visibilities  $\V^{\rm XX}_i(\nu)$ and $\V^{\rm YY}_i(\nu)$. Our procedure is presented in detail in Section 3 of \citetalias{Elahi2025}, and we briefly summarize this here.  First, we grid the visibilities in exactly the same way as described earlier  (eq.~\ref{eq:vcg}) to obtain  $\V_{cg}^{\rm XX}(\nu)$ and $\V_{cg}^{\rm YY}(\nu)$. Next, we 
estimate the MAPS using

\begin{align}
 C_{\ell_{\rm g}}( \nu_{\rm a},\nu_{\rm b})
 &= M^{-1}_{\rm g}(\nu_{\rm a},\nu_{b}) \,
 \mathcal{R}e \Big\langle \Big[
   \vcg^{\rm XX}(\nu_{\rm a}) \vcg^{*\rm YY}(\nu_{\rm b})   \nonumber \\
 &\qquad\qquad\qquad
 + \vcg^{\rm YY}(\nu_{\rm a}) \vcg^{*\rm XX}(\nu_{\rm b})
 \Big] \Big\rangle
 \label{eq:maps3}
\end{align}
which cross-correlates the two polarizations to avoid noise bias, and this also avoids contributions from polarization-dependent systematics that may be present in the data. 
Similar to eq.~(\ref{eq:a6}),  $M_{\rm g}(\nu_{\rm a},\nu_{\rm b})$ is a normalization factor whose value is estimated using simulations corresponding to unit MAPS $(C_{\ell}(\nu_a,\nu_b)=1)$. 
We assume the 21-cm signal to be ergodic along the line-of-sight (LoS) (\citealt{Mondal2018}), 
whereby the MAPS can be represented as a function of frequency separation $(\Delta \nu= |\nu_{\rm a}-\nu_{\rm b}|)$. Finally, we used the $C_{\rm \ell}(\Delta \nu)$ to estimate the cylindrical PS using \citep{Datta2007}
\begin{equation}
    \pk = r^2\,r^{\prime} \int_{-\infty}^{\infty} d(\Delta \nu) \, e^{-i \, \kpar r^{\prime} \Delta \nu} C_{\ell}(\Delta \nu) \,,
    \label{eq:pk-2} 
\end{equation}
where $ k_{\perp}$ and  $k_{\parallel}$ are components of $\kk$, perpendicular and parallel to LoS, respectively. The comoving distance $r = 9210$~Mpc and its frequency derivative $r^{\prime} = 16.99$~Mpc/MHz are evaluated at the reference redshift $z=8.2$.

\begin{figure}
    \includegraphics[width=\columnwidth]{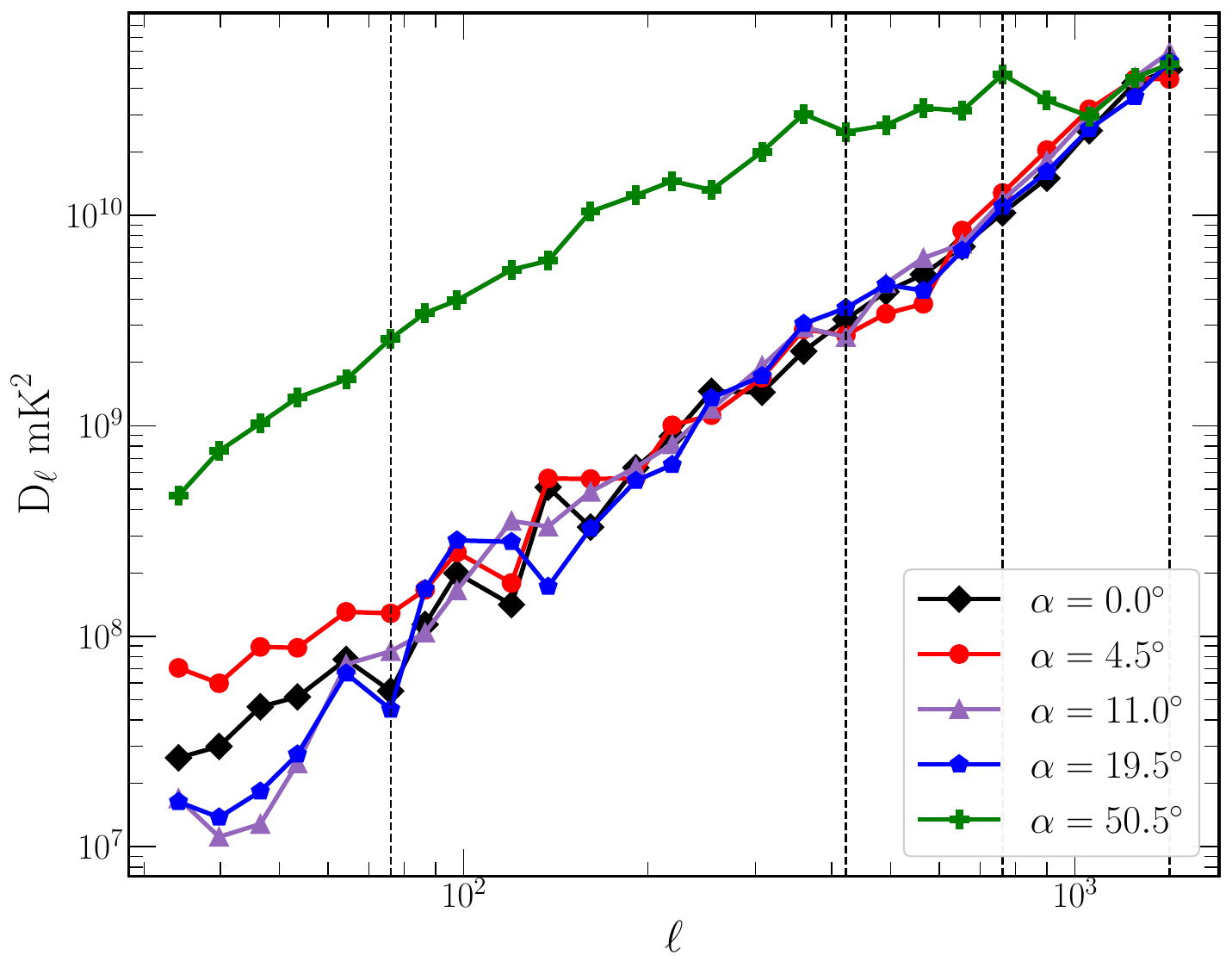}
    \caption{The measured $D_{\ell}$ vs $\ell$ for five PCs. The four vertical dotted lines mark the $\ell$ values $76, 422, 762$ and $1428$ respectively.}
    \label{fig:D_ell_3PC}
\end{figure}

\citetalias{Elahi2025} showed that due to the combined effect of foregrounds and the periodic missing channels in the MWA data, a naive implementation of the estimator leads to horizontal streaks in $P(k_\perp, \kpar)$. They proposed that the streaks can be mitigated by applying the Smooth Component Filtering (SCF; \citetalias{Elahi2025}) technique, which first filters out the spectrally smooth component from the gridded visibilities and then uses the filtered visibilities to estimate the power spectrum. Specifically, we use a Hann window
\begin{equation}
    H(n) = \frac{1}{4N} \left[ 1 + \cos \left( 2\pi \frac{n}{2N} \right)   \right] \, , \, \,  -N \leq n \leq N
    \label{eq:Hann}
\end{equation}
to calculate the smooth component of  $\vcg(\nu)$ using a convolution 
\begin{equation}
    \vcg^S(\nu_n) = (  \vcg * H )(\nu_n) = \sum_{m} \vcg(\nu_{m})\, H(n - m)\,.
    \label{eq:conv}
\end{equation}
In eq.~(\ref{eq:Hann}), the Hann window is of width $2N+1$, where $N$ determines the half-width of the smoothing kernel. We adopt $N=50$, which corresponds to a smoothing scale of $N\Delta\nu_c = 2\, {\rm MHz}$ for 
$\vcg(\nu)$. In \citetalias{Elahi2025} (Appendix B), we compare the results for different window size, also discuss the reason to choose the window size of  $2\, {\rm MHz}$. We then subtract the smooth component $\vcg^S(\nu)$ from $\vcg(\nu)$ to obtain the filtered component using  
\begin{equation}
    \vcg^F(\nu) =\vcg(\nu) - \vcg^S(\nu) \,,
    \label{eq:res}
\end{equation}
which we use  in eq.~(\ref{eq:maps3}) to estimate the MAPS, and subsequently $P(k_\perp, \kpar)$. We note that the convolution is not valid at the edges of the band, which leads us to discard $N$ channels from each edge. The effective number of channels is then $ N_c-2N = 668$, which corresponds to an effective bandwidth of $26.72$~MHz.  


We expect SCF to suppress the power at scales somewhere in between the full width $(2N +1)$ and the half width $(N)$ of the Hann window. In the Fourier domain, this leads to a suppression of power at a characteristic filtering scale $[k_{\parallel}]_F$, which should be somewhere in this range $2 \pi/(r^{\prime} \, 4~{\rm MHz}) < [k_{\parallel}]_F < 2 \pi/(r^{\prime} \, 2~{\rm MHz})$, i.e., $0.092  < [k_{\parallel}]_F < 0.185 \, {\rm  Mpc}^{-1}$. In \citetalias{Elahi2025}, we conservatively discarded the range  $k_\parallel < 0.185 \, {\rm Mpc}^{-1}$ for constraining the EoR 21-cm signal. However, simulations in Appendix~\ref{app:valid} show that it is possible to safely extend the range up to $[k_{\parallel}]_F = 0.135 \, {\rm  Mpc}^{-1}$, which allows us to probe the largest scales that were discarded previously. We find that this choice results in a signal loss of approximately $15.8\%$ and $1.24\%$ at the first two $k$-bins, $k=0.165\,{\rm Mpc}^{-1}$ and $k=0.211\,{\rm Mpc}^{-1}$, respectively, in the spherical power spectrum $P(k)$. We also refer the reader to \citetalias{Elahi2025}, who provided the validation of the entire methodology presented in this work.  They have demonstrated that the signal loss due to SCF is restricted to the range $k_\parallel < [k_{\parallel}]_F$, which is set by the choice of the smoothing scale. We `avoid' these $k_\parallel$ modes while placing constraints on the EoR 21-cm signal.  Considering the range $k_\parallel \ge [k_{\parallel}]_F$, which we use to constrain the EoR signal, the recovered signal is in reasonable agreement with  the input model. Therefore we do not perform any correction for signal loss in the final results. We also note that in Appendix~\ref{app:valid}, we have performed simulations with the exact choice of $(k_\perp, k_\parallel)$ modes and spherical binning scheme that we have considered in this work, and discussed how accurately we can recover the EoR signal in each $k$-mode that we probe in this analysis. 


\section{Results}
\label{sec:Results}
\subsection{Angular Power Spectrum}
\label{sub-sec:APS}

In this section, we present the measurements of $C_{\ell}$ for all $163$ PCs. These $C_{\rm \ell}$ measurements quantify the level of total foreground contamination in this region of sky. Figure~\ref{fig:D_ell_3PC} shows  $D_{\rm \ell} ={\ell(\ell+1)} \,C_{\rm \ell}\, /{2\pi}$, which measures the mean squared brightness temperature fluctuations of the sky signal, as a function of $\ell$. The available  $\ell$ range, $34 \leq \ell \leq 1428$, is divided into 25 equally spaced logarithmic bins.

\begin{figure*}
   \includegraphics[width=\textwidth]{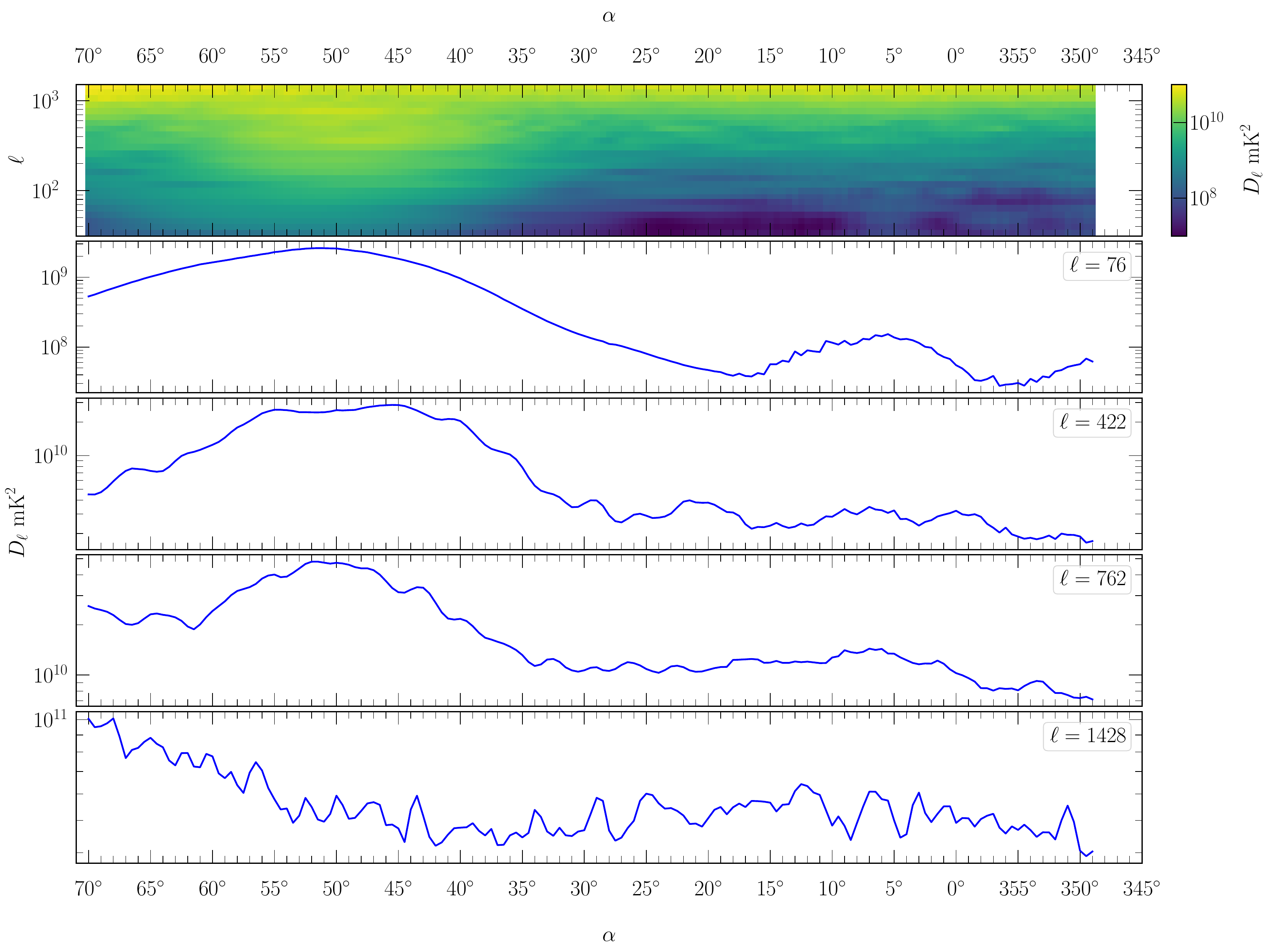}
    \caption{The top panel shows the variation of \( D_{\ell} \) as a function of $\alpha$ and $\ell$.The bottom four panels show the slices of \( D_{\ell} \) vs $\alpha$, for $\ell$ values $76, 422, 762$ and $1428$ respectively. At lower $\ell$ values, we see two prominent peaks in the variation of $D_{\ell}$. The peak at $\alpha \approx 50.0^{\circ}$ corresponds to the situation when Fornax~A is located in the main- lobe of the PB, whereas the peak at $\alpha \approx 5.0^{\circ}$ may plausibly arise from contamination by the emission from Galactic plane entering through the side-lobes of the MWA primary beam along with Fornax A.}
    \label{fig:D_l-unsub}
\end{figure*}
We have shown the measured $D_{\rm \ell}$ as a function of $\ell$ for five PCs centred at $\alpha=0.0^{\circ},4.5^{\circ}, 11.0^{\circ}, 19.5^{\circ}$ and $50.5^{\circ}$ respectively. The PC centred at $\alpha= 0.0^{\circ}$ is a well-studied MWA field, namely EoR0 \citep{Carroll2016,Trott2020, Nunhokee2025}. We observe that the first four PCs all exhibit very similar behaviour (along with other PCs at $\alpha \leq 22^{\circ}$)  where $D_{\ell}$  increase with $\ell$ as $D_{\ell}\propto \rm \ell^2$ for $\ell \ge 200$, indicating that the measured $D_{\ell}$ is dominated by the Poisson fluctuations due to bright point sources \citep{Ali2008, Gehlot2022}. In contrast, the $\ell$ dependence of $D_{\ell}$ is noticeably shallower for $\alpha = 50.5^{\circ}$. Furthermore,  for this PC,  the amplitude of $D_{\ell}$ is more than an order of magnitude larger as compared to the other four PCs in the $\ell$ range $< 400$.  For example,  at $\ell=76$, $D_{\ell}$ has values in the range $5 \times 10^{7} \, {\rm mK}^2$ to $\sim 10^{8} \, {\rm mK}^2$  for the first four PCs,  whereas $D_{\ell} \sim  2 \times 10^{9} \, {\rm mK}^2$ for $\alpha = 50.5^{\circ}$. As discussed later, these differences are primarily due to the very bright source  Fornax~A that is present in the main lobe of the MWA PB at $\alpha = 50.5^{\circ}$. The results from all five PCs shown here roughly match at $\ell>1055$  where the signal is dominated by the Poisson fluctuations of the point sources  \citep{Choudhuri2017, Choudhuri2020} for $\alpha = 50.5^{\circ}$ also. The DGSE contribution is subdominant in the $\ell$  range considered here.  As demonstrated in \citet{chatterjee2025},  it is necessary to model and subtract out the bright point sources in order to study the DGSE using these MWA  observations. 

\begin{figure*}
   \includegraphics[width=\textwidth]{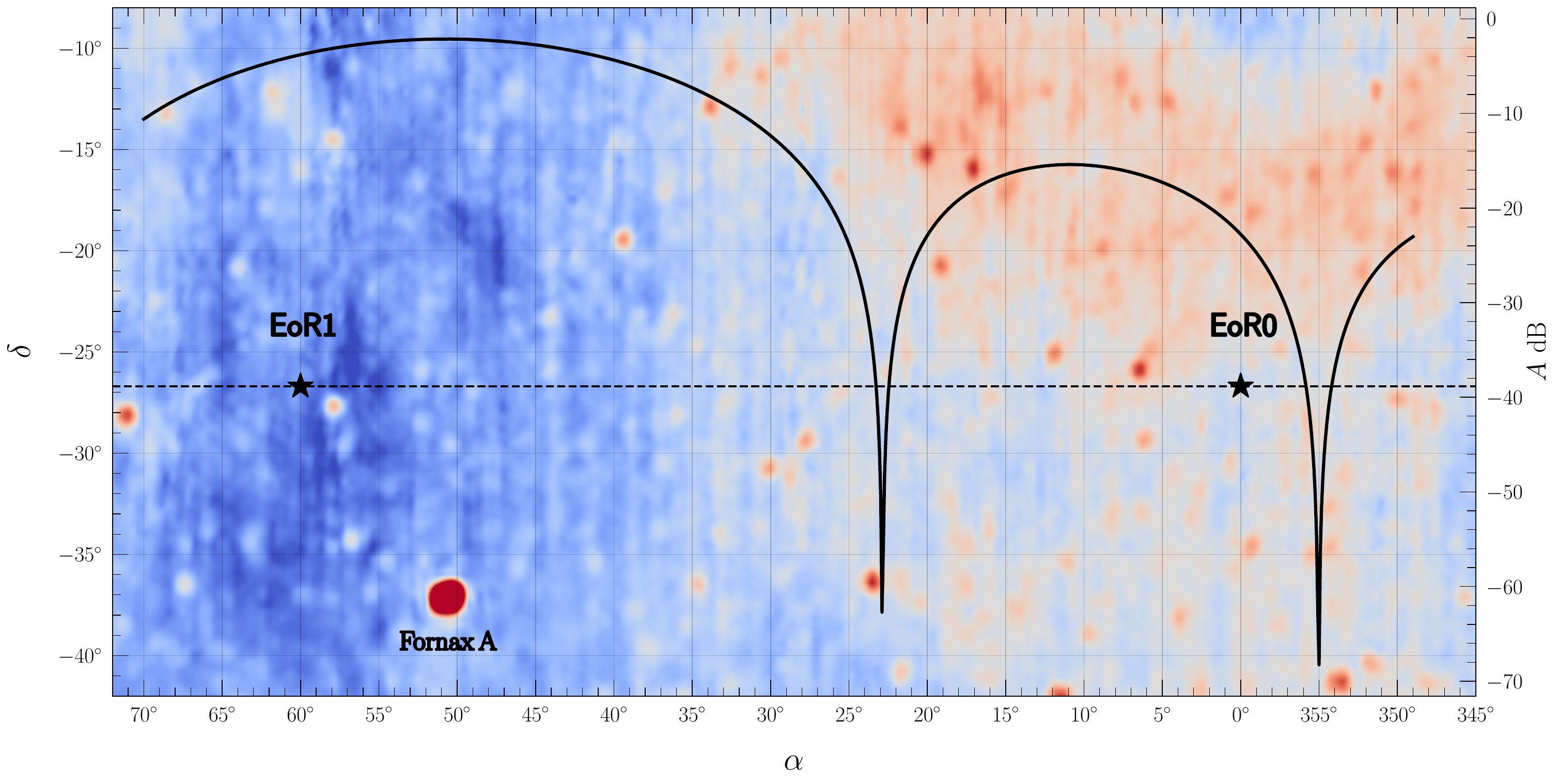}
    \caption{The background shows the Haslam 408 MHz map \citep{Haslam1982} scaled to 154 MHz, assuming a spectral index of \(\alpha = -2.52\) \citep{Rogers2008}. The solid black line shows $A(\alpha) \equiv {\cal A}(\alpha,\delta=-37.2) $, which corresponds to a section of the MWA PB centered at   $(\alpha,\delta)=(50.5^{\circ},-26.7^{\circ})$. The right-hand axis shows PB amplitude in dB. The black dashed line indicates the MWA declination ($\delta = -26.7^{\circ}$). The two well-studied MWA fields EoR0 ($0.0^{\circ}, -26.7^{\circ}$) and EoR1 ($60.0^{\circ}, -26.7^{\circ}$) are denoted by `$\star$' marks.}
    \label{fig:GLEAM}
\end{figure*}

The top panel of Figure~\ref{fig:D_l-unsub} shows the variation of $D_{\ell}$ as a function of $\alpha$ and $\ell$. Overall, the value of $D_{\ell}$ varies in the range $10^{7}\,{\rm mK}^{2}$ to $10^{11}\,{\rm mK}^{2}$, roughly consistent with the values in \citet{chatterjee2025}. For all $\alpha$,  the value of $D_{\ell}$  increases with $\ell$, similar to the behaviour seen  in Figure~\ref{fig:D_ell_3PC}. At the largest $\ell$ values $(\ell  \ge  10^3)$, $D_{\ell}$ shows  little variation with $\alpha$. However, we have a noticeable $\alpha$ dependence at smaller $\ell$, where the values of $D_{\ell} $  are relatively smaller to the right of $\alpha \sim 22.0^{\circ}$. The values of $D_{\ell}$ rise gradually for $\alpha > 22.0^{\circ}$, they peak at $\alpha \sim 50.0^{\circ}$. The lower four rows of  Figure~\ref{fig:D_l-unsub} show the variation of $D_{\ell}$ as a function of $\alpha$, where each panel corresponds to a different $\ell$. The $\ell$ values corresponding to the different rows are indicated by the four vertical lines as shown in Figure~\ref{fig:D_ell_3PC}. Considering $\ell=76$, we see two prominent peaks in the variation of $D_{\ell}$ as a function of $\alpha$. The first peak at $\alpha \approx 5.0^{\circ}$ is smaller in both height and width relative to the second peak at $\alpha \approx 50.0^{\circ}$. A similar $\alpha$ dependence is seen in the next two rows, $\ell=422$ and $762$ respectively, however, the relative amplitude of the first peak diminishes with increasing $\ell$. The two peaks are absent in the lowest row  ($\ell=1428$) where $D_{\ell}$  increases gradually with $\alpha$ for $\alpha \ge 50.0^{\circ}$.  Furthermore, $D_{\ell}$ exhibits a smooth $\alpha$ dependence at $\ell=76$, whereas we find some rapidly varying undulations in the next two rows, and these become more prominent in the lowest row. As shown in Figure~\ref{fig:D_ell_3PC}, at the largest $\ell$ bins ($\ge 10^3$),  the  $D_{\ell}$ curves for different $\alpha$  converge to the $D_{\ell} \propto \ell^2$ behaviour expected for the Poisson fluctuations from extra-galactic point sources. This picture is roughly consistent with the lowest row of Figure~\ref{fig:D_l-unsub} where we find $D_{\ell}$ to be nearly isotropic,  (independent of $\alpha$), barring the rapidly varying undulations and the gradual increase at $\alpha \ge 50.0^{\circ}$. We have included a table in the Supplementary Material containing the $D_{\ell}$ values for all $25$ $\ell$ bins and across all 163 PCs in machine-readable format, whose column format is given in Appendix \ref{app:D_ell_all_PC}.



We now consider the sky sources that could lead to the behaviour of $D_{\ell}$ seen in  Figures~\ref{fig:D_ell_3PC} and \ref{fig:D_l-unsub}. Figure~\ref{fig:GLEAM} shows the same region of the sky (\(349.0^{\circ}\leq\alpha\leq70.0^{\circ}\) and \(-40.0^{\circ} \leq \delta \leq -10.0^{\circ}\)) where the drift scan observation has been carried out. Here, the $\delta$ range roughly corresponds to the FWHM $(\approx 25^{\circ})$ of the MWA PB. The horizontal dashed black line shows the declination  \(\delta=-26.7^{\circ}\) of the zenith pointing observations, which is also the latitude of the MWA telescope. The background image shows the brightness temperature distribution at \(154\, \text{MHz}\),  scaled from the \(408 \, {\rm MHz}\) Haslam map \citep{Haslam1982} by assuming a spectral index of \(\alpha = -2.52\) \citep{Rogers2008}. In addition to the diffuse Galactic emission, we can also see several discrete sources in the image.  A visual inspection reveals that these discrete sources correspond to the bright sources with flux above \(15 \, \rm Jy\) in the GLEAM-X catalogue \citep{Ross2024}.  We see a relatively lower level of DGSE at  $\alpha>30.0^{\circ}$, and would expect this region to be better suited for EoR observations as compared to other parts of the sky. However,  we see that for most $\ell$, \(D_{\ell}\) increases with $\alpha$ at $\alpha \ge 22.0^{\circ}$  (Figure~\ref{fig:D_l-unsub}), exhibiting a prominent peak at  $\alpha \approx 50.0^{\circ}$. This can be attributed to the bright A-type source Fornax~A that is clearly visible at $(\alpha,\delta)=(50.5^{\circ},-37.2^{\circ})$. This is an extended radio source with a core and two radio lobes, and a total flux of  $750 \, \rm Jy$ at $154 \, \rm MHz$ \citep{Mckinley2015}, which makes this the brightest discrete source visible in Figure~\ref{fig:GLEAM}. Although the declination $\delta=-37.2^{\circ}$ is quite removed from $\delta=-26.7^{\circ}$, the center of the drift-scan observation, a combination of the wide MWA PB (FWHM$\approx 25^{\circ}$) and the large flux of  Fornax~A together results in Fornax~A  dominating the $D_{\ell}$  observed at low $\ell$  in the range  $\alpha \ge 22.0^{\circ}$. We see that $D_{\ell}$ peaks at $\alpha \approx 50.0^{\circ}$, which coincides with the $\alpha$ of Fornax~A. Figure~\ref{fig:GLEAM} also shows $A(\alpha) \equiv {\cal A}(\alpha,\delta=-37.2) $,   which corresponds to a section of the MWA PB centered at   $(\alpha,\delta)=(50.5^{\circ},-26.7^{\circ})$. Note that we have used an approximate analytical form of the same, as given in eq.~3 of \citet{Chatterjee2022}. We see that  at $\ell=76$,  $D_{\ell}$ closely resembles $A(\alpha)$. In addition to the peak at $\alpha \approx 50.0^{\circ}$, which corresponds to the situation when Fornax~A is located in the primary lobe, we also notice a smaller second peak around $\alpha \approx 5^{\circ}$.
This may plausibly arise from contamination by diffuse Galactic emission entering through the sidelobes of the MWA primary beam along with Fornax A. Previous studies have demonstrated that bright, large scale foregrounds such as the Galactic plane can significantly contaminate the 21 cm power spectrum, even when located outside the primary field of view, due to wide-field chromatic effects and sidelobe response \citep{Beardsley2016}. As noted earlier, these peaks are relatively less prominent at larger  $\ell$, and they cannot be made out at the largest $\ell$ values. It is possible that this is related to the tapering that is introduced (eq.~\ref{eq:vcg}) to suppress the response at large angles from the center of the PB. The tapering is expected to be more effective in regions of the $uv$ plane that have a high baseline density, which may explain why the relative contribution from Fornax~A is suppressed at large baselines or $\ell$.

\subsection{Individual PC Power Spectra}
\label{sub-sec:PS_est}

In this section, we consider the 21-cm power spectra of the $163$ individual PCs considered here.
The analysis is restricted to short baselines $6 \lambda  \le  U <  220 \lambda$, with the baselines $U < 6 \lambda$ being discarded due to the presence of strong RFIs (\citetalias{Chatterjee2024}).  The baselines correspond to the range $0.007\,  \le  \kpp \le 0.146\,\rm Mpc^{-1}$, which is divided into 20 equally spaced linear bins. The LoS Fourier modes $k_\parallel$ span the range $0\,  \le  k_\parallel \le 4.623\,\rm Mpc^{-1}$.

\begin{figure*}
    \includegraphics[width=0.9\textwidth]{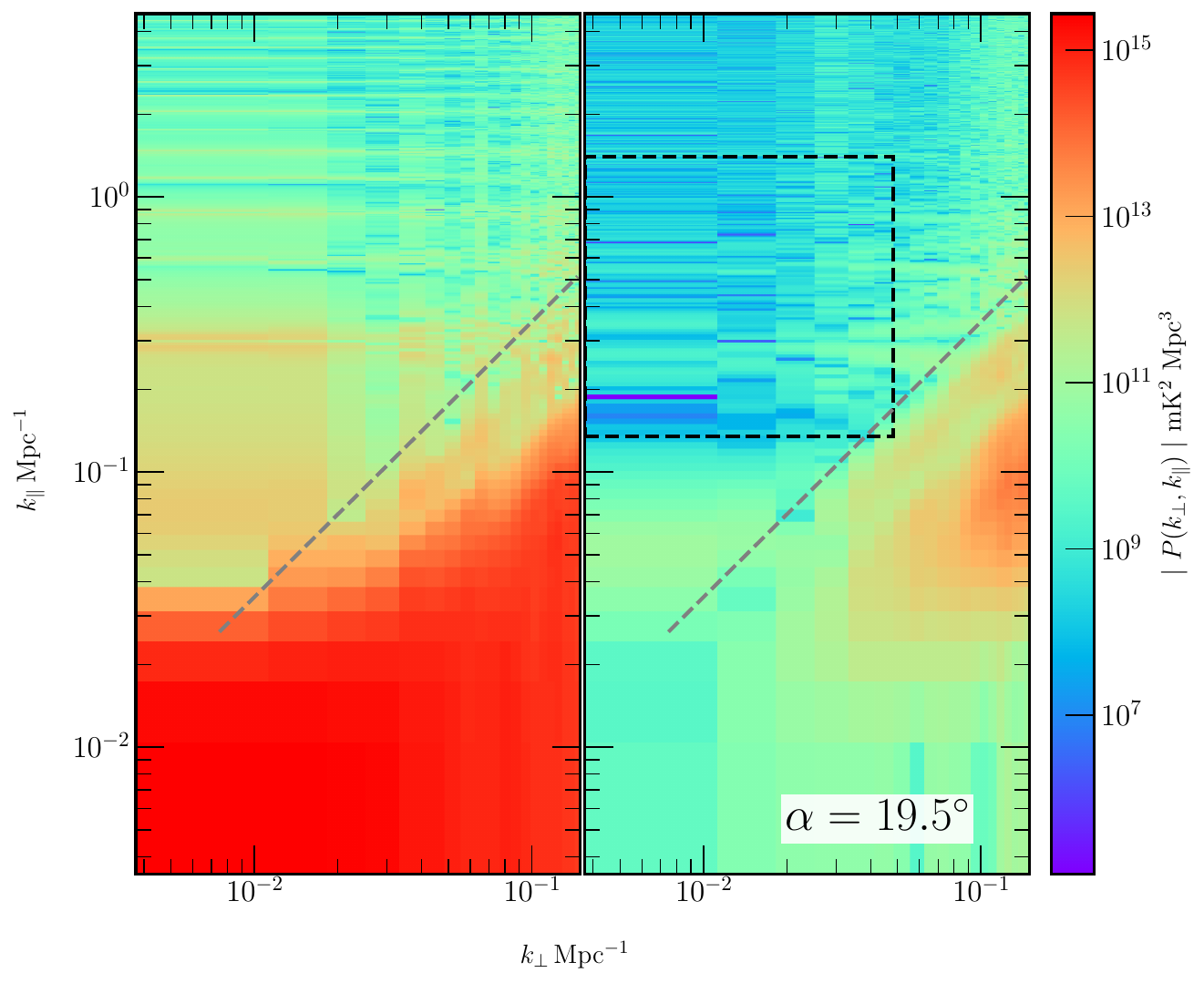}
    \caption{The estimated cylindrical power spectrum $|P(\kpp, \kpar)|$ for a PC centered at $\alpha = 19.5^{\circ}$ before (left panel) and after (right panel) applying SCF. The grey dashed line in both panels shows the theoretically predicted boundary of the foreground wedge. The region marked by the black dashed line has been identified as a suitable region for constraining the 21-cm PS.}
    \label{fig:cylps}
\end{figure*}

Figure~\ref{fig:cylps} shows the measured $|\pk|$ for a representative  PC ($\alpha = 19.5^{\circ}$) to explicitly demonstrate the effectiveness of SCF. The left panel shows the case when SCF is not applied to the data. Here, the values of $|\pk|$ are found to be $\sim 10^{15}\,{\rm mK^2 Mpc^3 }$ inside the theoretically predicted foreground wedge, whereas these are  $\sim 10^{11} - 10^{13}\,{\rm mK^2 Mpc^3 }$ in the EoR window. Although the values of $|\pk|$ are approximately $2-4$ orders of magnitude smaller in the EoR window, we observe a periodic pattern of horizontal streaks at specific $k_\parallel$ values where $|\pk|$ is relatively higher. The period of the streaks $\Delta k_{\parallel}=0.290\,{\rm Mpc}^{-1}$, corresponds to a frequency of $1.28 \, {\rm MHz}$, which matches the period of the frequency flagging pattern in the MWA data. We note here that the measured $\cl(\dnu)$ does not have missing values for any $\dnu$ due to the flagging, but the presence of strong spectral structure in the data and uneven sampling due to the missing channels leads to tiny ripples in  $\cl(\dnu)$, which then leads to these streaks in $|\pk|$ (\citetalias{Elahi2025}).
It is important to note that the amplitude of these streaks are considerably smaller than that we would obtain if we directly Fourier transform  $\V_{cg}(\nu)$ to obtain  $\V_{cg}(\tau)$ in delay space, and use this to estimate $\pk$ (\citealt{Morales2004, Parsons2009, Patwa2021}). Nevertheless, the amplitude of the streaks is much larger than the estimated noise level in the data, and it is necessary to mitigate these to put reliable constraints on the 21-cm PS.  

We have mitigated the amplitude of these streaks by using SCF on the gridded visibilities $\vcg(\nu)$ (eq.~\ref{eq:res}), and the right panel of Figure~\ref{fig:cylps} shows this case. Similar to \citetalias{Elahi2025}, here we have chosen a 2~MHz smoothing scale in SCF, which filters out the spectrally smooth components of the sky signal in the range $k_\parallel < 0.135 \, {\rm Mpc}^{-1}$.  We do not use this $k_\parallel$ range 
for PS estimation in the subsequent analysis.  Once the dominant smooth foreground components are removed, we see that the overall amplitude of $|\pk|$ as well as the amplitude of the periodic streaks is significantly reduced in the range $k_\parallel \ge 0.135 \, {\rm Mpc}^{-1}$. In the subsequent analysis we use the range $k_\parallel \ge 0.135 \, {\rm Mpc}^{-1}$ to estimate the 21-cm PS.  We have used simulations (Appendix~\ref{app:valid} of this paper and also in \citetalias{Elahi2025}) to validate that SCF does not cause a significant signal loss at $k_\parallel \ge 0.135 \, {\rm Mpc}^{-1}$.  In essence, SCF substantially reduces the foreground leakage into the EoR window. We note that SCF performs particularly well at small $\kpp$, whereas at large $k_{\perp}$ there is considerable residual foreground even after SCF. This can be attributed to the fact that the $k_{\parallel}$ value corresponding to the foreground wedge boundary exceeds the smoothing scale of $k_\parallel = 0.135 \, {\rm Mpc}^{-1}$.  In other words, at large baselines, the visibilities oscillate faster than  $2 \, {\rm MHz}$   due to baseline migration \citep{Vedantham2012, Pal2022} and SCF with a smoothing scale of  $2 \, {\rm MHz}$ is not very effective in mitigating the foreground contribution. We therefore restrict the subsequent analysis to $\kpp \le 0.045\,{\rm Mpc}^{-1}$ where SCF appears to work well.  We further notice that the region $k_\parallel > 1.399 \, {\rm Mpc}^{-1}$ is noisy, and any residual contamination from the combined effect of foregrounds and the missing channels is difficult to visualize in this range. The rest of the analysis thus focus on the region bounded by $0.007\,  \le  \kpp \le 0.045\,{\rm Mpc}^{-1}$ and $0.135\,  \le  \kpar \le 1.399\,{\rm Mpc}^{-1}$, which is demarcated by a black dashed line. In this selected region, the values of  $|\pk|$ vary between $\sim 10^{6} - 10^{11} {\rm mK}^{2}\, {\rm Mpc}^{3}$, therefore, the overall amplitude of $|\pk|$ is reduced by three orders of magnitude compared to the case where SCF is not applied.

\begin{figure}
\includegraphics[width=\columnwidth]{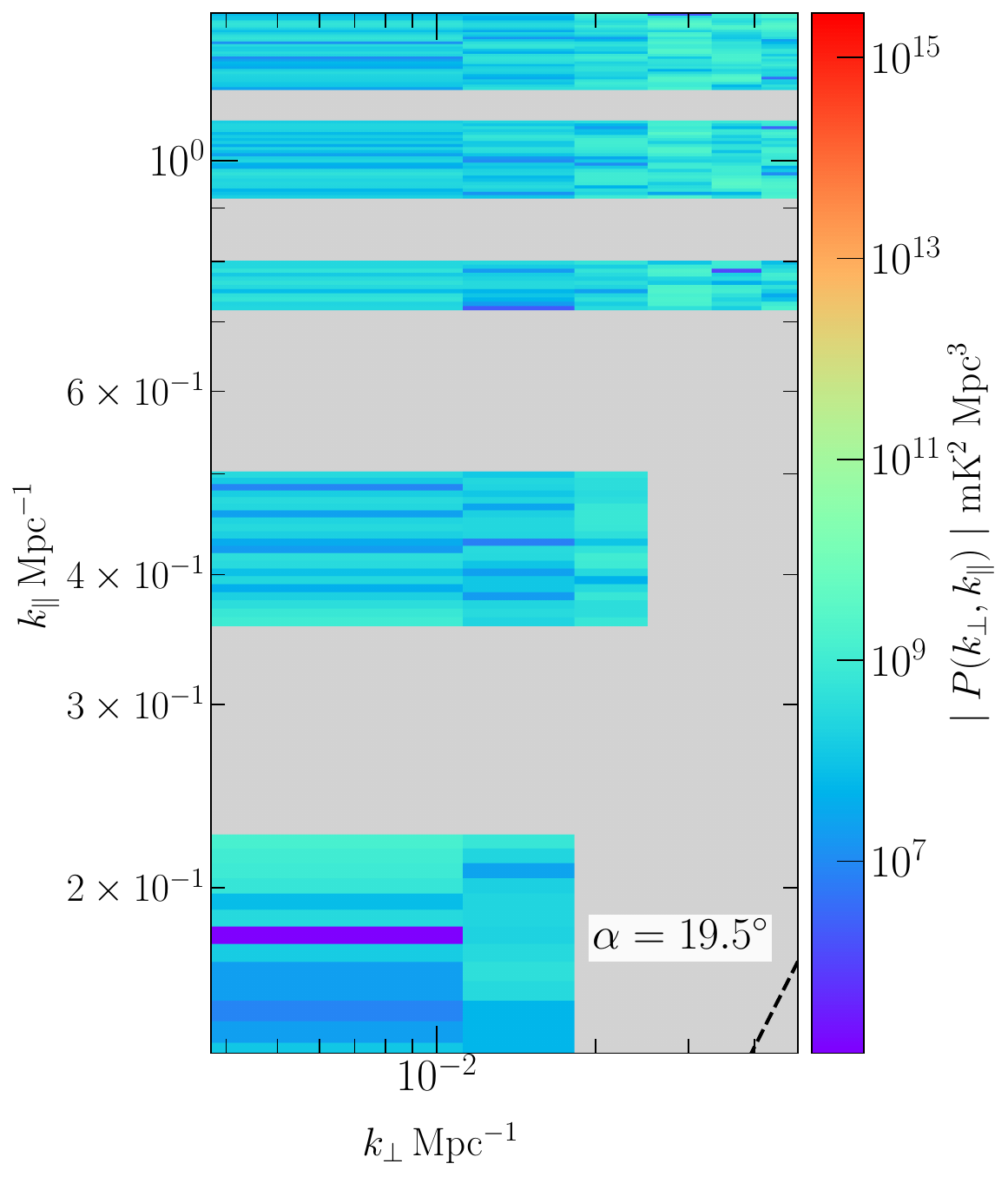}
\caption{Masking applied in the $(\kpp, \kpar)$ plane inside the selected region of Figure~\ref{fig:cylps}.}
\label{fig:cylps_mask}
\end{figure}

Figure~\ref{fig:cylps_mask} shows the region marked by the black dashed line in Figure~\ref{fig:cylps}. We notice some very faint streaks around some specific $\kpar$ values that survive SCF, and we find it to be advantageous to mask out these $\kpar$ ranges before calculating the spherically averaged PS $P(k)$. To estimate the 21-cm PS, the same mask has been applied to all the PCs considered in our analysis. The grey shaded region in Figure~\ref{fig:cylps_mask} explicitly shows this mask. We have used the $(k_{\perp},k_{\parallel})$ of the unmasked region to constrain the 21-cm signal. The unmasked $\kpar$ ranges for each $\kpp$ bin are provided in Table~\ref{table:mask}, also discussed in Appendix~\ref{app:masking}. We have validated the TTGE with masking in Appendix~\ref{app:valid}, and we found that we can recover the model 21-cm PS for the whole $k$ range considered here. However, we also found that the recovered power spectrum deviates from the input model at some intermediate $k$ bins, possibly due to the combination of SCF, the choice of a selected $k_{\perp}$ bins, and masking along $k_{\parallel}$. We find a maximum of $30.5\%$ deviation at $k=0.742{\rm\, Mpc^{-1}}$. The deviations are much smaller than the present uncertainties in noise estimation (e.g., we find an excess variance at several PCs), primary beam modelling, and calibration uncertainties. The deviations also vary with the choice of binning and masked regions. Furthermore, the final upper limits on the 21-cm PS are significantly higher than expected from standard theoretical models of the EoR, and therefore, we do not correct for these in the final results. In the subsequent sections, we have used only the unmasked $(k_{\perp},k_{\parallel})$ modes. 


\begin{figure*}
\includegraphics[width=\textwidth]{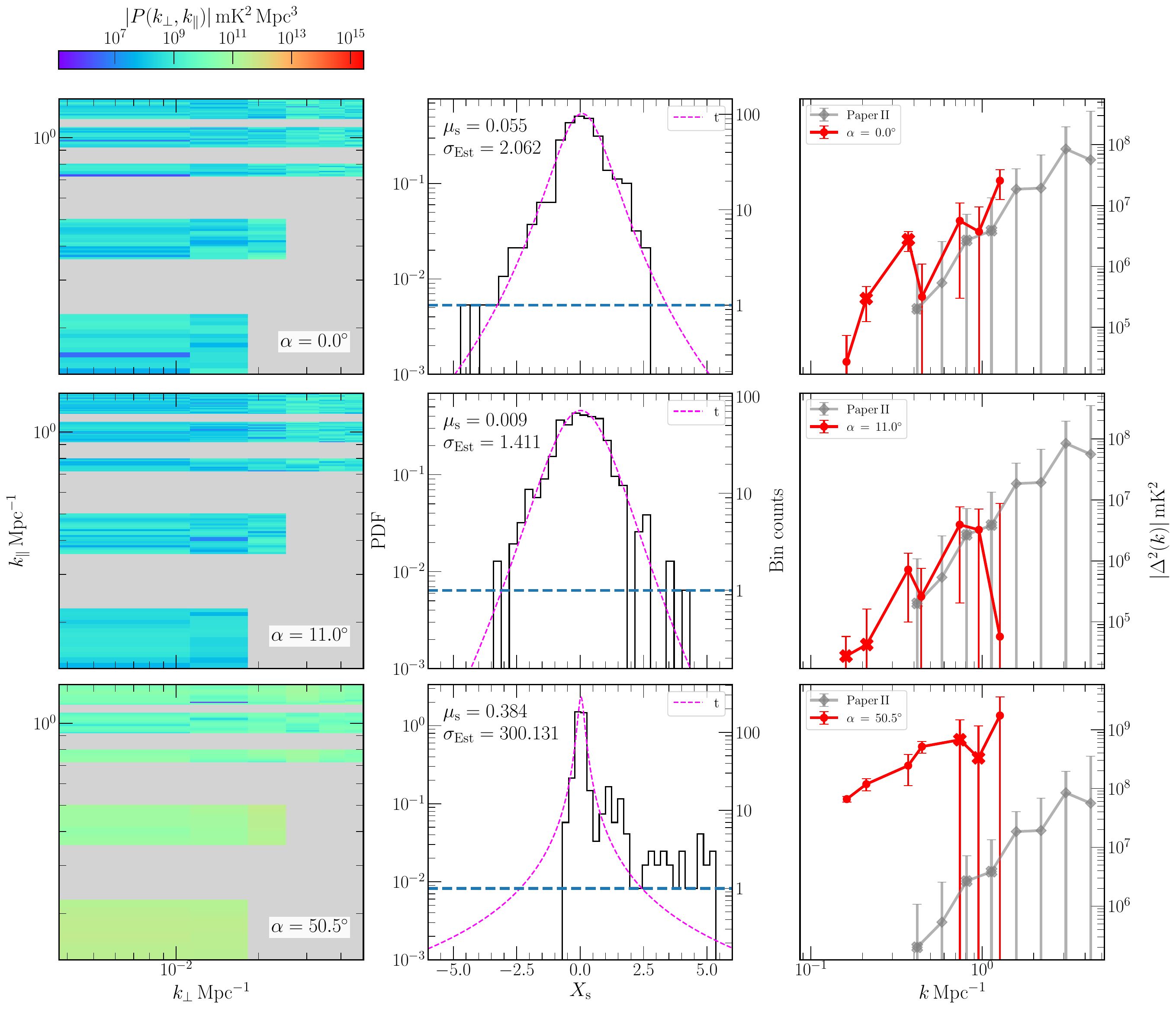}
\caption{ A comparison of three selected PCs. The left column shows $|\pk|$, where the grey shaded regions are masked to avoid contaminated modes. The middle column displays the histograms of $X_{\rm s}$, with the mean $\mu_{\rm s}$ and the estimated standard deviation $\sigma_{\rm Est}$ annotated. Here, the left and right axes indicate the PDF and the bin counts, respectively, with the blue dashed line marking a bin count of 1. The right column shows the estimated $|\Delta^{2}(k)|$ (red circles) along with the results from \citetalias{Elahi2025} (grey diamonds). The negative values are marked with crosses. The plots for all 163 PCs identical to each row are given in the Supplementary Material.}
\label{fig:CylXsp}
\end{figure*}

Figure~\ref{fig:CylXsp} shows a comparison of the estimated PS from three representative PCs at $\alpha=$ $0.0^{\circ}$ (top row), $11.0^{\circ}$ (middle row), and $50.5^{\circ}$ (bottom row). The results from all 163 PCs are given in the Supplementary Material.
The left column shows $|\pk|$, whereas the middle column shows $X_{\rm s}$ which quantifies the statistics of $\pk$. Following \citet{Pal2020}, we define $X_{\rm s}$ as :   
\begin{equation}
X_{\rm s}=\frac{\pk}{\delta P_{\rm N}^{\rm True}(\kpp, \kpar)} \,,
\label{eq:Xs}
\end{equation}
where $\delta P_{\rm N}^{\rm True}(\kpp, \kpar) = \sigma_{\rm Est} \times \delta P_{\rm N}(\kpp, \kpar)$ . Here, $ \delta P_{\rm N}(\kpp, \kpar)$ is the statistical uncertainty expected from the system noise and we use 50 realizations of simulated noise-only visibility data to estimate the same. The real and imaginary parts of the noise visibilities are assumed to be Gaussian random variables with r.m.s.  $\sigma_{\rm N} = {60 \,\rm Jy}/\sqrt{N_{\rm nights}}$  (Section~\ref{sec:data}). We estimate the PS of these noise visibility realisations identically to the data, and estimate $ \delta P_{\rm N}(\kpp, \kpar)$. The value of $\sigma_{\rm N}$ is expected to vary across the different PCs due to the variation in the sky temperature $T_{\rm sky}$ and residual systematics. To account for this variation, we correct the uncertainties with $\sigma^2_{\rm Est} = {\rm var}[\pk/\delta P_{\rm N}(\kpp, \kpar)]$, where ${\rm var}(x)$ is the variance of $x$. We expect $\delta P_{\rm N}^{\rm True}(\kpp, \kpar)$ to provide a more realistic estimate of the statistical fluctuations in the measured 21-cm PS, and we expect ${\rm var}(X_{\rm s})=1$. Ideally, if the data does not contain a significant foreground contribution, we expect $X_{\rm s}$ to follow a symmetric distribution, as well as $\sigma_{\rm Est} \approx 1$ and $\mu_s \equiv {\rm mean}(X_{\rm s}) \approx 0$.  We expect $\sigma_{\rm Est}$ and $\mu_s$ to have relatively large values if there is significant foreground contamination. The value of $\sigma_{\rm Est}$ varies with PC, and it is typically in the range $1-2$ for the PCs that can be used to meaningfully constrain the 21-cm PS.

We perform an inverse-variance weighted spherical binning using the unmasked $(\kpp,\kpar)$ modes to obtain the spherically averaged PS $P(k)$. We then obtain the mean squared brightness temperature $\Delta^{2}\left(k\right) \equiv k^{3}P(k)/2\pi^{2}$ as a function $k$, which is shown in the right column of Figure~\ref{fig:CylXsp} (red circles). 
The $2\sigma$ error bars on $\Delta^{2}\left(k\right)$ are also shown in the same color as the data points. 
The circles and crosses represent the positive and negative values of $\Delta^{2}\left(k\right)$, respectively. For reference, we also present the results from \citetalias{Elahi2025}  for the PC at $\alpha = 6.0^{\circ}$ (grey diamonds). Note that the minimum $k$ value probed in \citetalias{Elahi2025} was $0.418\,{\rm Mpc^{-1}}$ due to a conservative choice of $[k_\parallel]_F$ (Section~\ref{sec:methodology1}).
Table \ref{tab:ps_three_alphas}  presents the measured $\Delta^{2}(k)$,  along with the corresponding statistical uncertainty $\sigma(k)$, the SNR defined as $|\Delta^{2}(k)|/\sigma(k)$, and the $2\sigma$ upper limit for the positive values of $\Delta^{2}(k)$ defined as $\Delta^{2}_{\rm UL}(k)=\Delta^{2}(k)+2 \, \sigma(k)$ for 3 PCs. For the negative values of $\Delta^{2}(k)$, the upper limit is defined as $\Delta^{2}_{\rm UL}(k)=2 \, \sigma(k)$. Given these quantities, we now discuss the 3 PCs individually.

\subsubsection{PC at $\alpha=0.0^{\circ}$ (EoR0)}

The top row in Figure~\ref{fig:CylXsp} corresponds to an extensively analysed field named EoR0  \citep{Carroll2016,Trott2020, Nunhokee2025}. We find the values of $|\pk|$ to vary in the range $\sim 3\times10^{6} \, {\rm mK}^{2}\, {\rm Mpc}^{3}$ to $2 \times 10^{9} \,{\rm mK}^{2}\, {\rm Mpc}^{3}$. The distribution of $X_{\rm s}$ is symmetric around zero, and the estimated $\rm \mu_{s}$ and $\sigma_{\rm Est}$ are $0.055$ and $2.062$, respectively. The PC is relatively clean and appears to be largely noise-dominated. Similar to \citetalias{Elahi2025}, we notice that a t-distribution provides a good fit to the PDF of $X_{\rm s}$ (shown by dashed magenta lines). The values of $|\Delta^{2}\left(k\right)|$ vary in the range $4\,\times 10^{4}\, \rm mK^{2}$ to $3\,\times 10^{7}\, \rm mK^{2}$, which roughly matches with the values obtained \citetalias{Elahi2025} for the PC at $\alpha = 6.0^{\circ}$ at the overlapping $k$ range.

\subsubsection{PC at $\alpha=11.0^{\circ}$}
The middle row corresponds to the PC 
for which we obtain the lowest individual $2\sigma$ upper limit on the 21-cm brightness temperature fluctuations. Similar to EoR0 ($\alpha=0.0^{\circ}$), this PC is free from contamination. Here, the distribution of $X_{\rm s}$ is symmetric around zero with $\rm \mu_{s}=0.009$ and $\sigma_{\rm Est}=1.411$, and the t-distribution is able to fit $X_{\rm s}$. We find that $\sigma_{\rm Est}$ is slightly lower for this PC as compared to EoR0, and find this PC to be better than the EoR0 field .  
The values of $|\Delta^{2}\left(k\right)|$ vary in the range from $3\,\times 10^{4}\, \rm mK^{2}$ to $4\,\times 10^{6}\, \rm mK^{2}$, and these are very close to \citetalias{Elahi2025} for the overlapping $k$ range. We obtain the  $2\sigma$ upper limit $\rm (173.13)^{2}\,mK^{2}$ at $k=0.161\,{\rm Mpc^{-1}}$ from this PC, which is the tightest value amongst all the 163 PCs analyzed here.


\subsubsection{PC at $\alpha=50.5^{\circ}$}

The bottom row corresponds to a PC which coincides with the $\alpha$ of Fornax~A. Therefore, here $|\pk|$ attains the maximum value at $3.85\times10^{11} \, {\rm mK}^{2}\, {\rm Mpc}^{3}$, indicating that this PC is severely contaminated by Fornax~A. 
We find an asymmetric distribution of $X_{\rm s}$, with predominantly positive values. The estimated $\rm \mu_{s} = 0.384$ and $\sigma_{\rm Est}=300.131$ are much higher than other PCs. Unlike the other two PCs, here we notice that the t-distribution is not able to fit the $X_{\rm s}$ at the positive tail of the distribution. In the right column, the values of $|\Delta^{2}\left(k\right)|$ are approximately 2-3 orders of magnitude larger than those of the other two PCs shown here.

\subsubsection{Summary of the three PCs}

A brief summary of the comparison of the three PCs follows. We find that the PC at $\alpha=11.0^{\circ}$ is one of the best PCs, which provides the tightest $2\sigma$ upper limit (middle row). The distribution of $X_{\rm s}$ for this PC indicates noise-like statistics with ($\mu_{\rm s} \approx 0$, $\sigma_{\rm Est} \approx 1$). Therefore, we conclude that this is one of the best PCs for EoR observation. There are several other PCs like this one, for example, the EoR0 field (top row), which also shows a similar distribution. On the other extreme, we have several PCs that are similar to those shown in the bottom row of Figure~\ref{fig:CylXsp} at $(\alpha=50.5^{\circ})$, which are contaminated by Fornax~A and other bright sources. These PCs are not suitable for the EoR experiment. In the next sections, we study all the PCs based on these metrics. We identify the PCs that are relatively uncontaminated (referred to as `good' PCs) and combine them to tighten the constraints on the 21-cm PS.

\subsubsection{Analysis of all PCs}

\begin{figure*}
\includegraphics[width=0.95\textwidth]{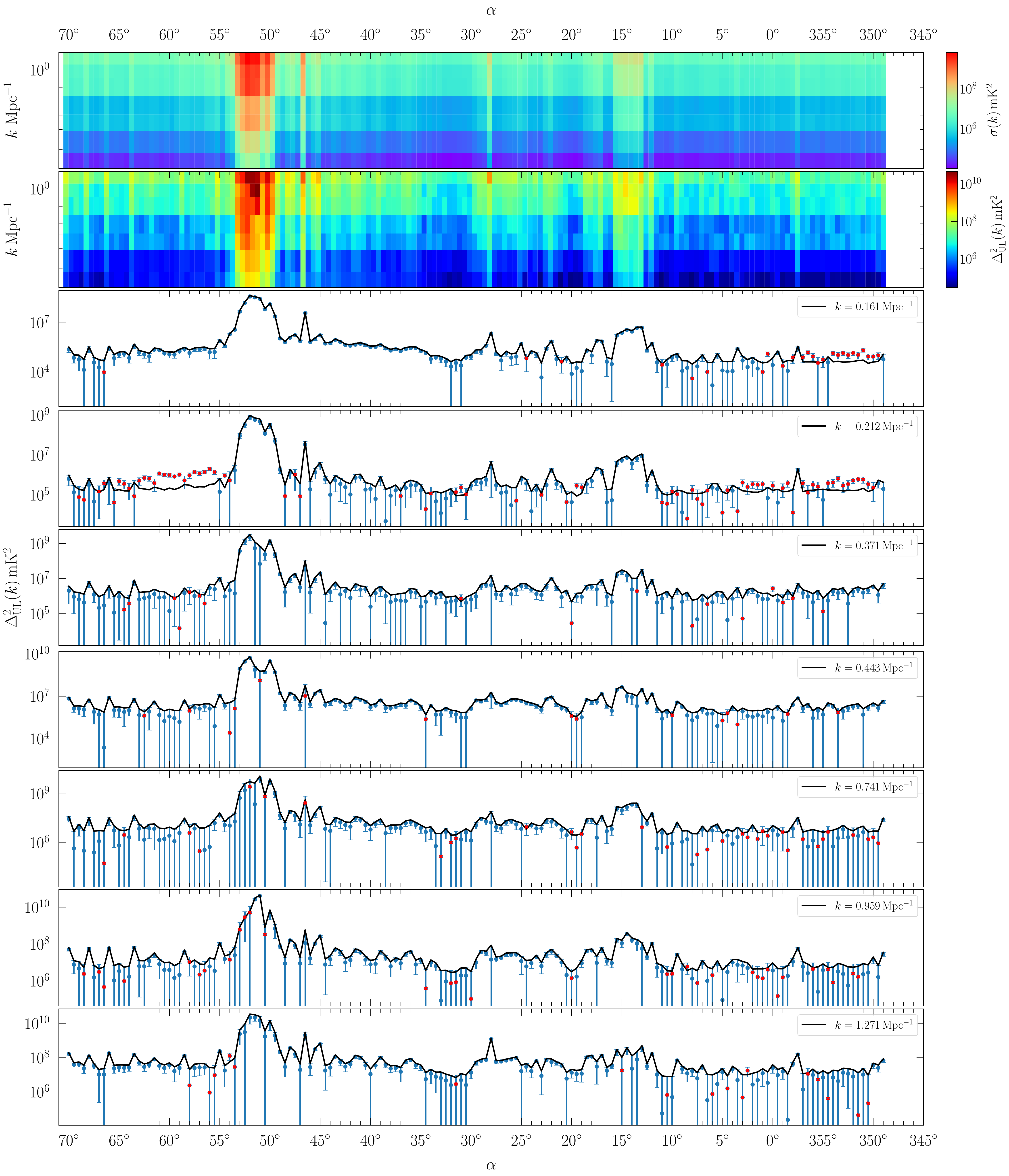}
\caption{The top two panels show the variation of $\sigma$ and $\Delta^{2}_{\rm UL}$ with $\alpha$ and $k$. The lower panels (from third onwards) show $\Delta^{2}(k)$, with $2\sigma(k)$ error bars, as a function of $\alpha$ for different $k$ bins mentioned in the panel. Negative $\Delta^{2}(k)$ values are marked with red crosses. The black curves represent $\Delta^{2}_{\rm UL}(k)$.}
\label{fig:uL-RA}
\end{figure*}

Figure~\ref{fig:uL-RA} presents a comprehensive view of the estimated $|\Delta^{2}(k)|$ for the entire $\alpha$ range covered by the drift scan observation here. The top two rows show the $\sigma(\alpha,k)$ and $\Delta^{2}_{\rm UL}(\alpha,k)$ values as a heatmap. We notice that the $\sigma$ varies across five orders of magnitude from $10^{4}\,{\rm mK^{2}}$ to $10^{9}\,{\rm mK^{2}}$ across the whole $\alpha$ and $k$ range considered here, whereas $\Delta^{2}_{\rm UL}$ varies from $10^{4}\,{\rm mK^{2}}$ to $10^{10}\,{\rm mK^{2}}$. Considering all the PCs, we observe that both $\sigma(\alpha,k)$ and $\Delta^{2}_{\rm UL}(\alpha,k)$ increase with increasing $k$, which is similar to the $k$ dependence seen in the right column of Figure~\ref{fig:CylXsp}. In general, we find that the first four $k$ bins are much cleaner (less noisy) than the last three $k$ bins. We see that both $\sigma(\alpha,k)$ and $\Delta^{2}_{\rm UL}(\alpha,k)$ show a considerable variation along $\alpha$. Near $\alpha \approx 50.0^{\circ}$, where the extended bright source Fornax~A is situated, we find approximately a two orders of magnitude increase in the values of $\sigma(\alpha,k)$ and $\Delta^{2}_{\rm UL}(\alpha,k)$. Another, somewhat smaller peak is seen at around $\alpha \approx 14.0^{\circ}$, where the measured values are slightly higher than the rest. These patterns are very similar to what we have earlier found in the measured APS (Section~\ref{sub-sec:APS}). These features can be clearly seen in the bottom panels (third row onwards), which show  $\Delta^{2}(\alpha,k)$, $2 \sigma(\alpha,k)$ and $\Delta^{2}_{\rm UL}(\alpha,k)$ as functions of $\alpha$, with each row showing the results for a fixed  $k$-bin. In these panels, the positive (negative) values of 
$\Delta^{2}\left(k\right)$ are denoted by blue circles (red crosses) with the corresponding 2$\sigma$ error bars. The black solid curve in each panel shows the $\Delta^{2}_{\rm UL}(\alpha,k)$ as a function of $\alpha$ for different $k$ values. Apart from those two peaks, we also find a few isolated values of $\alpha$ where we see small spikes in the values of  $\sigma(\alpha,k)$ and $\Delta^{2}_{\rm UL}(\alpha,k)$.


Table \ref{tab:ps_summary_alpha}  presents the values of $\Delta^{2}(k)$, $\sigma(k)$ and the associated $2\sigma$ upper limit $\Delta_{\rm UL}^{2}(k)$ for only the two PCs corresponding to $\alpha = 0.0^{\circ}$ and $11.0^{\circ}$ respectively,
and for four of the seven $k$-bins. The entire table covering all $163$ PCs and all seven $k$-bins is available in the online Supplementary Material along with a machine-readable form, also discussed in detail in Appendix~\ref{app:ps_summary}. We have further combined the good PCs incoherently to get the best upper limit of the 21-cm PS from this drift scan observation in the next section.

\renewcommand{\arraystretch}{1.4}

\begin{table*}
\centering
\caption{The measured $\Delta^2(k)$, corresponding errors $\sigma(k)$,
${\rm SNR} = |\Delta^2(k)|/\sigma(k)$, and the $2\sigma$ upper limits
$\Delta_{\rm UL}^{2}(k)$ for PCs at
$\alpha = 0.0^{\circ}, 11.0^{\circ}, 50.5^{\circ}$.}

\resizebox{\textwidth}{!}{%
\begin{tabular}{
c
@{\hspace{0.1cm}}|@{\hspace{0.1cm}}
cccc
!{\vrule width 1.4pt}
cccc
!{\vrule width 1.4pt}
cccc
}
\hline

& \multicolumn{4}{c}{$\alpha = 0.0^{\circ}$}
& \multicolumn{4}{c}{$\alpha = 11.0^{\circ}$}
& \multicolumn{4}{c}{$\alpha = 50.5^{\circ}$} \\

\hline

$k$
& $\Delta^2(k)$ & $\sigma(k)$ & SNR & $\Delta^2_{\rm UL}(k)$
& $\Delta^2(k)$ & $\sigma(k)$ & SNR & $\Delta^2_{\rm UL}(k)$
& $\Delta^2(k)$ & $\sigma(k)$ & SNR & $\Delta^2_{\rm UL}(k)$ \\

$\rm{Mpc}^{-1}$
& $\rm{mK}^2$ & $\rm{mK}^2$ &  & $\rm{mK}^2$
& $\rm{mK}^2$ & $\rm{mK}^2$ &  & $\rm{mK}^2$
& $\rm{mK}^2$ & $\rm{mK}^2$ &  & $\rm{mK}^2$ \\

\hline

$0.161$
& $(164.60)^2$ & $(151.26)^2$ & $1.18$ & $(269.91)^2$
& $-(165.39)^2$ & $(122.42)^2$ & $1.83$ & $\mathbf{(173.13)^2}$
& $(8128.49)^2$ & $(1868.80)^2$ & $18.9$ & $(8547.35)^2$ \\

$0.212$
& $-(544.25)^2$ & $(293.26)^2$ & $3.44$ & $(414.74)^2$
& $-(205.69)^2$ & $(245.10)^2$ & $0.71$ & $(346.62)^2$
& $(10886.56)^2$ & $(3717.83)^2$ & $8.58$ & $(12089.74)^2$ \\

$0.371$
& $-(1658.25)^2$ & $(704.96)^2$ & $5.53$ & $(996.96)^2$
& $(847.32)^2$ & $(556.11)^2$ & $2.32$ & $(1156.05)^2$
& $(15652.15)^2$ & $(8135.58)^2$ & $3.70$ & $(19425.88)^2$ \\

$0.443$
& $(562.69)^2$ & $(620.56)^2$ & $0.82$ & $(1042.50)^2$
& $(509.17)^2$ & $(496.74)^2$ & $1.05$ & $(867.62)^2$
& $(22690.27)^2$ & $(7566.04)^2$ & $9.00$ & $(25086.62)^2$ \\

$0.741$
& $(2368.39)^2$ & $(1629.27)^2$ & $2.11$ & $(3304.29)^2$
& $(1980.79)^2$ & $(1363.55)^2$ & $2.11$ & $(2764.43)^2$
& $-(25959.29)^2$ & $(19901.43)^2$ & $1.70$ & $(28144.88)^2$ \\

$0.959$
& $(1929.73)^2$ & $(1698.34)^2$ & $1.29$ & $(3081.00)^2$
& $(1797.20)^2$ & $(1383.99)^2$ & $1.69$ & $(2657.21)^2$
& $-(18215.96)^2$ & $(20302.24)^2$ & $0.80$ & $(28711.70)^2$ \\

$1.271$
& $(5059.95)^2$ & $(2563.36)^2$ & $3.90$ & $(6224.53)^2$
& $(238.92)^2$ & $(2081.15)^2$ & $0.01$ & $(2952.88)^2$
& $(41684.63)^2$ & $(30527.83)^2$ & $1.87$ & $(60012.54)^2$ \\

\hline
\end{tabular}
}

\label{tab:ps_three_alphas}
\end{table*}

\subsection{Incoherent combination of multiple PCs}
\label{sec:Incoherent-averaging}

In this section, we incoherently combine the estimates of the $\pk$ from multiple PCs to place tighter constraints on the EoR 21-cm PS. We find that combining data from all PCs is not meaningful, as some are clearly foreground-dominated. As discussed earlier, the PCs centered at $\alpha \sim 14.0^\circ$ and $\alpha \sim 50.0^\circ$ show strong residual foreground, whereas, the $\alpha$ range  from $358.5^{\circ}$ to $11.5^{\circ}$ is relatively cleaner compared to other parts of the sky, and the measured $\Delta^2(k)$ are consistent with noise for these PCs (Figure~\ref{fig:uL-RA}). There are also several other PCs for which the measured $\Delta^{2}(k)$ values are consistent with the noise. To identify the best PCs, for a fixed $k$-bin, we first sort the PCs in  ascending  order of $|\Delta^2(k)|$ .   We then cumulatively add the values of $\pk$  with inverse-variance weights. We find that the resulting $2 \sigma$ upper limit  $|\Delta^2_{\rm UL}(k)|$ initially goes down as we combine more PCs, indicating that the incoherently combined estimate is consistent with the expected noise predictions. However, we see that $ \Delta^2_{\rm UL}(k) $  starts to increase  beyond 
a certain point, where the resulting foregrounds become comparable to the noise, which goes down as we combine more PCs. We finally choose the PCs that combine to give the lowest value of  $\Delta^2_{\rm UL}(k)$ to obtain the tightest upper limit.   We have discussed the PC selection  further in Appendix~\ref{appdx:PC_selection_caseI}.  

We have chosen two different regions of the $(k_\perp, \kpar)$ space for sorting the PCs. In \textbf{Case~I}, we divide the low $\kpar$ region $(\kpar \leq 0.228 \, \rm Mpc^{-1})$ into two logarithmic bins, and sort the values of $|\Delta^{2}(k)|$ based on the lowest bin at $k= 0.156 {\rm\, Mpc^{-1}}$. This results in $23$ PCs, which we combine to get the tightest upper limit. In \textbf{Case~II}, we divide the high $\kpar$ region $(\kpar \geq 0.360 \, \rm Mpc^{-1})$ region into six logarithmic bins, and sort the values of $|\Delta^{2}(k)|$ based on  the lowest bin $k= 0.406 {\rm\, Mpc^{-1}}$, which results in $14$ PCs. The main motivation for choosing these two regions is to obtain the best upper limits on both the smallest and the largest scales accessible from this observation. All selected PCs for both cases are presented in Table~\ref{tab:RA_cases}. We note that $7$ of the chosen PCs are common between the two cases,  while the rest are different. The histograms of $X_{\rm s}$ after combining the chosen PCs for both \textbf{Case~I} and \textbf{Case~II} are shown in Figure~\ref{fig:X_comb}. 





\begin{figure}
\includegraphics[width=\columnwidth]{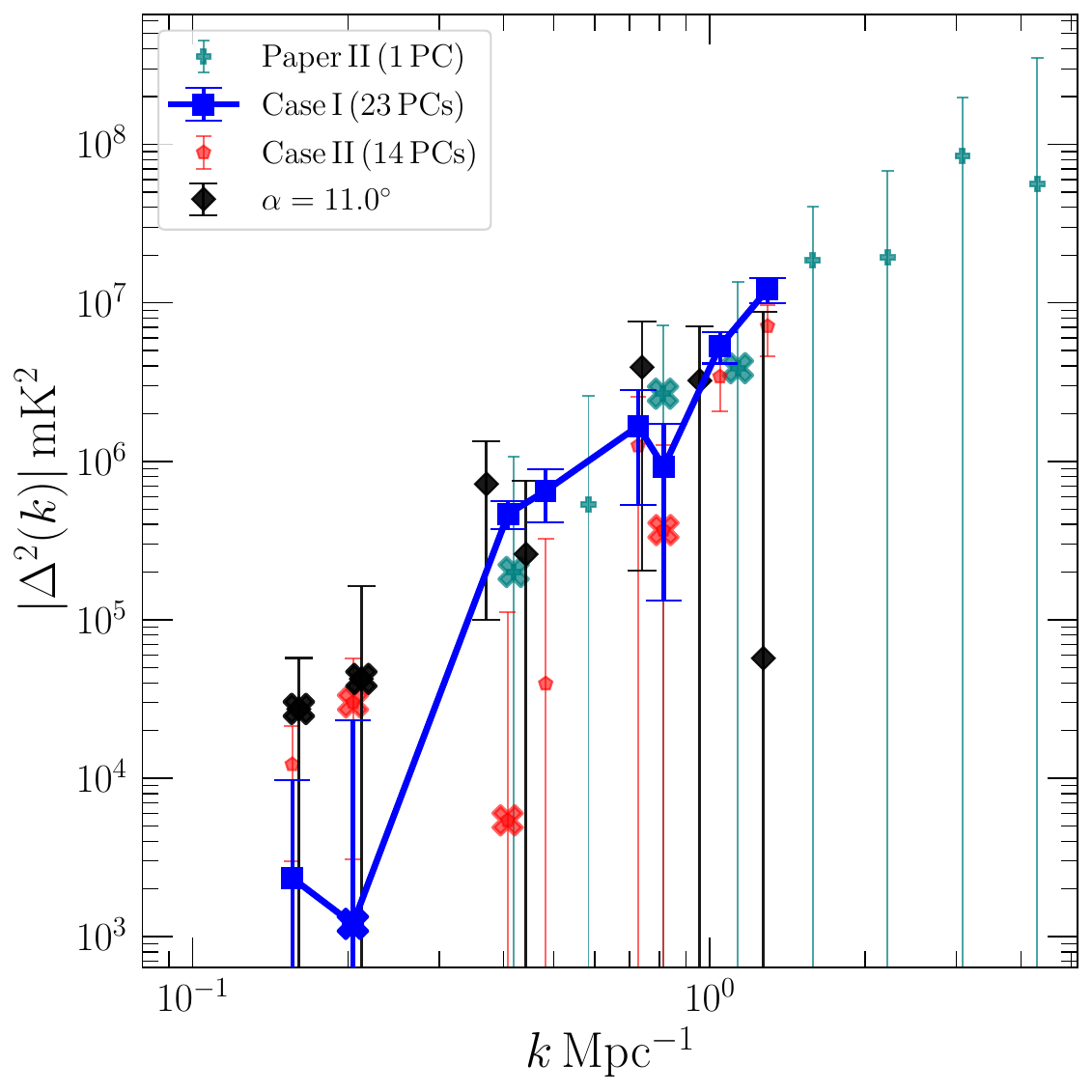}
\caption{The blue curve and the red points show the measured $|\Delta^2(k)|$ and  $2\sigma$ uncertainties for Case~I (II). The results from \citetalias{Elahi2025} are shown in teal points for reference. Black points show the same, but for the best PC centered at $\alpha=11.0^{\circ}$. Negative values of $\Delta^2(k)$ are indicated using cross (x) marks.} 
\label{fig:ps_inco}
\end{figure}

\begin{figure}
    \includegraphics[width=\columnwidth]{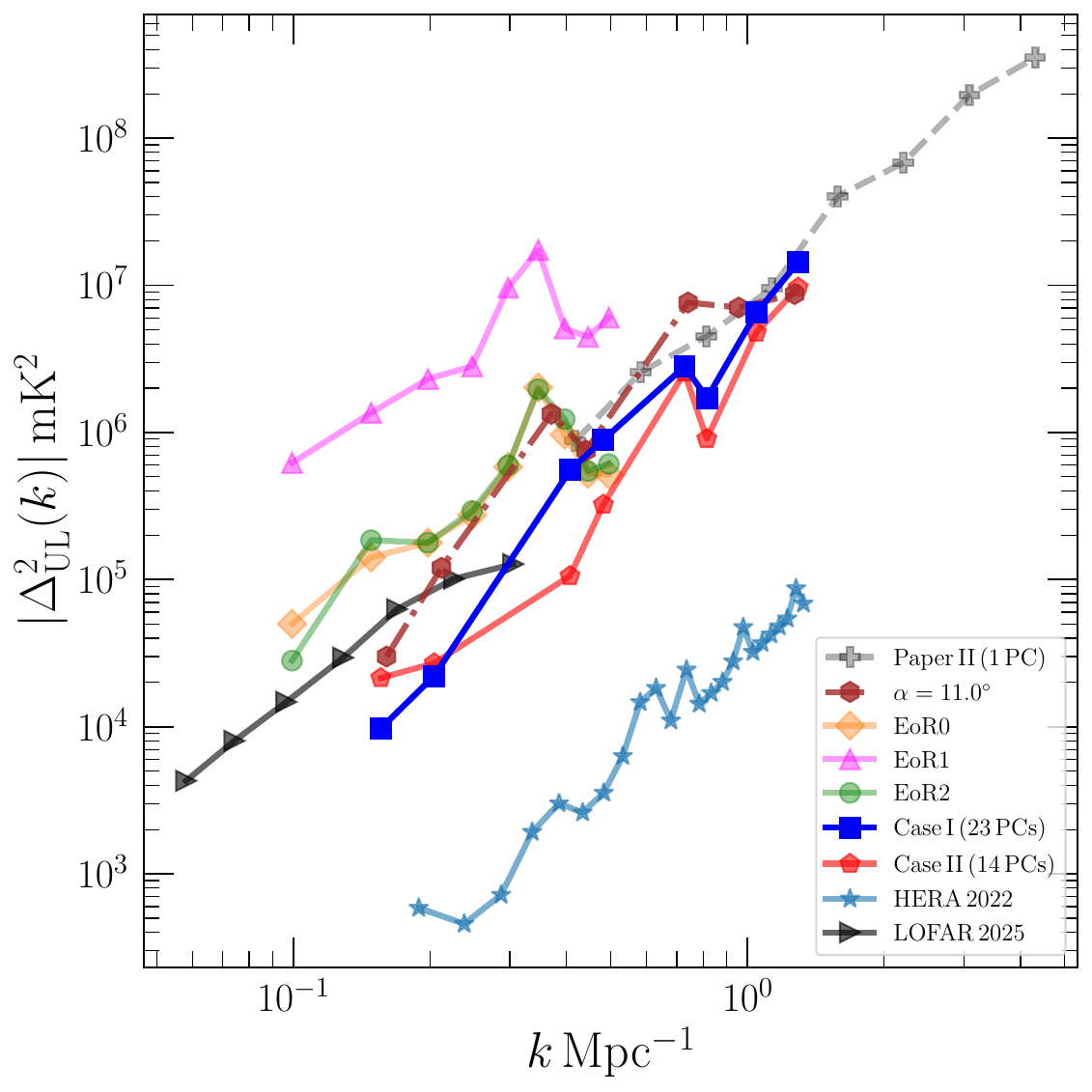}
    \caption{The 2$\sigma$ upper limits in units of $\rm mK^{2}$ . The deep blue (red) curve shows the upper limit from Case~I (Case~II). The orange, magenta, and green curves show the measured $|\Delta^{2}_{\rm UL}(k)|$ from \citealt{Trott2020} for the EoR0, EoR1, and EoR2 fields, respectively. We also include the same for \citetalias{Elahi2025} and the PC at $\alpha = 11.0^{\circ}$ with a dashed black and a dot-dashed brown curve.} 
    \label{fig:uL_inco}
\end{figure}

Figure~\ref{fig:ps_inco} shows the values of $|\Delta^2(k)|$ for both \textbf{Case~I} and \textbf{Case~II}, along with the respective $2\sigma$ error bars. These results are tabulated in Table~\ref{tab:ul_combined}, which also shows the SNR and $2 \sigma$ upper limits $\Delta^{2}_{\rm UL}(k)$ (Figure~\ref{fig:uL_inco}).  For comparison, Figure~\ref{fig:ps_inco} also shows the results from \citetalias{Elahi2025} and  our  best individual PC  at 
 $\alpha=11.0^{\circ}$ (middle row of Figure~\ref{fig:CylXsp}).  Considering \citetalias{Elahi2025}, the $k$ values cover the range  $0.418 \, {\rm Mpc}^{-1}  
 \le k \le 4.305 \, {\rm Mpc}^{-1}$, and the tightest constrain obtained there is  $\Delta^{2}_{\rm UL}(k) = (934 . 60)^2 \, {\rm mK}^2 $ at the lowest $k$ bin. The estimates there are all within the predicted $2 \sigma$ statistical fluctuations, implying that they are consistent with zero.  Considering our best single PC at $\alpha=11.0^{\circ}$, we 
note that the $k$ bins used for the individual PC analysis are slightly different from those used for the incoherent combination of PCs, and the $k$ range there spans $0.161 \, {\rm Mpc}^{-1}   \le k \le 1.271 \, {\rm Mpc}^{-1}$. The  tightest constraint obtained there is  $\Delta^{2}_{\rm UL}(k) = (173.13)^2 \, {\rm mK}^2 $ at the lowest $k$ bin.  The estimates there are mostly within the predicted $2 \sigma$ statistical fluctuations, except for 2 $k$ bins where $2 < {\rm SNR} \le 2.4$. For both \citetalias{Elahi2025} and the single PC at 
$\alpha=11.0^{\circ}$, we may interpret the results to be consistent with noise, with no indication of foreground contamination. Although many of the individual PCs are consistent with noise, the predicted noise variance reduces as $N_{\rm PC}^{-1}$ as we increase $N_{\rm PC}$, the number of PCs that we combine, and the residual foregrounds become visible in some of the combined estimate presented in  \textbf{Case~I} and \textbf{Case~II}.

We see that \textbf{Case~I} performs best at small $k$ ($ < 0.4\,\mathrm{Mpc}^{-1}$),  where the estimated $|\Delta^2(k)|$ in the two $k$ bins are both consistent with the predicted $2 \sigma$ statistical fluctuations, and the $2\sigma$ upper limits $\Delta^2_{\rm UL}(k)$  are considerably smaller than  \textbf{Case~II} and the best single PC. 
However, \textbf{Case~I} does not perform well at large $k$ ($\ge 0.4\mathrm{Mpc}^{-1}$) where the estimates  of  $|\Delta^2(k)|$  exceed the predicted $2 \sigma$ statistical fluctuations, and the $2 \sigma$ upper limits are comparable to those from  both \citetalias{Elahi2025} and our best single PC. This behaviour is expected, as \textbf{Case~I} targets the smallest $k$ bin in choosing the PCs to combine. Considering 
\textbf{Case~II}, we see that it provides the best estimates in the range  $0.4 \leq k \leq 1\,\mathrm{Mpc}^{-1}$ where   $|\Delta^2(k)|$ and $\Delta^2_{\rm UL}(k)$ are considerably smaller than \textbf{Case~I} and the two single PC analyses, and  ${\rm SNR <2}$, which indicates that the estimates are consistent with noise.  
This behaviour is expected, as \textbf{Case~II} targets the $k$ bin at $0.406 \, {\rm Mpc}^{-1}$ to choose the PCs to combine.  \textbf{Case~II} also performs reasonably well at small $k$ ($< 0.4\,\mathrm{Mpc}^{-1}$), where the results are between those for the best single PC and  \textbf{Case~I}, and we have  $2 < {\rm SNR < 3}$, which we may interpret as being marginally consistent with noise. At $k \ge  1 \, \mathrm{Mpc}^{-1}$, the estimates from 
\textbf{Case~I} and \textbf{Case~II} are comparable to those from our best single PC and \citetalias{Elahi2025}. Furthermore, we find ${\rm SNR > 5}$, which indicates that the estimates become foreground dominated when we combine multiple PCs. Note however, as discussed earlier, a large number of the individual PCs are consistent with noise.

\renewcommand{\arraystretch}{1.3}  

\begin{table*}

\centering
\caption{The measured $\Delta^2(k)$, corresponding errors $\sigma(k)$, ${\rm SNR} = |\Delta^2(k)|/\sigma(k)$, and the $2\sigma$ upper limits $\Delta_{\rm UL}^{2}(k)$ for Case I and II.}
\begin{tabular}{
        c
        @{\hspace{0.5cm}}|@{\hspace{0.5cm}}
        c c c c
        !{\vrule width 1.4pt}
        c c c c
    }
        \hline

        & \multicolumn{4}{c}{Case~I} 
        & \multicolumn{4}{c}{Case~II} \\

        \hline

        $k$ & $\Delta^2(k)$ & $\sigma(k)$ & SNR & $\Delta_{\rm UL}^{2}(k)$
            & $\Delta^2(k)$ & $\sigma(k)$ & SNR & $\Delta_{\rm UL}^{2}(k)$ \\

        $\rm{Mpc}^{-1}$ & $\rm{mK}^2$ & $\rm{mK}^2$ & & $\rm{mK}^2$
            & $\rm{mK}^2$ & $\rm{mK}^2$ & & $\rm{mK}^2$ \\

        \hline

        $0.156$ & $(48.43)^2$ & $(60.79)^2$ & $0.63$ & $\mathbf{(98.67)^2}$
                 & $(110.52)^2$ & $(67.95)^2$ & $2.64$ & $(146.45)^2$ \\

        $0.204$ & $-(34.69)^2$ & $(105.09)^2$ & $0.10$ & $(148.63)^2$
                 & $-(173.43)^2$ & $(116.23)^2$ & $2.22$ & $(164.37)^2$ \\

        $0.406$ & $(681.60)^2$ & $(215.31)^2$ & $10.02$ & $(746.52)^2$
                 & $-(73.64)^2$ & $(230.04)^2$ & $0.10$ & $(325.32)^2$ \\

        $0.482$ & $(806.51)^2$ & $(346.45)^2$ & $5.42$ & $(943.66)^2$
                 & $(198.72)^2$ & $(376.80)^2$ & $0.28$ & $(568.73)^2$ \\

        $0.728$ & $(1290.58)^2$ & $(754.46)^2$ & $2.93$ & $(1674.52)^2$
                 & $(1119.48)^2$ & $(806.32)^2$ & $1.93$ & $(1597.98)^2$ \\

        $0.817$ & $(960.24)^2$ & $(628.71)^2$ & $2.33$ & $(1308.67)^2$
                 & $-(606.57)^2$ & $(672.25)^2$ & $0.81$ & $(950.71)^2$ \\

        $1.048$ & $(2307.45)^2$ & $(769.43)^2$ & $8.99$ & $(2551.15)^2$
                 & $(1848.90)^2$ & $(825.16)^2$ & $5.02$ & $(2186.37)^2$ \\

        $1.294$ & $(3483.14)^2$ & $(1047.61)^2$ & $11.05$ & $(3785.13)^2$
                 & $(2668.65)^2$ & $(1126.08)^2$ & $5.62$ & $(3107.71)^2$ \\

        \hline
    \end{tabular}
    \label{tab:ul_combined}
\end{table*}

Figure~\ref{fig:uL_inco} shows a comparison of the upper limits obtained from this paper with the results from \citetalias{Elahi2025}, and also those from earlier deep MWA tracking observation at redshift $z\,= 8.2$ \citep{Trott2020} for three different fields, namely EoR0 $[(\alpha, \delta) = (0.0^{\circ}, -26.7^{\circ})]$, EoR1 $ (60.0^{\circ}, -26.7^{\circ})$ and EoR2 $ (154.5^{\circ}, -10.0^{\circ})$. We see that, across the entire overlapping $ k$-region $k < 0.5 \,\mathrm{Mpc}^{-1}$, our upper limits are either comparable to or lower than those from the earlier MWA results by \citet{Trott2020}.  We note that the  EoR1 upper limits are considerably larger than EoR0 and EoR2, and we do not include them in the subsequent discussion.  Considering \citetalias{Elahi2025}, we see that only the smallest $k$ bin at  $k \sim  0.4 \, \mathrm{Mpc}^{-1}$  overlaps, and in this bin the upper limits are consistent with EoR0 and EoR2. 
The best upper limits from \citet{Trott2020} are  $(223.5)^{2}\,{\rm mK}^{2}$ and $(167.7)^{2}\,{\rm mK}^{2}$ respectively for EoR0 and EoR2, at their smallest $k$ bin $k = 0.099\,{\rm Mpc^{-1}}$, which is  outside our $k$ range. Considering our best single PC, 
 we have the best upper limit $\Delta^2_{\rm UL}(k)=(173.13)^2 \, {\rm mK}^2$ at the smallest bin $k=0.161 \, {\rm Mpc}^{-1}$, which is between the best upper limits from EoR0 and EoR2. 
 In fact, we have a total of $34$ individual PCs (see the Supplementary Material) for which the best upper limits are between those from EoR0 and EoR2. Considering the entire overlapping $k$ range, the upper limits from our best single PC  are lower than those from EoR0 and EoR2. Considering \textbf{Case~I}, the upper limits are well below those from EoR0 and EoR2 at $k <0.4 \, {\rm Mpc}^{-1}$, however, it exceeds EoR0 and EoR2 at the bin at $k =0.482 \, {\rm Mpc}^{-1}$. For \textbf{Case~II}, the upper limits are well below those from EoR0 and EoR2 through the entire $k$ range.

In this work, the best $2\sigma$ upper limit is found to be $(98.67)^{2}\,{\rm mK}^{2}$ at $k = 0.156\,{\rm Mpc}^{-1}$ for Case~I. This is the tightest upper limit to date obtained from MWA data at the redshift $z=8.2$. This upper limit is almost 90 times tighter than that of \citetalias{Elahi2025}, which was restricted to a larger $k$ range. 
 After incoherent averaging of $N_{\rm PC}=23$ PCs, the upper limit  $\Delta_{\rm UL}^{2}(k)$  improves by a factor of 3  compared with the best individual  PC at $\alpha=11.0^{\circ}$.
This improvement is smaller than the expected  $\sqrt{N_{\rm PC}}=4.8$ reduction, possibly due to the presence of residual foregrounds.  At $k = 0.148\,{\rm Mpc}^{-1}$, \cite{Trott2020} quote an upper limit of  $(376.3)^{2} \,{\rm mK}^{2}$, whereas, at a similar $k=0.156\,{\rm Mpc}^{-1}$ we find the limit of $(98.67)^{2}\,{\rm mK}^{2}$ which is 15 times lower. Our best upper limit at $k = 0.156\,{\rm Mpc}^{-1}$ is $\sim3$ times smaller than the best upper limit from EoR2 at $k = 0.099\,{\rm Mpc^{-1}}$. In Figure~\ref{fig:uL_inco}, we also compare our upper limits with that from LOFAR \citep{Mertens2025} and HERA \citep{HERA2023} at $z = 8.3$ and $7.9$ respectively. We note that \cite{Mertens2025} were able to probe much larger scales at $k = 0.0581\,{\rm Mpc^{-1}}$ where their upper limit of $\Delta^{2}_{\rm UL} = (65.5)^{2}\,{\rm mK^{2}}$ is $\approx 2$ times lower than our lowest $k$ bin. However, within the overlapping $k$ range, their upper limits are higher than ours; for example, in our two lowest 
$k$ bins, our upper limits are approximately 6 and 5 times lower, respectively. Furthermore, our upper limit is $\approx 21$ times higher than that found by \cite{HERA2023} for Field C at $z=7.9$ using the HERA telescope at a value of $\Delta^{2}_{\rm UL} = (21.4)^{2}\,{\rm mK^{2}}$ for $k = 0.238\,{\rm Mpc^{-1}}$. Although our upper limit is $\approx 3$ orders of magnitude larger than the expected EoR 21-cm signal \citep{Mondal2017}, we can still use this result to rule out some exotic reionization models. For example, we can constrain the ionisation and thermal state of the IGM along with constraints on the strength of a possible excess radio background at this redshift \citep{Ghara2021}.

\section{Summary and Conclusions}
\label{sec:summary}

We analyze zenith pointing $(\delta=-26.7^{\circ})$ drift scan data from the Phase II compact configuration of the MWA, spanning  $\sim81.0^{\circ}$ along $\alpha$ in the range $(349.0^{\circ} \leq \alpha \leq 70.0^{\circ})$, sampled at an interval of $0.5^{\circ}$. We have used the TGE to measure the angular power spectrum $C_{\ell}$   $(34 \le \ell \le 1428)$, 
as a function of $\alpha$, covering all $163$ PCs, at the central frequency $\nu_c=154.2 \, {\rm MHz}$. 
We study the variation of $D_{\ell}=\ell(\ell+1) C_{\ell}/(2 \pi)$, the 2D mean-squared brightness temperature fluctuations as a function of both $\alpha$ and $\ell$.
The most prominent feature we found from the aforementioned study is an enhancement in the values of $D_{\ell}$ that starts at $\alpha \ge 22^{\circ}$, peaks at $\alpha \approx 50^{\circ}$, which corresponds to the passage of the bright extended source Fornax~A  through the main lobe of the MWA PB. We also note a minor enhancement at $\alpha \approx 5^{\circ}$, which possibly corresponds to the passage of Fornax~A through the first side lobe along with contributions from the Galactic plane emission.  
The enhancements are most prominent at low $\ell$, and they are not seen in the largest $\ell$ bin. 
At $\alpha < 22^{\circ}$, we find $D_{\rm \ell} \propto \rm \ell^2$ for $\ell \ge 200$, which indicates that the measured $D_{\ell}$ is dominated by the Poisson fluctuations due to bright point sources. In this $\alpha$ range, we find $D_{\ell} \approx 5 \times 10^8 \, {\rm mK}^2$ at $\ell \approx 200$, whereas the value of $D_{\ell}$ is one order of magnitude larger for $\alpha=50.5^{\circ}$ which is the RA of Fornax~A where the large enhancement peaks.   Although Fornax~A  $(50.5^{\circ},-37.2^{\circ})$ is $\sim 10^{\circ}$ to the south of the zenith, it severely contaminates the measured sky brightness temperature fluctuations over a large range $\alpha \ge 22^{\circ}$ of the drift scan observations analyzed here. 

We have  estimated $\pk$ the cylindrical PS and $P(k)$ the spherical PS for each of the 163 PCs separately. To mitigate foreground leakage into the EoR window 
due to the strong spectral dependence of the foregrounds and the periodic pattern of missing frequency channels present in the MWA data, we have applied Smooth Component Filtering (SCF), which subtracts out the spectrally smooth component from the gridded visibility data. Furthermore, we have used the TTGE to measure the Multi-frequency Angular Power Spectrum (MAPS) $C_{\ell}(\Delta \nu)$, which has no missing values of $\Delta \nu$, despite the individual frequency channels being flagged. We estimate $\pk$ by performing 
a Fourier transform of $C_{\ell}(\Delta \nu)$ along $\Delta \nu$. This approach to $\pk$ estimations considerably reduces foreground leakage.  Using simulations (Appendix~\ref{app:valid}), and a close inspection of the cylindrical PS after applying SCF, we selected the region bounded by $0.007\,  \le  \kpp \le 0.045\,{\rm Mpc}^{-1}$ and $0.135\,  \le  \kpar \le 1.399\,{\rm Mpc}^{-1}$ to estimate the EoR 21-cm PS. To avoid any low-level contamination, we further mask out certain $(\kpp,\kpar)$ ranges
which correspond to the predicted positions of the periodic spike pattern caused by the missing frequency channels. We see that the unmasked region is relatively clean and the power here varies in the range  $10^{6} \sim 10^{11} \, {\rm mK^{2}\,{\rm Mpc^{3}}}$. We have validated the TTGE with masking and found that we can recover the 21-cm PS across the entire $k$ range considered here. Note that the choice of the masked region allows us 
to access lower $k$ values than what was possible in our earlier works \citetalias{Chatterjee2024} and \citetalias{Elahi2025}.

We applied the above methodology to all 163 PCs in the drift scan dataset. Further we note that our best $2 \sigma$ upper limit from an individual PC comes from $\alpha = 11.0^{\circ}$ where we obtain  $\Delta^{2}_{\rm UL}(k) = (173.13)^{2}\,{\rm mK^{2}}$ at the smallest $k$ bin with $k = 0.161\,{\rm Mpc^{-1}}$. For this PC, the estimates of  $\Delta^{2}(k)$ in all the $k$ bins are consistent with noise with no trace of foreground contamination. 
We further note that the PC coinciding with that of Fornax~A has one of the highest values of the $2\sigma$ upper limit $\Delta^{2}_{\rm UL}(k) = (8547.35)^{2}\,{\rm mK^{2}}$ at the smallest $k$ bin, which is three orders of magnitude higher that our best upper limit. For this PC, the estimates 
of  $\Delta^{2}(k)$ are all severely contaminated by Fornax~A. 


Similar to our study of the variation of $D_{\ell}$, which gave us information about the foregrounds, we also studied how the 3D brightness temperature fluctuations $\Delta^{2}(k)$ behave for the whole drift scan duration. We find that $\Delta^{2}(k)$ and $\Delta^{2}_{\rm UL}(k)$ exhibit a peak of width $\Delta \alpha=\pm 5^{\circ}$ around $\alpha \approx 50^{\circ}$, where the values increase by more than three orders of magnitude. This corresponds to the enhanced foreground level as Fornax~A traverses the main lobe of the PB. We also notice a smaller peak around $\alpha \approx 14.0^{\circ}$, and several more isolated peaks. The exact cause for these peaks is not understood at present. We further note that the region $358.5 ^{\circ} \leq \alpha \leq 11.5^{\circ}$ appears to be relatively clean in terms of the value of $\Delta^{2}(k)$ and $\Delta^{2}_{\rm UL}(k)$.
Our findings are in line with \cite{Jong2025}, who found that the sky around $\alpha=0.0^{\circ}$ is suitable for future EoR observations.

In order to place tighter limits on the EoR 21-cm signal,  we incoherently combine the results from multiple PCs. We consider two separate strategies to choose the PCs to combine, namely \textbf{Case~I}  and \textbf{Case~II}.  The tightest upper limit comes from \textbf{Case~I}, which combines $N_{\rm PC}=23$ PCs to obtain $\Delta^{2}_{\rm UL}(k)=(98.67)^2 \rm \,mK^{2}$ at $k = 0.156\rm\, Mpc^{-1}$ (Figure~\ref{fig:uL_inco}).  Currently, this is the best upper limit on the $z=8.2$ EoR 21-cm signal from MWA. This is $\sim 90$ times better than the best upper limit at  $k = 0.418\,{\rm Mpc^{-1}}$ from \citetalias{Elahi2025},  $\sim 3$ times better than the best upper limit at $k=0.161\rm\, Mpc^{-1}$ from a single PC in this work (Table \ref{tab:ps_three_alphas}), and also  $\sim 3$ times better than the best  $z \approx 8.2$ upper limit at $k=0.099\rm\, Mpc^{-1}$ 
from \cite{Trott2020}. We note that our best upper limit is still $\approx 2$ times higher than the LOFAR result at $k = 0.0581 \, {\rm Mpc^{-1}}$ \citep{Mertens2025}, and $\approx 21$ times higher than the upper limit obtained from the HERA at $k = 0.238\,{\rm Mpc^{-1}}$ \citep{HERA2023} at similar redshift.
Although our upper limit remain $\approx 3$ orders of magnitude higher than the theoretically expected 21-cm signal \citep{Mondal2017}, the improvement from incoherent addition of noise-limited upper limits represents considerable progress. Coherent addition of the signal from different PCs holds the promise to push the upper limits even lower. We also plan to apply some of the techniques presented here to deep tracking observations. 
The upcoming SKA-Low will observe the same region of sky, and our study is expected to provide useful guidance for selecting target fields to detect the EoR 21-cm signal.  

\section*{Acknowledgements}
AE and S. Choudhuri acknowledges the support from the CoE research grant, IIT Madras. S. Choudhuri would also like to acknowledge SERB-Start-up Research Grant (SRG) for providing financial support. S. Choudhuri would also like to acknowledge SERB-MATRICS for providing financial support. S. Chatterjee acknowledges support from the South African National Research Foundation (Grant No. 84156) and the Inter-University Institute for Data Intensive Astronomy (IDIA). IDIA is a partnership of the University of Cape Town, the University of Pretoria and the University of the Western Cape. 

\section*{Data Availability} 
The data sets were derived from sources in the public domain (the MWA Data Archive: project ID G0031) at \url{https://asvo.mwatelescope.org/}.



\bibliographystyle{mnras}
\bibliography{mylist} 

@ARTICLE{Gill2025b,
       author = {{Gill}, Sukhdeep Singh and {Bharadwaj}, Somnath},
        title = "{A Visibility-based 21 cm Bispectrum Estimator for Radio-interferometric Data}",
      journal = {\apj},
         year = 2025,
        month = dec,
       volume = {995},
       number = {2},
          eid = {175},
        pages = {175},
          doi = {10.3847/1538-4357/ae160c},
archivePrefix = {arXiv},
       eprint = {2506.10526},
 primaryClass = {astro-ph.CO},
       adsurl = {https://ui.adsabs.harvard.edu/abs/2025ApJ...995..175G}
}

@ARTICLE{Gill2025a,
       author = {{Gill}, Sukhdeep Singh and {Bharadwaj}, Somnath and {Ali}, Sk. Saiyad and {Elahi}, Khandakar Md Asif},
        title = "{A Visibility-based Angular Bispectrum Estimator for Radio-interferometric Data}",
      journal = {\apj},
         year = 2025,
        month = jan,
       volume = {979},
       number = {1},
          eid = {25},
        pages = {25},
          doi = {10.3847/1538-4357/ad9b20},
archivePrefix = {arXiv},
       eprint = {2412.02246},
 primaryClass = {astro-ph.CO},
       adsurl = {https://ui.adsabs.harvard.edu/abs/2025ApJ...979...25G}
}

@ARTICLE{Gill2025,
       author = {{Gill}, Sukhdeep Singh and {Bharadwaj}, Somnath and {Elahi}, Khandakar Md Asif and {Sethi}, Shiv K. and {Patwa}, Akash Kumar},
        title = "{The Epoch of Reionization 21 cm Bispectrum at z = 8.2 from MWA Data. I. Foregrounds and Preliminary Upper Limits}",
      journal = {\apj},
         year = 2025,
        month = nov,
       volume = {993},
       number = {1},
          eid = {56},
        pages = {56},
          doi = {10.3847/1538-4357/ae0463},
archivePrefix = {arXiv},
       eprint = {2507.04964},
 primaryClass = {astro-ph.CO},
       adsurl = {https://ui.adsabs.harvard.edu/abs/2025ApJ...993...56G}
}

@ARTICLE{Gill2024,
       author = {{Gill}, Sukhdeep Singh and {Pramanick}, Suman and {Bharadwaj}, Somnath and {Shaw}, Abinash Kumar and {Majumdar}, Suman},
        title = "{The monopole and quadrupole moments of the epoch of reionization (EoR) 21-cm bispectrum}",
      journal = {\mnras},
         year = 2024,
        month = jan,
       volume = {527},
       number = {1},
        pages = {1135-1140},
          doi = {10.1093/mnras/stad3273},
archivePrefix = {arXiv},
       eprint = {2310.15579},
 primaryClass = {astro-ph.CO},
       adsurl = {https://ui.adsabs.harvard.edu/abs/2024MNRAS.527.1135G}
}

@ARTICLE{Majumdar2018,
       author = {{Majumdar}, Suman and {Pritchard}, Jonathan R. and {Mondal}, Rajesh and {Watkinson}, Catherine A. and {Bharadwaj}, Somnath and {Mellema}, Garrelt},
        title = "{Quantifying the non-Gaussianity in the EoR 21-cm signal through bispectrum}",
      journal = {\mnras},
         year = 2018,
        month = may,
       volume = {476},
       number = {3},
        pages = {4007-4024},
          doi = {10.1093/mnras/sty535},
archivePrefix = {arXiv},
       eprint = {1708.08458},
 primaryClass = {astro-ph.CO},
       adsurl = {https://ui.adsabs.harvard.edu/abs/2018MNRAS.476.4007M}
}

@INPROCEEDINGS{casa07,
       author = {{McMullin}, J.~P. and {Waters}, B. and {Schiebel}, D. and {Young}, W. and {Golap}, K.},
        title = "{CASA Architecture and Applications}",
    booktitle = {Astronomical Data Analysis Software and Systems XVI},
         year = 2007,
       editor = {{Shaw}, R.~A. and {Hill}, F. and {Bell}, D.~J.},
       series = {Astronomical Society of the Pacific Conference Series},
       volume = {376},
        month = oct,
        pages = {127},
       adsurl = {https://ui.adsabs.harvard.edu/abs/2007ASPC..376..127M}
}

@ARTICLE{Prabu2015,
       author = {{Prabu}, Thiagaraj and {Srivani}, K.~S. and {Roshi}, D. Anish and {Kamini}, P.~A. and {Madhavi}, S. and {Emrich}, David and {Crosse}, Brian and {Williams}, Andrew J. and {Waterson}, Mark and {Deshpande}, Avinash A. and {Shankar}, N. Udaya and {Subrahmanyan}, Ravi and {Briggs}, Frank H. and {Goeke}, Robert F. and {Tingay}, Steven J. and {Johnston-Hollitt}, Melanie and {R}, Gopalakrishna M. and {Morgan}, Edward H. and {Pathikulangara}, Joseph and {Bunton}, John D. and {Hampson}, Grant and {Williams}, Christopher and {Ord}, Stephen M. and {Wayth}, Randall B. and {Kumar}, Deepak and {Morales}, Miguel F. and {deSouza}, Ludi and {Kratzenberg}, Eric and {Pallot}, D. and {McWhirter}, Russell and {Hazelton}, Bryna J. and {Arcus}, Wayne and {Barnes}, David G. and {Bernardi}, Gianni and {Booler}, T. and {Bowman}, Judd D. and {Cappallo}, Roger J. and {Corey}, Brian E. and {Greenhill}, Lincoln J. and {Herne}, David and {Hewitt}, Jacqueline N. and {Kaplan}, David L. and {Kasper}, Justin C. and {Kincaid}, Barton B. and {Koenig}, Ronald and {Lonsdale}, Colin J. and {Lynch}, Mervyn J. and {Mitchell}, Daniel A. and {Oberoi}, Divya and {Remillard}, Ronald A. and {Rogers}, Alan E. and {Salah}, Joseph E. and {Sault}, Robert J. and {Stevens}, Jamie B. and {Tremblay}, S. and {Webster}, Rachel L. and {Whitney}, Alan R. and {Wyithe}, Stuart B.},
        title = "{A digital-receiver for the MurchisonWidefield Array}",
      journal = {Experimental Astronomy},
         year = 2015,
        month = mar,
       volume = {39},
       number = {1},
        pages = {73-93},
          doi = {10.1007/s10686-015-9444-3},
archivePrefix = {arXiv},
       eprint = {1502.05015},
 primaryClass = {astro-ph.IM},
       adsurl = {https://ui.adsabs.harvard.edu/abs/2015ExA....39...73P}
}

@ARTICLE{Ali2008,
   author = {{Ali}, S.~S. and {Bharadwaj}, S. and {Chengalur}, J.~N.},
    title = "{Foregrounds for redshifted 21-cm studies of reionization: Giant Meter Wave Radio Telescope 153-MHz observations}",
  journal = {\mnras},
archivePrefix = "arXiv",
   eprint = {0801.2424},
     year = 2008,
    month = apr,
   volume = 385,
    pages = {2166-2174},
      doi = {10.1111/j.1365-2966.2008.12984.x},
   adsurl = {http://adsabs.harvard.edu/abs/2008MNRAS.385.2166A}
}

@ARTICLE{Beardsley2016,
   author = {{Beardsley}, A.~P. and {Hazelton}, B.~J. and {Sullivan}, I.~S. and 
	{Carroll}, P. and {Barry}, N. and {Rahimi}, M. and {Pindor}, B. and 
	{Trott}, C.~M. and {Line}, J. and {Jacobs}, D.~C. and {Morales}, M.~F. and 
	{Pober}, J.~C. and {Bernardi}, G. and {Bowman}, J.~D. and {Busch}, M.~P. and 
	{Briggs}, F. and {Cappallo}, R.~J. and {Corey}, B.~E. and {de Oliveira-Costa}, A. and 
	{Dillon}, J.~S. and {Emrich}, D. and {Ewall-Wice}, A. and {Feng}, L. and 
	{Gaensler}, B.~M. and {Goeke}, R. and {Greenhill}, L.~J. and 
	{Hewitt}, J.~N. and {Hurley-Walker}, N. and {Johnston-Hollitt}, M. and 
	{Kaplan}, D.~L. and {Kasper}, J.~C. and {Kim}, H.~S. and {Kratzenberg}, E. and 
	{Lenc}, E. and {Loeb}, A. and {Lonsdale}, C.~J. and {Lynch}, M.~J. and 
	{McKinley}, B. and {McWhirter}, S.~R. and {Mitchell}, D.~A. and 
	{Morgan}, E. and {Neben}, A.~R. and {Thyagarajan}, N. and {Oberoi}, D. and 
	{Offringa}, A.~R. and {Ord}, S.~M. and {Paul}, S. and {Prabu}, T. and 
	{Procopio}, P. and {Riding}, J. and {Rogers}, A.~E.~E. and {Roshi}, A. and 
	{Udaya Shankar}, N. and {Sethi}, S.~K. and {Srivani}, K.~S. and 
	{Subrahmanyan}, R. and {Tegmark}, M. and {Tingay}, S.~J. and 
	{Waterson}, M. and {Wayth}, R.~B. and {Webster}, R.~L. and {Whitney}, A.~R. and 
	{Williams}, A. and {Williams}, C.~L. and {Wu}, C. and {Wyithe}, J.~S.~B.
	},
    title = "{First Season MWA EoR Power spectrum Results at Redshift 7}",
  journal = {\apj},
archivePrefix = "arXiv",
   eprint = {1608.06281},
 primaryClass = "astro-ph.IM",
     year = 2016,
    month = dec,
   volume = 833,
      eid = {102},
    pages = {102},
      doi = {10.3847/1538-4357/833/1/102},
   adsurl = {http://adsabs.harvard.edu/abs/2016ApJ...833..102B}
}

@ARTICLE{Bernardi2009,
   author = {{Bernardi}, G. and {de Bruyn}, A.~G. and {Brentjens}, M.~A. and 
	{Ciardi}, B. and {Harker}, G. and {Jeli{\'c}}, V. and {Koopmans}, L.~V.~E. and 
	{Labropoulos}, P. and {Offringa}, A. and {Pandey}, V.~N. and 
	{Schaye}, J. and {Thomas}, R.~M. and {Yatawatta}, S. and {Zaroubi}, S.
	},
    title = "{Foregrounds for observations of the cosmological 21 cm line. I. First Westerbork measurements of Galactic emission at 150 MHz in a low latitude field}",
  journal = {\aap},
archivePrefix = "arXiv",
   eprint = {0904.0404},
     year = 2009,
    month = jun,
   volume = 500,
    pages = {965-979},
      doi = {10.1051/0004-6361/200911627},
   adsurl = {http://adsabs.harvard.edu/abs/2009A%26A...500..965B}
}

@article{Bharadwaj2018,
    author = {Bharadwaj, Somnath and Pal, Srijita and Choudhuri, Samir and Dutta, Prasun},
    title = "{A Tapered Gridded Estimator (TGE) for the multifrequency angular power spectrum (MAPS) and the cosmological H i 21-cm power spectrum}",
    journal = {\mnras},
    volume = {483},
    number = {4},
    pages = {5694-5700},
    year = {2018},
    month = {12},
    issn = {0035-8711},
    doi = {10.1093/mnras/sty3501},
    url = {https://doi.org/10.1093/mnras/sty3501},
    eprint = {https://academic.oup.com/mnras/article-pdf/483/4/5694/27503428/sty3501.pdf},
}

@article{Carroll2016,
    author = {Carroll, P. A. and Line, J. and Morales, M. F. and Barry, N. and Beardsley, A. P. and Hazelton, B. J. and Jacobs, D. C. and Pober, J. C. and Sullivan, I. S. and Webster, R. L. and Bernardi, G. and Bowman, J. D. and Briggs, F. and Cappallo, R. J. and Corey, B. E. and de Oliveira-Costa, A. and Dillon, J. S. and Emrich, D. and Ewall-Wice, A. and Feng, L. and Gaensler, B. M. and Goeke, R. and Greenhill, L. J. and Hewitt, J. N. and Hurley-Walker, N. and Johnston-Hollitt, M. and Kaplan, D. L. and Kasper, J. C. and Kim, HS. and Kratzenberg, E. and Lenc, E. and Loeb, A. and Lonsdale, C. J. and Lynch, M. J. and McKinley, B. and McWhirter, S. R. and Mitchell, D. A. and Morgan, E. and Neben, A. R. and Oberoi, D. and Offringa, A. R. and Ord, S. M. and Paul, S. and Pindor, B. and Prabu, T. and Procopio, P. and Riding, J. and Rogers, A. E. E. and Roshi, A. and Shankar, N. Udaya and Sethi, S. K. and Srivani, K. S. and Subrahmanyan, R. and Tegmark, M. and Thyagarajan, Nithyanandan and Tingay, S. J. and Trott, C. M. and Waterson, M. and Wayth, R. B. and Whitney, A. R. and Williams, A. and Williams, C. L. and Wu, C. and Wyithe, J. S. B.},
    title = {A high reliability survey of discrete Epoch of Reionization foreground sources in the MWA EoR0 field},
    journal = {Monthly Notices of the Royal Astronomical Society},
    volume = {461},
    number = {4},
    pages = {4151-4175},
    year = {2016},
    month = {07},
    issn = {0035-8711},
    doi = {10.1093/mnras/stw1599},
    url = {https://doi.org/10.1093/mnras/stw1599},
    eprint = {https://academic.oup.com/mnras/article-pdf/461/4/4151/13773950/stw1599.pdf},
}

@ARTICLE{Chapman2012,
   author = {{Chapman}, E. and {Abdalla}, F.~B. and {Harker}, G. and {Jeli{\'c}}, V. and 
	{Labropoulos}, P. and {Zaroubi}, S. and {Brentjens}, M.~A. and 
	{de Bruyn}, A.~G. and {Koopmans}, L.~V.~E.},
    title = "{Foreground removal using FASTICA: a showcase of LOFAR-EoR}",
  journal = {\mnras},
archivePrefix = "arXiv",
   eprint = {1201.2190},
 primaryClass = "astro-ph.CO",
     year = 2012,
    month = jul,
   volume = 423,
    pages = {2518-2532},
      doi = {10.1111/j.1365-2966.2012.21065.x},
   adsurl = {http://adsabs.harvard.edu/abs/2012MNRAS.423.2518C}
}

@article{Chatterjee2022,
    author = {Chatterjee, Suman and Bharadwaj, Somnath and Choudhuri, Samir and Sethi, Shiv and Patwa, Akash K},
    title = "{The tracking tapered gridded estimator for the power spectrum from drift scan observations}",
    journal = {\mnras},
    volume = {519},
    number = {2},
    pages = {2410-2425},
    year = {2022},
    month = {12},
    issn = {0035-8711},
    doi = {10.1093/mnras/stac3576},
    url = {https://doi.org/10.1093/mnras/stac3576},
    eprint = {https://academic.oup.com/mnras/article-pdf/519/2/2410/48480335/stac3576.pdf}
}

@ARTICLE{Chatterjee2024,
       author = {{Chatterjee}, Suman and {Elahi}, Khandakar Md Asif and {Bharadwaj}, Somnath and {Sarkar}, Shouvik and {Choudhuri}, Samir and {Sethi}, Shiv K. and {Patwa}, Akash Kumar},
        title = "{The Tracking Tapered Gridded Estimator for the 21-cm power spectrum from MWA drift scan observations I: Validation and preliminary results}",
      journal = {\pasa},
         year = 2024,
        month = oct,
       volume = {41},
          eid = {e077},
        pages = {e077},
          doi = {10.1017/pasa.2024.45},
archivePrefix = {arXiv},
       eprint = {2405.10080},
 primaryClass = {astro-ph.CO},
       adsurl = {https://ui.adsabs.harvard.edu/abs/2024PASA...41...77C}
}

@ARTICLE{Chatterjee2025,
       author = {{Chatterjee}, Suman and {Sarkar}, Shouvik and {Choudhuri}, Samir and {Elahi}, Khandakar Md Asif and {Bharadwaj}, Somnath and {Sethi}, Shiv K. and {Patwa}, Akash Kumar},
        title = "{A measurement of Galactic synchrotron emission using MWA drift scan observations}",
      journal = {\pasa},
         year = 2025,
        month = jul,
       volume = {42},
          eid = {e103},
        pages = {e103},
          doi = {10.1017/pasa.2025.10065},
archivePrefix = {arXiv},
       eprint = {2506.14310},
 primaryClass = {astro-ph.GA},
       adsurl = {https://ui.adsabs.harvard.edu/abs/2025PASA...42..103C}
}

@article{Choudhuri2016b,
author = {Choudhuri, Samir and Bharadwaj, Somnath and Chatterjee, Suman and Ali, Sk. Saiyad and Roy, Nirupam and Ghosh, Abhik},
title = {The visibility-based tapered gridded estimator (TGE) for the redshifted 21-cm power spectrum},
journal = {\mnras},
volume = {463},
number = {4},
pages = {4093},
year = {2016},
doi = {10.1093/mnras/stw2254},
URL = { + http://dx.doi.org/10.1093/mnras/stw2254},
eprint = {/oup/backfile/Content_public/Journal/mnras/463/4/10.1093_mnras_stw2254/2/stw2254.pdf}
}

@article{Choudhuri2017,
author = {Choudhuri, Samir and Bharadwaj, Somnath and Ali, Sk. Saiyad and Roy, Nirupam and Intema, Huib. T. and Ghosh, Abhik},
title = {The angular power spectrum measurement of the Galactic synchrotron emission in two fields of the TGSS survey},
journal = {\mnras: Letters},
volume = {470},
number = {1},
pages = {L11-L15},
year = {2017},
doi = {10.1093/mnrasl/slx066},
URL = {http://dx.doi.org/10.1093/mnrasl/slx066},
eprint = {/oup/backfile/content_public/journal/mnrasl/470/1/10.1093_mnrasl_slx066/1/slx066.pdf}
}

@ARTICLE{Choudhuri2014,
   author = {{Choudhuri}, S. and {Bharadwaj}, S. and {Ghosh}, A. and {Ali}, S.~S.
	},
    title = "{Visibility-based angular power spectrum estimation in low-frequency radio interferometric observations}",
  journal = {\mnras},
archivePrefix = "arXiv",
   eprint = {1409.7789},
     year = 2014,
    month = dec,
   volume = 445,
    pages = {4351-4365},
      doi = {10.1093/mnras/stu2027},
   adsurl = {http://adsabs.harvard.edu/abs/2014MNRAS.445.4351C}
}

@article{Choudhuri2020,
    author = {Choudhuri, Samir and Ghosh, Abhik and Roy, Nirupam and Bharadwaj, Somnath and Intema, Huib T and Ali, Sk Saiyad},
    title = "{All-sky angular power spectrum – I. Estimating brightness temperature fluctuations using the 150-MHz TGSS survey}",
    journal = {\mnras},
    volume = {494},
    number = {2},
    pages = {1936-1945},
    year = {2020},
    month = {04},
    issn = {0035-8711},
    doi = {10.1093/mnras/staa762},
    url = {https://doi.org/10.1093/mnras/staa762},
    eprint = {https://academic.oup.com/mnras/article-pdf/494/2/1936/33051118/staa762.pdf},
}

@ARTICLE{Datta2010,
   author = {{Datta}, A. and {Bowman}, J.~D. and {Carilli}, C.~L.},
    title = "{Bright Source Subtraction Requirements for Redshifted 21 cm Measurements}",
  journal = {\apj},
archivePrefix = "arXiv",
   eprint = {1005.4071},
     year = 2010,
    month = nov,
   volume = 724,
    pages = {526-538},
      doi = {10.1088/0004-637X/724/1/526},
   adsurl = {http://adsabs.harvard.edu/abs/2010ApJ...724..526D}
}

@ARTICLE{Datta2007,
   author = {{Datta}, K.~K. and {Choudhury}, T.~R. and {Bharadwaj}, S.},
    title = "{The multifrequency angular power spectrum of the epoch of reionization 21-cm signal}",
  journal = {\mnras},
   eprint = {astro-ph/0605546},
     year = 2007,
    month = jun,
   volume = 378,
    pages = {119-128},
      doi = {10.1111/j.1365-2966.2007.11747.x},
   adsurl = {http://adsabs.harvard.edu/abs/2007MNRAS.378..119D}
}

@article{Deboer2017,
  title={Hydrogen epoch of reionization array (HERA)},
  author={DeBoer, David R and Parsons, Aaron R and Aguirre, James E and Alexander, Paul and Ali, Zaki S and Beardsley, Adam P and Bernardi, Gianni and Bowman, Judd D and Bradley, Richard F and Carilli, Chris L and others},
  journal={Publications of the Astronomical Society of the Pacific},
  volume={129},
  number={974},
  pages={045001},
  year={2017},
  publisher={IOP Publishing}
}

@ARTICLE{Elahi2023,
       author = {{Elahi}, Kh Md Asif and {Bharadwaj}, Somnath and {Ghosh}, Abhik and {Pal}, Srijita and {Ali}, Sk Saiyad and {Choudhuri}, Samir and {Chakraborty}, Arnab and {Datta}, Abhirup and {Roy}, Nirupam and {Choudhury}, Madhurima and {Dutta}, Prasun},
        title = "{Towards 21-cm intensity mapping at z = 2.28 with uGMRT using the tapered gridded estimator - II. Cross-polarization power spectrum}",
      journal = {\mnras},
         year = 2023,
        month = apr,
       volume = {520},
       number = {2},
        pages = {2094-2108},
          doi = {10.1093/mnras/stad191},
archivePrefix = {arXiv},
       eprint = {2301.06677},
 primaryClass = {astro-ph.CO},
       adsurl = {https://ui.adsabs.harvard.edu/abs/2023MNRAS.520.2094E}
}

@ARTICLE{Elahi2023b,
       author = {{Elahi}, Kh Md Asif and {Bharadwaj}, Somnath and {Pal}, Srijita and {Ghosh}, Abhik and {Ali}, Sk Saiyad and {Choudhuri}, Samir and {Chakraborty}, Arnab and {Datta}, Abhirup and {Roy}, Nirupam and {Choudhury}, Madhurima and {Dutta}, Prasun},
        title = "{Towards 21-cm intensity mapping at z = 2.28 with uGMRT using the tapered gridded estimator - III. Foreground removal}",
      journal = {\mnras},
         year = 2023,
        month = nov,
       volume = {525},
       number = {3},
        pages = {3439-3454},
          doi = {10.1093/mnras/stad2495},
archivePrefix = {arXiv},
       eprint = {2308.08284},
 primaryClass = {astro-ph.CO},
       adsurl = {https://ui.adsabs.harvard.edu/abs/2023MNRAS.525.3439E}
}

@ARTICLE{Elahi2024,
       author = {{Elahi}, Khandakar Md Asif and {Bharadwaj}, Somnath and {Pal}, Srijita and {Ghosh}, Abhik and {Ali}, Sk Saiyad and {Choudhuri}, Samir and {Chakraborty}, Arnab and {Datta}, Abhirup and {Roy}, Nirupam and {Choudhury}, Madhurima and {Dutta}, Prasun},
        title = "{Towards 21-cm intensity mapping at z = 2.28 with uGMRT using the tapered gridded estimator - IV. Wide-band analysis}",
      journal = {\mnras},
         year = 2024,
        month = apr,
       volume = {529},
       number = {4},
        pages = {3372-3386},
          doi = {10.1093/mnras/stae740},
archivePrefix = {arXiv},
       eprint = {2403.06736},
 primaryClass = {astro-ph.CO},
       adsurl = {https://ui.adsabs.harvard.edu/abs/2024MNRAS.529.3372E}
}

@ARTICLE{Elahi2025,
       author = {{Elahi}, Khandakar Md Asif and {Bharadwaj}, Somnath and {Chatterjee}, Suman and {Sarkar}, Shouvik and {Choudhuri}, Samir and {Sethi}, Shiv and {Patwa}, Akash Kumar},
        title = "{The Tracking Tapered Gridded Estimator for the 21-cm power spectrum from MWA drift scan observations - II. The missing frequency channels}",
      journal = {\mnras},
         year = 2025,
        month = jul,
       volume = {540},
       number = {3},
        pages = {2745-2761},
          doi = {10.1093/mnras/staf896},
archivePrefix = {arXiv},
       eprint = {2410.11380},
 primaryClass = {astro-ph.CO},
       adsurl = {https://ui.adsabs.harvard.edu/abs/2025MNRAS.540.2745E}
}

@article{Gehlot2022,
	author = {{Gehlot, B. K.} and {Koopmans, L. V. E.} and {Offringa, A. R.} and {Gan, H.} and {Ghara, R.} and {Giri, S. K.} and {Kuiack, M.} and {Mertens, F. G.} and {Mevius, M.} and {Mondal, R.} and {Pandey, V. N.} and {Shulevski, A.} and {Wijers, R. A. M. J.} and {Yatawatta, S.}},
	title = {Degree-scale galactic radio emission at 122 MHz around the North Celestial Pole with LOFAR-AARTFAAC},
	DOI= "10.1051/0004-6361/202142939",
	url= "https://doi.org/10.1051/0004-6361/202142939",
	journal = {A&A},
	year = 2022,
	volume = 662,
	pages = "A97",
}

@article{Ghara2021,
    author = {Ghara, Raghunath and Giri, Sambit K and Ciardi, Benedetta and Mellema, Garrelt and Zaroubi, Saleem},
    title = {Constraining the state of the intergalactic medium during the Epoch of Reionization using MWA 21-cm signal observations},
    journal = {Monthly Notices of the Royal Astronomical Society},
    volume = {503},
    number = {3},
    pages = {4551-4562},
    year = {2021},
    month = {03},
    issn = {0035-8711},
    doi = {10.1093/mnras/stab776},
    url = {https://doi.org/10.1093/mnras/stab776},
    eprint = {https://academic.oup.com/mnras/article-pdf/503/3/4551/36997975/stab776.pdf},
}

@ARTICLE{Ghosh2012,
   author = {{Ghosh}, A. and {Prasad}, J. and {Bharadwaj}, S. and {Ali}, S.~S. and 
	{Chengalur}, J.~N.},
    title = "{Characterizing foreground for redshifted 21 cm radiation: 150 MHz Giant Metrewave Radio Telescope observations}",
  journal = {\mnras},
archivePrefix = "arXiv",
   eprint = {1208.1617},
 primaryClass = "astro-ph.CO",
     year = 2012,
    month = nov,
   volume = 426,
    pages = {3295-3314},
      doi = {10.1111/j.1365-2966.2012.21889.x},
   adsurl = {http://adsabs.harvard.edu/abs/2012MNRAS.426.3295G}
}

@article{Gupta2017,
  title={The upgraded GMRT: opening new windows on the radio Universe},
  author={Gupta, Y and Ajithkumar, B and Kale, HS and Nayak, S and Sabhapathy, S and Sureshkumar, S and Swami, RV and Chengalur, JN and Ghosh, SK and Ishwara-Chandra, CH and others},
  journal={CURRENT SCIENCE},
  volume={113},
  number={4},
  pages={707},
  year={2017},
  publisher={INDIAN ACAD SCIENCES CV RAMAN AVENUE, SADASHIVANAGAR, PB\# 8005, BANGALORE 560 080, INDIA}
}

@ARTICLE{Haslam1982,
   author = {{Haslam}, C.~G.~T. and {Salter}, C.~J. and {Stoffel}, H. and 
	{Wilson}, W.~E.},
    title = "{A 408 MHz all-sky continuum survey. II - The atlas of contour maps}",
  journal = {\aaps},
     year = 1982,
    month = jan,
   volume = 47,
    pages = {1},
   adsurl = {http://adsabs.harvard.edu/abs/1982A%26AS...47....1H}
}

@ARTICLE{HERA2023,
       author = {{HERA Collaboration} and {Abdurashidova}, Zara and {Adams}, Tyrone and {Aguirre}, James E. and {Alexander}, Paul and {Ali}, Zaki S. and {Baartman}, Rushelle and {Balfour}, Yanga and {Barkana}, Rennan and {Beardsley}, Adam P. and {Bernardi}, Gianni and {Billings}, Tashalee S. and {Bowman}, Judd D. and {Bradley}, Richard F. and {Breitman}, Daniela and {Bull}, Philip and {Burba}, Jacob and {Carey}, Steve and {Carilli}, Chris L. and {Cheng}, Carina and {Choudhuri}, Samir and {DeBoer}, David R. and {de Lera Acedo}, Eloy and {Dexter}, Matt and {Dillon}, Joshua S. and {Ely}, John and {Ewall-Wice}, Aaron and {Fagnoni}, Nicolas and {Fialkov}, Anastasia and {Fritz}, Randall and {Furlanetto}, Steven R. and {Gale-Sides}, Kingsley and {Garsden}, Hugh and {Glendenning}, Brian and {Gorce}, Ad{\'e}lie and {Gorthi}, Deepthi and {Greig}, Bradley and {Grobbelaar}, Jasper and {Halday}, Ziyaad and {Hazelton}, Bryna J. and {Heimersheim}, Stefan and {Hewitt}, Jacqueline N. and {Hickish}, Jack and {Jacobs}, Daniel C. and {Julius}, Austin and {Kern}, Nicholas S. and {Kerrigan}, Joshua and {Kittiwisit}, Piyanat and {Kohn}, Saul A. and {Kolopanis}, Matthew and {Lanman}, Adam and {La Plante}, Paul and {Lewis}, David and {Liu}, Adrian and {Loots}, Anita and {Ma}, Yin-Zhe and {MacMahon}, David H.~E. and {Malan}, Lourence and {Malgas}, Keith and {Malgas}, Cresshim and {Maree}, Matthys and {Marero}, Bradley and {Martinot}, Zachary E. and {McBride}, Lisa and {Mesinger}, Andrei and {Mirocha}, Jordan and {Molewa}, Mathakane and {Morales}, Miguel F. and {Mosiane}, Tshegofalang and {Mu{\~n}oz}, Julian B. and {Murray}, Steven G. and {Nagpal}, Vighnesh and {Neben}, Abraham R. and {Nikolic}, Bojan and {Nunhokee}, Chuneeta D. and {Nuwegeld}, Hans and {Parsons}, Aaron R. and {Pascua}, Robert and {Patra}, Nipanjana and {Pieterse}, Samantha and {Qin}, Yuxiang and {Razavi-Ghods}, Nima and {Robnett}, James and {Rosie}, Kathryn and {Santos}, Mario G. and {Sims}, Peter and {Singh}, Saurabh and {Smith}, Craig and {Swarts}, Hilton and {Tan}, Jianrong and {Thyagarajan}, Nithyanandan and {Wilensky}, Michael J. and {Williams}, Peter K.~G. and {van Wyngaarden}, Pieter and {Zheng}, Haoxuan},
        title = "{Improved Constraints on the 21 cm EoR Power Spectrum and the X-Ray Heating of the IGM with HERA Phase I Observations}",
      journal = {\apj},
         year = 2023,
        month = mar,
       volume = {945},
       number = {2},
          eid = {124},
        pages = {124},
          doi = {10.3847/1538-4357/acaf50},
archivePrefix = {arXiv},
       eprint = {2210.04912},
 primaryClass = {astro-ph.CO},
       adsurl = {https://ui.adsabs.harvard.edu/abs/2023ApJ...945..124H}
}

@article{Jacobs_2013,
doi = {10.1088/0004-637X/776/2/108},
url = {https://doi.org/10.1088/0004-637X/776/2/108},
year = {2013},
month = {oct},
publisher = {The American Astronomical Society},
volume = {776},
number = {2},
pages = {108},
author = {Jacobs, Daniel C. and Parsons, Aaron R. and Aguirre, James E. and Ali, Zaki and Bowman, Judd and Bradley, Richard F. and Carilli, Chris L. and DeBoer, David R. and Dexter, Matthew R. and Gugliucci, Nicole E. and Klima, Pat and MacMahon, Dave H. E. and Manley, Jason R. and Moore, David F. and Pober, Jonathan C. and Stefan, Irina I. and Walbrugh, William P.},
title = {A FLUX SCALE FOR SOUTHERN HEMISPHERE 21 cm EPOCH OF REIONIZATION EXPERIMENTS},
journal = {The Astrophysical Journal}
}

@article{Jong2025, 
title={Determining ideal fields for Epoch of Reionisation science using the 21 cm line}, 
volume={42}, 
DOI={10.1017/pasa.2025.10100}, 
journal={Publications of the Astronomical Society of Australia}, 
author={Jong, Eric and Trott, Cathryn and Nunhokee, Chuneeta D. and Zheng, Qian}, year={2025}, pages={e135}}

@article{Kolopanis2019,
	doi = {10.3847/1538-4357/ab3e3a},
	url = {https://doi.org/10.3847%2F1538-4357%2Fab3e3a},
	year = 2019,
	month = {sep},
	publisher = {American Astronomical Society},
	volume = {883},
	number = {2},
	pages = {133},
	author = {Matthew Kolopanis and Daniel C. Jacobs and Carina Cheng and Aaron R. Parsons and Saul A. Kohn and Jonathan C. Pober and James E. Aguirre and Zaki S. Ali and Gianni Bernardi and Richard F. Bradley and Chris L. Carilli and David R. DeBoer and Matthew R. Dexter and Joshua S. Dillon and Joshua Kerrigan and Pat Klima and Adrian Liu and David H. E. MacMahon and David F. Moore and Nithyanandan Thyagarajan and Chuneeta D. Nunhokee and William P. Walbrugh and Andre Walker},
	title = {A Simplified, Lossless Reanalysis of {PAPER}-64},
	journal = {ApJ}
}

@ARTICLE{Koopmans2015,
   author = {{Koopmans}, L. and {Pritchard}, J. and {Mellema}, G. and {Aguirre}, J. and 
	{Ahn}, K. and {Barkana}, R. and {van Bemmel}, I. and {Bernardi}, G. and 
	{Bonaldi}, A. and {Briggs}, F. and {de Bruyn}, A.~G. and {Chang}, T.~C. and 
	{Chapman}, E. and {Chen}, X. and {Ciardi}, B. and {Dayal}, P. and 
	{Ferrara}, A. and {Fialkov}, A. and {Fiore}, F. and {Ichiki}, K. and 
	{Illiev}, I.~T. and {Inoue}, S. and {Jelic}, V. and {Jones}, M. and 
	{Lazio}, J. and {Maio}, U. and {Majumdar}, S. and {Mack}, K.~J. and 
	{Mesinger}, A. and {Morales}, M.~F. and {Parsons}, A. and {Pen}, U.~L. and 
	{Santos}, M. and {Schneider}, R. and {Semelin}, B. and {de Souza}, R.~S. and 
	{Subrahmanyan}, R. and {Takeuchi}, T. and {Vedantham}, H. and 
	{Wagg}, J. and {Webster}, R. and {Wyithe}, S. and {Datta}, K.~K. and 
	{Trott}, C.},
    title = "{The Cosmic Dawn and Epoch of Reionisation with SKA}",
  journal = {Advancing Astrophysics with the Square Kilometre Array (AASKA14)},
archivePrefix = "arXiv",
   eprint = {1505.07568},
     year = 2015,
    month = apr,
      eid = {1},
    pages = {1},
   adsurl = {http://adsabs.harvard.edu/abs/2015aska.confE...1K}
}

@Article{Mellema2013,
author="Mellema, Garrelt
and Koopmans, L{\'e}on V. E.
and Abdalla, Filipe A.
and Bernardi, Gianni
and Ciardi, Benedetta
and Daiboo, Soobash
and de Bruyn, A. G.
and Datta, Kanan K.
and Falcke, Heino
and Ferrara, Andrea
and Iliev, Ilian T.
and Iocco, Fabio
and Jeli{\'{c}}, Vibor
and Jensen, Hannes
and Joseph, Ronniy
and Labroupoulos, Panos
and Meiksin, Avery
and Mesinger, Andrei
and Offringa, Andr{\'e} R.
and Pandey, V. N.
and Pritchard, Jonathan R.
and Santos, Mario G.
and Schwarz, Dominik J.
and Semelin, Benoit
and Vedantham, Harish
and Yatawatta, Sarod
and Zaroubi, Saleem",
title="Reionization and the Cosmic Dawn with the Square Kilometre Array",
journal="Experimental Astronomy",
year="2013",
month="Aug",
day="01",
volume="36",
number="1",
pages="235--318",
issn="1572-9508",
doi="10.1007/s10686-013-9334-5",
url="https://doi.org/10.1007/s10686-013-9334-5"
}

@ARTICLE{McKinley2015,
       author = {{McKinley}, B. and {Yang}, R. and {L{\'o}pez-Caniego}, M. and {Briggs}, F. and {Hurley-Walker}, N. and {Wayth}, R.~B. and {Offringa}, A.~R. and {Crocker}, R. and {Bernardi}, G. and {Procopio}, P. and {Gaensler}, B.~M. and {Tingay}, S.~J. and {Johnston-Hollitt}, M. and {McDonald}, M. and {Bell}, M. and {Bhat}, N.~D.~R. and {Bowman}, J.~D. and {Cappallo}, R.~J. and {Corey}, B.~E. and {Deshpande}, A.~A. and {Emrich}, D. and {Ewall-Wice}, A. and {Feng}, L. and {Goeke}, R. and {Greenhill}, L.~J. and {Hazelton}, B.~J. and {Hewitt}, J.~N. and {Hindson}, L. and {Jacobs}, D. and {Kaplan}, D.~L. and {Kasper}, J.~C. and {Kratzenberg}, E. and {Kudryavtseva}, N. and {Lenc}, E. and {Lonsdale}, C.~J. and {Lynch}, M.~J. and {McWhirter}, S.~R. and {Mitchell}, D.~A. and {Morales}, M.~F. and {Morgan}, E. and {Oberoi}, D. and {Ord}, S.~M. and {Pindor}, B. and {Prabu}, T. and {Riding}, J. and {Rogers}, A.~E.~E. and {Roshi}, D.~A. and {Udaya Shankar}, N. and {Srivani}, K.~S. and {Subrahmanyan}, R. and {Waterson}, M. and {Webster}, R.~L. and {Whitney}, A.~R. and {Williams}, A. and {Williams}, C.~L.},
        title = "{Modelling of the spectral energy distribution of Fornax A: leptonic and hadronic production of high-energy emission from the radio lobes}",
      journal = {\mnras},
         year = 2015,
        month = feb,
       volume = {446},
       number = {4},
        pages = {3478-3491},
          doi = {10.1093/mnras/stu2310},
archivePrefix = {arXiv},
       eprint = {1411.1487},
 primaryClass = {astro-ph.HE},
       adsurl = {https://ui.adsabs.harvard.edu/abs/2015MNRAS.446.3478M}
}

@article{Mertens2018,
    author = {Mertens, F G and Ghosh, A and Koopmans, L V E},
    title = "{Statistical 21-cm signal separation via Gaussian Process Regression analysis}",
    journal = {\mnras},
    volume = {478},
    number = {3},
    pages = {3640-3652},
    year = {2018},
    month = {06},
    issn = {0035-8711},
    doi = {10.1093/mnras/sty1207},
    url = {https://doi.org/10.1093/mnras/sty1207},
    eprint = {https://academic.oup.com/mnras/article-pdf/478/3/3640/25072259/sty1207.pdf},
}

@ARTICLE{Mertens2025,
       author = {{Mertens}, F.~G. and {Mevius}, M. and {Koopmans}, L.~V.~E. and {Offringa}, A.~R. and {Zaroubi}, S. and {Acharya}, A. and {Brackenhoff}, S.~A. and {Ceccotti}, E. and {Chapman}, E. and {Chege}, K. and {Ciardi}, B. and {Ghara}, R. and {Ghosh}, S. and {Giri}, S.~K. and {Hothi}, I. and {H{\"o}fer}, C. and {Iliev}, I.~T. and {Jeli{\'c}}, V. and {Ma}, Q. and {Mellema}, G. and {Munshi}, S. and {Pandey}, V.~N. and {Yatawatta}, S.},
        title = "{Deeper multi-redshift upper limits on the epoch of reionisation 21 cm signal power spectrum from LOFAR between z = 8.3 and z = 10.1}",
      journal = {\aap},
         year = 2025,
        month = jun,
       volume = {698},
          eid = {A186},
        pages = {A186},
          doi = {10.1051/0004-6361/202554158},
archivePrefix = {arXiv},
       eprint = {2503.05576},
 primaryClass = {astro-ph.CO},
       adsurl = {https://ui.adsabs.harvard.edu/abs/2025A&A...698A.186M}
}

@ARTICLE{Mondal2017,
       author = {{Mondal}, Rajesh and {Bharadwaj}, Somnath and {Majumdar}, Suman},
        title = "{Statistics of the epoch of reionization (EoR) 21-cm signal - II. The evolution of the power-spectrum error-covariance}",
      journal = {\mnras},
         year = 2017,
        month = jan,
       volume = {464},
       number = {3},
        pages = {2992-3004},
          doi = {10.1093/mnras/stw2599},
archivePrefix = {arXiv},
       eprint = {1606.03874},
 primaryClass = {astro-ph.CO},
       adsurl = {https://ui.adsabs.harvard.edu/abs/2017MNRAS.464.2992M}
}

@ARTICLE{Mondal2018,
       author = {{Mondal}, Rajesh and {Bharadwaj}, Somnath and {Datta}, Kanan K.},
        title = "{Towards simulating and quantifying the light-cone EoR 21-cm signal}",
      journal = {\mnras},
         year = 2018,
        month = feb,
       volume = {474},
       number = {1},
        pages = {1390-1397},
          doi = {10.1093/mnras/stx2888},
archivePrefix = {arXiv},
       eprint = {1706.09449},
 primaryClass = {astro-ph.CO},
       adsurl = {https://ui.adsabs.harvard.edu/abs/2018MNRAS.474.1390M}
}

@article{Mondal2019,
    author = {Mondal, Rajesh and Bharadwaj, Somnath and Iliev, Ilian T and Datta, Kanan K and Majumdar, Suman and Shaw, Abinash K and Sarkar, Anjan K},
    title = "{A method to determine the evolution history of the mean neutral Hydrogen fraction}",
    journal = {MNRAS: Letters},
    volume = {483},
    number = {1},
    pages = {L109-L113},
    year = {2018},
    month = {11},
    issn = {1745-3925},
    doi = {10.1093/mnrasl/sly226},
    url = {https://doi.org/10.1093/mnrasl/sly226},
    eprint = {https://academic.oup.com/mnrasl/article-pdf/483/1/L109/27201242/sly226.pdf},
}

@ARTICLE{Morales2004,
   author = {{Morales}, M.~F. and {Hewitt}, J.},
    title = "{Toward Epoch of Reionization Measurements with Wide-Field Radio Observations}",
  journal = {\apj},
   eprint = {astro-ph/0312437},
     year = 2004,
    month = nov,
   volume = 615,
    pages = {7-18},
      doi = {10.1086/424437},
   adsurl = {http://adsabs.harvard.edu/abs/2004ApJ...615....7M}
}

@ARTICLE{Morales2012,
   author = {{Morales}, M.~F. and {Hazelton}, B. and {Sullivan}, I. and {Beardsley}, A.
	},
    title = "{Four Fundamental Foreground Power Spectrum Shapes for 21 cm Cosmology Observations}",
  journal = {\apj},
archivePrefix = "arXiv",
   eprint = {1202.3830},
 primaryClass = "astro-ph.IM",
     year = 2012,
    month = jun,
   volume = 752,
      eid = {137},
    pages = {137},
      doi = {10.1088/0004-637X/752/2/137},
   adsurl = {http://adsabs.harvard.edu/abs/2012ApJ...752..137M}
}

@ARTICLE{Nunhokee2025,
       author = {{Nunhokee}, C.~D. and {Null}, D. and {Trott}, C.~M. and {Barry}, N. and {Qin}, Y. and {Wayth}, R.~B. and {Line}, J.~L.~B. and {Jordan}, C.~H. and {Pindor}, B. and {Cook}, J.~H. and {Bowman}, J. and {Chokshi}, A. and {Ducharme}, J. and {Elder}, K. and {Guo}, Q. and {Hazelton}, B. and {Hidayat}, W. and {Ito}, T. and {Jacobs}, D. and {Jong}, E. and {Kolopanis}, M. and {Kunicki}, T. and {Lilleskov}, E. and {Morales}, M.~F. and {Pober}, J.~C. and {Selvaraj}, A. and {Shi}, R. and {Takahashi}, K. and {Tingay}, S.~J. and {Webster}, R.~L. and {Yoshiura}, S. and {Zheng}, Q.},
        title = "{Limits on the 21 cm Power Spectrum at z = 6.5─7.0 from Murchison Widefield Array Observations}",
      journal = {\apj},
         year = 2025,
        month = aug,
       volume = {989},
       number = {1},
          eid = {57},
        pages = {57},
          doi = {10.3847/1538-4357/adda45},
archivePrefix = {arXiv},
       eprint = {2505.09097},
 primaryClass = {astro-ph.CO},
       adsurl = {https://ui.adsabs.harvard.edu/abs/2025ApJ...989...57N}
}

@misc{Offringa2010,
  author = {Offringa, A. R.},
  title = {AOFlagger: RFI Software},
  howpublished = {Astrophysics Source Code Library, ascl:1010.017},
  year = {2010}
}

@article{Offringa2015, 
title={The Low-Frequency Environment of the Murchison Widefield Array: Radio-Frequency Interference Analysis and Mitigation}, 
volume={32}, 
DOI={10.1017/pasa.2015.7}, 
journal={Publications of the Astronomical Society of Australia}, 
publisher={Cambridge University Press}, 
author={Offringa, A. R. and Wayth, R. B. and Hurley-Walker, N. and Kaplan, D. L. and Barry, N. and Beardsley, A. P. and Bell, M. E. and Bernardi, G. and Bowman, J. D. and Briggs, F. and et al.}, 
year={2015}, 
pages={e008}
}

@ARTICLE{Paciga2011,
   author = {{Paciga}, G. and {Chang}, T.-C. and {Gupta}, Y. and {Nityanada}, R. and 
	{Odegova}, J. and {Pen}, U.-L. and {Peterson}, J.~B. and {Roy}, J. and 
	{Sigurdson}, K.},
    title = "{The GMRT Epoch of Reionization experiment: a new upper limit on the neutral hydrogen power spectrum at $z \approx 8.6$}",
  journal = {\mnras},
archivePrefix = "arXiv",
   eprint = {1006.1351},
     year = 2011,
    month = may,
   volume = 413,
    pages = {1174-1183},
      doi = {10.1111/j.1365-2966.2011.18208.x},
   adsurl = {http://adsabs.harvard.edu/abs/2011MNRAS.413.1174P}
}

@ARTICLE{Paciga2013,
   author = {{Paciga}, G. and {Albert}, J.~G. and {Bandura}, K. and {Chang}, T.-C. and 
	{Gupta}, Y. and {Hirata}, C. and {Odegova}, J. and {Pen}, U.-L. and 
	{Peterson}, J.~B. and {Roy}, J. and {Shaw}, J.~R. and {Sigurdson}, K. and 
	{Voytek}, T.},
    title = "{A simulation-calibrated limit on the H I power spectrum from the GMRT Epoch of Reionization experiment}",
  journal = {\mnras},
archivePrefix = "arXiv",
   eprint = {1301.5906},
     year = 2013,
    month = jul,
   volume = 433,
    pages = {639-647},
      doi = {10.1093/mnras/stt753},
   adsurl = {http://adsabs.harvard.edu/abs/2013MNRAS.433..639P}
}

@article{Pal2020,
       author = {{Pal}, Srijita and {Bharadwaj}, Somnath and {Ghosh}, Abhik and {Choudhuri}, Samir},
        title = "{Demonstrating the Tapered Gridded Estimator (TGE) for the cosmological H I 21-cm power spectrum using 150-MHz GMRT observations}",
      journal = {\mnras},
         year = 2021,
        month = mar,
       volume = {501},
       number = {3},
        pages = {3378-3391},
          doi = {10.1093/mnras/staa3831},
archivePrefix = {arXiv},
       eprint = {2012.04998},
 primaryClass = {astro-ph.CO},
       adsurl = {https://ui.adsabs.harvard.edu/abs/2021MNRAS.501.3378P}
}

@article{Pal2022,
    author = {Pal, Srijita and Elahi, Kh Md Asif and Bharadwaj, Somnath and Ali, Sk Saiyad and Choudhuri, Samir and Ghosh, Abhik and Chakraborty, Arnab and Datta, Abhirup and Roy, Nirupam and Choudhury, Madhurima and Dutta, Prasun},
    title = "{Towards 21-cm intensity mapping at z = 2.28 with uGMRT using the tapered gridded estimator I: Foreground avoidance}",
    journal = {\mnras},
    volume = {516},
    number = {2},
    pages = {2851-2863},
    year = {2022},
    month = {08},
    issn = {0035-8711},
    doi = {10.1093/mnras/stac2419},
    url = {https://doi.org/10.1093/mnras/stac2419},
    eprint = {https://academic.oup.com/mnras/article-pdf/516/2/2851/45821097/stac2419.pdf}
}

@ARTICLE{Parsons2012b,
   author = {{Parsons}, A.~R. and {Pober}, J.~C. and {Aguirre}, J.~E. and 
	{Carilli}, C.~L. and {Jacobs}, D.~C. and {Moore}, D.~F.},
    title = "{A Per-baseline, Delay-spectrum Technique for Accessing the 21 cm Cosmic Reionization Signature}",
  journal = {\apj},
archivePrefix = "arXiv",
   eprint = {1204.4749},
 primaryClass = "astro-ph.IM",
     year = 2012,
    month = sep,
   volume = 756,
      eid = {165},
    pages = {165},
      doi = {10.1088/0004-637X/756/2/165},
   adsurl = {http://adsabs.harvard.edu/abs/2012ApJ...756..165P}
}

@ARTICLE{Parsons2014,
   author = {{Parsons}, A.~R. and {Liu}, A. and {Aguirre}, J.~E. and {Ali}, Z.~S. and 
	{Bradley}, R.~F. and {Carilli}, C.~L. and {DeBoer}, D.~R. and 
	{Dexter}, M.~R. and {Gugliucci}, N.~E. and {Jacobs}, D.~C. and 
	{Klima}, P. and {MacMahon}, D.~H.~E. and {Manley}, J.~R. and 
	{Moore}, D.~F. and {Pober}, J.~C. and {Stefan}, I.~I. and {Walbrugh}, W.~P.
	},
    title = "{New Limits on 21 cm Epoch of Reionization from PAPER-32 Consistent with an X-Ray Heated Intergalactic Medium at z = 7.7}",
  journal = {\apj},
archivePrefix = "arXiv",
   eprint = {1304.4991},
     year = 2014,
    month = jun,
   volume = 788,
      eid = {106},
    pages = {106},
      doi = {10.1088/0004-637X/788/2/106},
   adsurl = {http://adsabs.harvard.edu/abs/2014ApJ...788..106P}
}

@article{Patil2017,
  author={A. H. Patil and S. Yatawatta and L. V. E. Koopmans and A. G. de Bruyn and M. A. Brentjens and S. Zaroubi and K. M. B.
Asad and M. Hatef and V. Jelić and M. Mevius and A. R. Offringa and V. N. Pandey and H. Vedantham and F. B. Abdalla and W. N.
Brouw and E. Chapman and B. Ciardi and B. K. Gehlot and A. Ghosh and G. Harker and I. T. Iliev and K. Kakiichi and S. Majumdar and G.
Mellema and M. B. Silva and J. Schaye and D. Vrbanec and S. J. Wijnholds},
  title={Upper Limits on the 21 cm Epoch of Reionization Power Spectrum from One Night with LOFAR},
  journal={ApJ},
  volume={838},
  number={1},
  pages={65},
  url={http://stacks.iop.org/0004-637X/838/i=1/a=65},
  year={2017}
}

@article{Patwa2021,
    author = {Patwa, Akash Kumar and Sethi, Shiv and Dwarakanath, K S},
    title = "{Extracting the 21 cm EoR signal using MWA drift scan data}",
    journal = {\mnras},
    volume = {504},
    number = {2},
    pages = {2062-2072},
    year = {2021},
    month = {04},
    issn = {0035-8711},
    doi = {10.1093/mnras/stab989},
    url = {https://doi.org/10.1093/mnras/stab989},
    eprint = {https://academic.oup.com/mnras/article-pdf/504/2/2062/37518347/stab989.pdf},
}

@article{Planck2020f,
       author = {{Planck Collaboration} and {Aghanim}, N. and {Akrami}, Y. and {Ashdown}, M. and {Aumont}, J. and {Baccigalupi}, C. and {Ballardini}, M. and {Banday}, A.~J. and {Barreiro}, R.~B. and {Bartolo}, N. and {Basak}, S. and {Battye}, R. and {Benabed}, K. and {Bernard}, J. -P. and {Bersanelli}, M. and {Bielewicz}, P. and {Bock}, J.~J. and {Bond}, J.~R. and {Borrill}, J. and {Bouchet}, F.~R. and {Boulanger}, F. and {Bucher}, M. and {Burigana}, C. and {Butler}, R.~C. and {Calabrese}, E. and {Cardoso}, J. -F. and {Carron}, J. and {Challinor}, A. and {Chiang}, H.~C. and {Chluba}, J. and {Colombo}, L.~P.~L. and {Combet}, C. and {Contreras}, D. and {Crill}, B.~P. and {Cuttaia}, F. and {de Bernardis}, P. and {de Zotti}, G. and {Delabrouille}, J. and {Delouis}, J. -M. and {Di Valentino}, E. and {Diego}, J.~M. and {Dor{\'e}}, O. and {Douspis}, M. and {Ducout}, A. and {Dupac}, X. and {Dusini}, S. and {Efstathiou}, G. and {Elsner}, F. and {En{\ss}lin}, T.~A. and {Eriksen}, H.~K. and {Fantaye}, Y. and {Farhang}, M. and {Fergusson}, J. and {Fernandez-Cobos}, R. and {Finelli}, F. and {Forastieri}, F. and {Frailis}, M. and {Fraisse}, A.~A. and {Franceschi}, E. and {Frolov}, A. and {Galeotta}, S. and {Galli}, S. and {Ganga}, K. and {G{\'e}nova-Santos}, R.~T. and {Gerbino}, M. and {Ghosh}, T. and {Gonz{\'a}lez-Nuevo}, J. and {G{\'o}rski}, K.~M. and {Gratton}, S. and {Gruppuso}, A. and {Gudmundsson}, J.~E. and {Hamann}, J. and {Handley}, W. and {Hansen}, F.~K. and {Herranz}, D. and {Hildebrandt}, S.~R. and {Hivon}, E. and {Huang}, Z. and {Jaffe}, A.~H. and {Jones}, W.~C. and {Karakci}, A. and {Keih{\"a}nen}, E. and {Keskitalo}, R. and {Kiiveri}, K. and {Kim}, J. and {Kisner}, T.~S. and {Knox}, L. and {Krachmalnicoff}, N. and {Kunz}, M. and {Kurki-Suonio}, H. and {Lagache}, G. and {Lamarre}, J. -M. and {Lasenby}, A. and {Lattanzi}, M. and {Lawrence}, C.~R. and {Le Jeune}, M. and {Lemos}, P. and {Lesgourgues}, J. and {Levrier}, F. and {Lewis}, A. and {Liguori}, M. and {Lilje}, P.~B. and {Lilley}, M. and {Lindholm}, V. and {L{\'o}pez-Caniego}, M. and {Lubin}, P.~M. and {Ma}, Y. -Z. and {Mac{\'\i}as-P{\'e}rez}, J.~F. and {Maggio}, G. and {Maino}, D. and {Mandolesi}, N. and {Mangilli}, A. and {Marcos-Caballero}, A. and {Maris}, M. and {Martin}, P.~G. and {Martinelli}, M. and {Mart{\'\i}nez-Gonz{\'a}lez}, E. and {Matarrese}, S. and {Mauri}, N. and {McEwen}, J.~D. and {Meinhold}, P.~R. and {Melchiorri}, A. and {Mennella}, A. and {Migliaccio}, M. and {Millea}, M. and {Mitra}, S. and {Miville-Desch{\^e}nes}, M. -A. and {Molinari}, D. and {Montier}, L. and {Morgante}, G. and {Moss}, A. and {Natoli}, P. and {N{\o}rgaard-Nielsen}, H.~U. and {Pagano}, L. and {Paoletti}, D. and {Partridge}, B. and {Patanchon}, G. and {Peiris}, H.~V. and {Perrotta}, F. and {Pettorino}, V. and {Piacentini}, F. and {Polastri}, L. and {Polenta}, G. and {Puget}, J. -L. and {Rachen}, J.~P. and {Reinecke}, M. and {Remazeilles}, M. and {Renzi}, A. and {Rocha}, G. and {Rosset}, C. and {Roudier}, G. and {Rubi{\~n}o-Mart{\'\i}n}, J.~A. and {Ruiz-Granados}, B. and {Salvati}, L. and {Sandri}, M. and {Savelainen}, M. and {Scott}, D. and {Shellard}, E.~P.~S. and {Sirignano}, C. and {Sirri}, G. and {Spencer}, L.~D. and {Sunyaev}, R. and {Suur-Uski}, A. -S. and {Tauber}, J.~A. and {Tavagnacco}, D. and {Tenti}, M. and {Toffolatti}, L. and {Tomasi}, M. and {Trombetti}, T. and {Valenziano}, L. and {Valiviita}, J. and {Van Tent}, B. and {Vibert}, L. and {Vielva}, P. and {Villa}, F. and {Vittorio}, N. and {Wandelt}, B.~D. and {Wehus}, I.~K. and {White}, M. and {White}, S.~D.~M. and {Zacchei}, A. and {Zonca}, A.},
        title = "{Planck 2018 results. VI. Cosmological parameters}",
      journal = {\aap},
         year = 2020,
        month = sep,
       volume = {641},
          eid = {A6},
        pages = {A6},
          doi = {10.1051/0004-6361/201833910},
archivePrefix = {arXiv},
       eprint = {1807.06209},
 primaryClass = {astro-ph.CO},
       adsurl = {https://ui.adsabs.harvard.edu/abs/2020A&A...641A...6P}
}

@article{Pober2016,
	doi = {10.3847/0004-637x/819/1/8},
	url = {https://doi.org/10.3847%2F0004-637x%2F819%2F1%2F8},
	year = 2016,
	month = {feb},
	publisher = {American Astronomical Society},
	volume = {819},
	number = {1},
	pages = {8},
	author = {J. C. Pober and B. J. Hazelton and A. P. Beardsley and N. A. Barry and Z. E. Martinot and I. S. Sullivan and M. F. Morales and M. E. Bell and G. Bernardi and N. D. R. Bhat and J. D. Bowman and F. Briggs and R. J. Cappallo and P. Carroll and B. E. Corey and A. de Oliveira-Costa and A. A. Deshpande and Joshua. S. Dillon and D. Emrich and A. M. Ewall-Wice and L. Feng and R. Goeke and L. J. Greenhill and J. N. Hewitt and L. Hindson and N. Hurley-Walker and D. C. Jacobs and M. Johnston-Hollitt and D. L. Kaplan and J. C. Kasper and Han-Seek Kim and P. Kittiwisit and E. Kratzenberg and N. Kudryavtseva and E. Lenc and J. Line and A. Loeb and C. J. Lonsdale and M. J. Lynch and B. McKinley and S. R. McWhirter and D. A. Mitchell and E. Morgan and A. R. Neben and D. Oberoi and A. R. Offringa and S. M. Ord and Sourabh Paul and B. Pindor and T. Prabu and P. Procopio and J. Riding and A. E. E. Rogers and A. Roshi and Shiv K. Sethi and N. Udaya Shankar and K. S. Srivani and R. Subrahmanyan and M. Tegmark and Nithyanandan Thyagarajan and S. J. Tingay and C. M. Trott and M. Waterson and R. B. Wayth and R. L. Webster and A. R. Whitney and A. Williams and C. L. Williams and J. S. B. Wyithe},
	title = {{THE} {IMPORTANCE} {OF} {WIDE}-{FIELD} {FOREGROUND} {REMOVAL} {FOR} 21 cm {COSMOLOGY}: A {DEMONSTRATION} {WITH} {EARLY} {MWA} {EPOCH} {OF} {REIONIZATION} {OBSERVATIONS}},
	journal = {\apj}
}

@article{Ross2024, 
title={GaLactic and Extragalactic All-sky Murchison Widefield Array eXtended (GLEAM-X) survey II: Second Data Release}, 
volume={41}, 
DOI={10.1017/pasa.2024.57}, 
journal={Publications of the Astronomical Society of Australia}, 
author={Ross, Kathryn and Hurley-Walker, Natasha and Galvin, Timothy James and Venville, Brandon and Duchesne, Stefan William and Morgan, John and An, Tao and Gürkan, Gulay and Hancock, Paul J. and Heald, George and et al.}, 
year={2024}, 
pages={e054}}

@ARTICLE{Rogers2008,
   author = {{Rogers}, A.~E.~E. and {Bowman}, J.~D.},
    title = "{Spectral Index of the Diffuse Radio Background Measured from 100 to 200 MHz}",
  journal = {\aj},
archivePrefix = "arXiv",
   eprint = {0806.2868},
     year = 2008,
    month = aug,
   volume = 136,
    pages = {641-648},
      doi = {10.1088/0004-6256/136/2/641},
   adsurl = {http://adsabs.harvard.edu/abs/2008AJ....136..641R}
}

@article{Saha2019,
    author = {Saha, Preetha and Bharadwaj, Somnath and Roy, Nirupam and Choudhuri, Samir and Chattopadhyay, Debatri},
    title = "{A study of Kepler supernova remnant: angular power spectrum estimation from radio frequency data}",
    journal = {\mnras},
    volume = {489},
    number = {4},
    pages = {5866-5875},
    year = {2019},
    month = {09},
    issn = {0035-8711},
    doi = {10.1093/mnras/stz2528},
    url = {https://doi.org/10.1093/mnras/stz2528},
    eprint = {https://academic.oup.com/mnras/article-pdf/489/4/5866/30097722/stz2528.pdf},
}

@ARTICLE{Swarup1991,
   author = {{Swarup}, G. and {Ananthakrishnan}, S. and {Kapahi}, V.~K. and 
	{Rao}, A.~P. and {Subrahmanya}, C.~R. and {Kulkarni}, V.~K.},
    title = "{The Giant Metre-Wave Radio Telescope}",
  journal = {Current Science, Vol.~60, NO.2/JAN25, P.~95, 1991},
     year = 1991,
    month = jan,
   volume = 60,
    pages = {95},
   adsurl = {http://adsabs.harvard.edu/abs/1991CuSc...60...95S}
}

@ARTICLE{Tingay2013,
   author = {{Tingay}, S.~J. and {Goeke}, R. and {Bowman}, J.~D. and {Emrich}, D. and 
	{Ord}, S.~M. and {Mitchell}, D.~A. and {Morales}, M.~F. and 
	{Booler}, T. and {Crosse}, B. and {Wayth}, R.~B. and {Lonsdale}, C.~J. and 
	{Tremblay}, S. and {Pallot}, D. and {Colegate}, T. and {Wicenec}, A. and 
	{Kudryavtseva}, N. and {Arcus}, W. and {Barnes}, D. and {Bernardi}, G. and 
	{Briggs}, F. and {Burns}, S. and {Bunton}, J.~D. and {Cappallo}, R.~J. and 
	{Corey}, B.~E. and {Deshpande}, A. and {Desouza}, L. and {Gaensler}, B.~M. and 
	{Greenhill}, L.~J. and {Hall}, P.~J. and {Hazelton}, B.~J. and 
	{Herne}, D. and {Hewitt}, J.~N. and {Johnston-Hollitt}, M. and 
	{Kaplan}, D.~L. and {Kasper}, J.~C. and {Kincaid}, B.~B. and 
	{Koenig}, R. and {Kratzenberg}, E. and {Lynch}, M.~J. and {Mckinley}, B. and 
	{Mcwhirter}, S.~R. and {Morgan}, E. and {Oberoi}, D. and {Pathikulangara}, J. and 
	{Prabu}, T. and {Remillard}, R.~A. and {Rogers}, A.~E.~E. and 
	{Roshi}, A. and {Salah}, J.~E. and {Sault}, R.~J. and {Udaya-Shankar}, N. and 
	{Schlagenhaufer}, F. and {Srivani}, K.~S. and {Stevens}, J. and 
	{Subrahmanyan}, R. and {Waterson}, M. and {Webster}, R.~L. and 
	{Whitney}, A.~R. and {Williams}, A. and {Williams}, C.~L. and 
	{Wyithe}, J.~S.~B.},
    title = "{The Murchison Widefield Array: The Square Kilometre Array Precursor at Low Radio Frequencies}",
  journal = {\pasa},
archivePrefix = "arXiv",
   eprint = {1206.6945},
 primaryClass = "astro-ph.IM",
     year = 2013,
    month = jan,
   volume = 30,
      eid = {e007},
    pages = {e007},
      doi = {10.1017/pasa.2012.007},
   adsurl = {http://adsabs.harvard.edu/abs/2013PASA...30....7T}
}

@article{Trott2012,
  author={Cathryn M. Trott and Randall B. Wayth and Steven J. Tingay},
  title={The Impact of Point-source Subtraction Residuals on 21 cm Epoch of Reionization Estimation},
  journal={\apj},
  volume={757},
  number={1},
  pages={101},
  url={http://stacks.iop.org/0004-637X/757/i=1/a=101},
  year={2012}
}

@article{Trott2016a,
doi = {10.3847/0004-637X/818/2/139},
url = {https://dx.doi.org/10.3847/0004-637X/818/2/139},
year = {2016},
month = {feb},
publisher = {The American Astronomical Society},
volume = {818},
number = {2},
pages = {139},
author = {C. M. Trott and B. Pindor and P. Procopio and R. B. Wayth and D. A. Mitchell and B. McKinley and S. J. Tingay and N. Barry and A. P. Beardsley and G. Bernardi and Judd D. Bowman and F. Briggs and R. J. Cappallo and P. Carroll and A. de Oliveira-Costa and Joshua S. Dillon and A. Ewall-Wice and L. Feng and L. J. Greenhill and B. J. Hazelton and J. N. Hewitt and N. Hurley-Walker and M. Johnston-Hollitt and Daniel C. Jacobs and D. L. Kaplan and H. S. Kim and E. Lenc and J. Line and A. Loeb and C. J. Lonsdale and M. F. Morales and E. Morgan and A. R. Neben and Nithyanandan Thyagarajan and D. Oberoi and A. R. Offringa and S. M. Ord and S. Paul and J. C. Pober and T. Prabu and J. Riding and N. Udaya Shankar and Shiv K. Sethi and K. S. Srivani and R. Subrahmanyan and I. S. Sullivan and M. Tegmark and R. L. Webster and A. Williams and C. L. Williams and C. Wu and J. S. B. Wyithe},
title = {CHIPS: THE COSMOLOGICAL H i POWER SPECTRUM ESTIMATOR},
journal = {The Astrophysical Journal}
}

@article{Trott2020,
    author = {Trott, Cathryn M and Jordan, C H and Midgley, S and Barry, N and Greig, B and Pindor, B and Cook, J H and Sleap, G and Tingay, S J and Ung, D and Hancock, P and Williams, A and Bowman, J and Byrne, R and Chokshi, A and Hazelton, B J and Hasegawa, K and Jacobs, D and Joseph, R C and Li, W and Line, J L B and Lynch, C and McKinley, B and Mitchell, D A and Morales, M F and Ouchi, M and Pober, J C and Rahimi, M and Takahashi, K and Wayth, R B and Webster, R L and Wilensky, M and Wyithe, J S B and Yoshiura, S and Zhang, Z and Zheng, Q},
    title = "{Deep multiredshift limits on Epoch of Reionization 21 cm power spectra from four seasons of Murchison Widefield Array observations}",
    journal = {\mnras},
    volume = {493},
    number = {4},
    pages = {4711-4727},
    year = {2020},
    month = {02},
    issn = {0035-8711},
    doi = {10.1093/mnras/staa414},
    url = {https://doi.org/10.1093/mnras/staa414},
    eprint = {https://academic.oup.com/mnras/article-pdf/493/4/4711/32927265/staa414.pdf},
}

@ARTICLE{vanHarlem2013,
   author = {{van Haarlem}, M.~P. and {Wise}, M.~W. and {Gunst}, A.~W. and 
	{Heald}, G. and {McKean}, J.~P. and {Hessels}, J.~W.~T. and 
	{de Bruyn}, A.~G. and {Nijboer}, R. and {Swinbank}, J. and {Fallows}, R. and 
	{Brentjens}, M. and {Nelles}, A. and {Beck}, R. and {Falcke}, H. and 
	{Fender}, R. and {H{\"o}randel}, J. and {Koopmans}, L.~V.~E. and 
	{Mann}, G. and {Miley}, G. and {R{\"o}ttgering}, H. and {Stappers}, B.~W. and 
	{Wijers}, R.~A.~M.~J. and {Zaroubi}, S. and {van den Akker}, M. and 
	{Alexov}, A. and {Anderson}, J. and {Anderson}, K. and {van Ardenne}, A. and 
	{Arts}, M. and {Asgekar}, A. and {Avruch}, I.~M. and {Batejat}, F. and 
	{B{\"a}hren}, L. and {Bell}, M.~E. and {Bell}, M.~R. and {van Bemmel}, I. and 
	{Bennema}, P. and {Bentum}, M.~J. and {Bernardi}, G. and {Best}, P. and 
	{B{\^i}rzan}, L. and {Bonafede}, A. and {Boonstra}, A.-J. and 
	{Braun}, R. and {Bregman}, J. and {Breitling}, F. and {van de Brink}, R.~H. and 
	{Broderick}, J. and {Broekema}, P.~C. and {Brouw}, W.~N. and 
	{Br{\"u}ggen}, M. and {Butcher}, H.~R. and {van Cappellen}, W. and 
	{Ciardi}, B. and {Coenen}, T. and {Conway}, J. and {Coolen}, A. and 
	{Corstanje}, A. and {Damstra}, S. and {Davies}, O. and {Deller}, A.~T. and 
	{Dettmar}, R.-J. and {van Diepen}, G. and {Dijkstra}, K. and 
	{Donker}, P. and {Doorduin}, A. and {Dromer}, J. and {Drost}, M. and 
	{van Duin}, A. and {Eisl{\"o}ffel}, J. and {van Enst}, J. and 
	{Ferrari}, C. and {Frieswijk}, W. and {Gankema}, H. and {Garrett}, M.~A. and 
	{de Gasperin}, F. and {Gerbers}, M. and {de Geus}, E. and {Grie{\ss}meier}, J.-M. and 
	{Grit}, T. and {Gruppen}, P. and {Hamaker}, J.~P. and {Hassall}, T. and 
	{Hoeft}, M. and {Holties}, H.~A. and {Horneffer}, A. and {van der Horst}, A. and 
	{van Houwelingen}, A. and {Huijgen}, A. and {Iacobelli}, M. and 
	{Intema}, H. and {Jackson}, N. and {Jelic}, V. and {de Jong}, A. and 
	{Juette}, E. and {Kant}, D. and {Karastergiou}, A. and {Koers}, A. and 
	{Kollen}, H. and {Kondratiev}, V.~I. and {Kooistra}, E. and 
	{Koopman}, Y. and {Koster}, A. and {Kuniyoshi}, M. and {Kramer}, M. and 
	{Kuper}, G. and {Lambropoulos}, P. and {Law}, C. and {van Leeuwen}, J. and 
	{Lemaitre}, J. and {Loose}, M. and {Maat}, P. and {Macario}, G. and 
	{Markoff}, S. and {Masters}, J. and {McFadden}, R.~A. and {McKay-Bukowski}, D. and 
	{Meijering}, H. and {Meulman}, H. and {Mevius}, M. and {Middelberg}, E. and 
	{Millenaar}, R. and {Miller-Jones}, J.~C.~A. and {Mohan}, R.~N. and 
	{Mol}, J.~D. and {Morawietz}, J. and {Morganti}, R. and {Mulcahy}, D.~D. and 
	{Mulder}, E. and {Munk}, H. and {Nieuwenhuis}, L. and {van Nieuwpoort}, R. and 
	{Noordam}, J.~E. and {Norden}, M. and {Noutsos}, A. and {Offringa}, A.~R. and 
	{Olofsson}, H. and {Omar}, A. and {Orr{\'u}}, E. and {Overeem}, R. and 
	{Paas}, H. and {Pandey-Pommier}, M. and {Pandey}, V.~N. and 
	{Pizzo}, R. and {Polatidis}, A. and {Rafferty}, D. and {Rawlings}, S. and 
	{Reich}, W. and {de Reijer}, J.-P. and {Reitsma}, J. and {Renting}, G.~A. and 
	{Riemers}, P. and {Rol}, E. and {Romein}, J.~W. and {Roosjen}, J. and 
	{Ruiter}, M. and {Scaife}, A. and {van der Schaaf}, K. and {Scheers}, B. and 
	{Schellart}, P. and {Schoenmakers}, A. and {Schoonderbeek}, G. and 
	{Serylak}, M. and {Shulevski}, A. and {Sluman}, J. and {Smirnov}, O. and 
	{Sobey}, C. and {Spreeuw}, H. and {Steinmetz}, M. and {Sterks}, C.~G.~M. and 
	{Stiepel}, H.-J. and {Stuurwold}, K. and {Tagger}, M. and {Tang}, Y. and 
	{Tasse}, C. and {Thomas}, I. and {Thoudam}, S. and {Toribio}, M.~C. and 
	{van der Tol}, B. and {Usov}, O. and {van Veelen}, M. and {van der Veen}, A.-J. and 
	{ter Veen}, S. and {Verbiest}, J.~P.~W. and {Vermeulen}, R. and 
	{Vermaas}, N. and {Vocks}, C. and {Vogt}, C. and {de Vos}, M. and 
	{van der Wal}, E. and {van Weeren}, R. and {Weggemans}, H. and 
	{Weltevrede}, P. and {White}, S. and {Wijnholds}, S.~J. and 
	{Wilhelmsson}, T. and {Wucknitz}, O. and {Yatawatta}, S. and 
	{Zarka}, P. and {Zensus}, A. and {van Zwieten}, J.},
    title = "{LOFAR: The LOw-Frequency ARray}",
  journal = {\aap},
archivePrefix = "arXiv",
   eprint = {1305.3550},
 primaryClass = "astro-ph.IM",
     year = 2013,
    month = aug,
   volume = 556,
      eid = {A2},
    pages = {A2},
      doi = {10.1051/0004-6361/201220873},
   adsurl = {http://adsabs.harvard.edu/abs/2013A%26A...556A...2V}
}

@ARTICLE{Vedantham2012,
   author = {{Vedantham}, H. and {Udaya Shankar}, N. and {Subrahmanyan}, R.
	},
    title = "{Imaging the Epoch of Reionization: Limitations from Foreground Confusion and Imaging Algorithms}",
  journal = {\apj},
archivePrefix = "arXiv",
   eprint = {1106.1297},
 primaryClass = "astro-ph.IM",
     year = 2012,
    month = feb,
   volume = 745,
      eid = {176},
    pages = {176},
      doi = {10.1088/0004-637X/745/2/176},
   adsurl = {http://adsabs.harvard.edu/abs/2012ApJ...745..176V}
}

@article{Wayth2018, 
    title={The Phase II Murchison Widefield Array: Design overview}, 
    volume={35}, 
    DOI={10.1017/pasa.2018.37}, 
    journal={Publications of the Astronomical Society of Australia}, 
    publisher={Cambridge University Press}, 
    author={Wayth, Randall B. and Tingay, Steven J. and Trott, Cathryn M. and Emrich, David and Johnston-Hollitt, Melanie and McKinley, Ben and Gaensler, B. M. and Beardsley, A. P. and Booler, T. and Crosse, B. and et al.}, 
    year={2018}, 
    pages={e033}
}

@ARTICLE{Ewall-Wice2021,
       author = {{Ewall-Wice}, Aaron and {Kern}, Nicholas and {Dillon}, Joshua S. and {Liu}, Adrian and {Parsons}, Aaron and {Singh}, Saurabh and {Lanman}, Adam and {La Plante}, Paul and {Fagnoni}, Nicolas and {Acedo}, Eloy de Lera and {DeBoer}, David R. and {Nunhokee}, Chuneeta and {Bull}, Philip and {Chang}, Tzu-Ching and {Lazio}, T. Joseph W. and {Aguirre}, James and {Weinberg}, Sean},
        title = "{DAYENU: a simple filter of smooth foregrounds for intensity mapping power spectra}",
      journal = {\mnras},
         year = 2021,
        month = jan,
       volume = {500},
       number = {4},
        pages = {5195-5213},
          doi = {10.1093/mnras/staa3293},
archivePrefix = {arXiv},
       eprint = {2004.11397},
 primaryClass = {astro-ph.CO},
       adsurl = {https://ui.adsabs.harvard.edu/abs/2021MNRAS.500.5195E}
}

@ARTICLE{Kennedy2023,
       author = {{Kennedy}, Fraser and {Bull}, Philip and {Wilensky}, Michael J. and {Burba}, Jacob and {Choudhuri}, Samir},
        title = "{Statistical Recovery of 21 cm Visibilities and Their Power Spectra with Gaussian-constrained Realizations and Gibbs Sampling}",
      journal = {\apjs},
         year = 2023,
        month = jun,
       volume = {266},
       number = {2},
          eid = {23},
        pages = {23},
          doi = {10.3847/1538-4365/acc324},
archivePrefix = {arXiv},
       eprint = {2211.05088},
 primaryClass = {astro-ph.CO},
       adsurl = {https://ui.adsabs.harvard.edu/abs/2023ApJS..266...23K}
}

@ARTICLE{Saha2021,
       author = {{Saha}, Preetha and {Bharadwaj}, Somnath and {Chakravorty}, Susmita and {Roy}, Nirupam and {Choudhuri}, Samir and {G{\"u}nther}, Hans Moritz and {Smith}, Randall K.},
        title = "{The auto- and cross-angular power spectrum of the Cas A supernova remnant in radio and X-ray}",
      journal = {\mnras},
         year = 2021,
        month = apr,
       volume = {502},
       number = {4},
        pages = {5313-5324},
          doi = {10.1093/mnras/stab446},
archivePrefix = {arXiv},
       eprint = {2102.06093},
 primaryClass = {astro-ph.GA},
       adsurl = {https://ui.adsabs.harvard.edu/abs/2021MNRAS.502.5313S}
}

@ARTICLE{Parsons2009,
       author = {{Parsons}, Aaron R. and {Backer}, Donald C.},
        title = "{Calibration of Low-Frequency, Wide-Field Radio Interferometers Using Delay/Delay-Rate Filtering}",
      journal = {\aj},
         year = 2009,
        month = jul,
       volume = {138},
       number = {1},
        pages = {219-226},
          doi = {10.1088/0004-6256/138/1/219},
archivePrefix = {arXiv},
       eprint = {0901.2575},
 primaryClass = {astro-ph.IM},
       adsurl = {https://ui.adsabs.harvard.edu/abs/2009AJ....138..219P}
}

@book{taylor1999synthesis,
  editor    = {Taylor, G. B. and Carilli, C. L. and Perley, R. A.},
  title     = {Synthesis Imaging in Radio Astronomy II},
  series    = {Astronomical Society of the Pacific Conference Series},
  volume    = {180},
  year      = {1999},
  publisher = {Astronomical Society of the Pacific},
  address   = {San Francisco}
}


\appendix

\section{\texorpdfstring{$D_{\ell}$}{Dl}  values for all PCs}
\label{app:D_ell_all_PC}
In this section, we present the detailed summary of the $D_{\ell}$ values obtained for all 25 $\ell$ bins and across all $\alpha$. Table \ref{tab:D_ell_allPC} represents the column definitions of the machine-readable table given in the Supplementary Material.

\begin{table*}
	\centering
	\caption{Definition of all columns in the supplementary table in machine-readable format for $D_{\ell}$ across 25 $\ell$ bins for all 163 PCs. Column number, label, units, and a short explanation are provided. The first row, commented with $\#$, contains the values of the $\ell$ bins}
	\label{tab:D_ell_allPC}
	
	\renewcommand{\arraystretch}{2.0} 
	
	\begin{tabular}{cccc} 
		\hline
		Column number & Column label & Units & Explanation \\
		\hline
		1 & $\alpha$ & degrees ($^{\circ}$) & Right Ascension of the PC \\
        2 & $D_{\ell}$ & $\rm mK^{2}$ & $D_{\ell}$ for the first $\ell$ bin at $\ell = 34$ \\
        3 & $D_{\ell}$ & $\rm mK^{2}$ & $D_{\ell}$ for the second $\ell$ bin at $\ell = 40$ \\
        4 & $D_{\ell}$ & $\rm mK^{2}$ & $D_{\ell}$ for the third $\ell$ bin at $\ell = 47$ \\
        5 & $D_{\ell}$ & $\rm mK^{2}$ & $D_{\ell}$ for the fourth $\ell$ bin at $\ell = 53$ \\
        6 & $D_{\ell}$ & $\rm mK^{2}$ & $D_{\ell}$ for the fifth $\ell$ bin at $\ell = 64$ \\
        7 & $D_{\ell}$ & $\rm mK^{2}$ & $D_{\ell}$ for the sixth $\ell$ bin at $\ell = 76$ \\
        8 & $D_{\ell}$ & $\rm mK^{2}$ & $D_{\ell}$ for the seventh $\ell$ bin at $\ell = 87$ \\
        9 & $D_{\ell}$ & $\rm mK^{2}$ & $D_{\ell}$ for the eighth $\ell$ bin at $\ell = 98$ \\
        10 & $D_{\ell}$ & $\rm mK^{2}$ & $D_{\ell}$ for the ninth $\ell$ bin at $\ell = 120$ \\
        11 & $D_{\ell}$ & $\rm mK^{2}$ & $D_{\ell}$ for the tenth $\ell$ bin at $\ell = 138$ \\
        12 & $D_{\ell}$ & $\rm mK^{2}$ & $D_{\ell}$ for the eleventh $\ell$ bin at $\ell = 161$ \\
        13 & $D_{\ell}$ & $\rm mK^{2}$ & $D_{\ell}$ for the twelfth $\ell$ bin at $\ell = 191$ \\
        14 & $D_{\ell}$ & $\rm mK^{2}$ & $D_{\ell}$ for the thirteenth $\ell$ bin at $\ell = 219$ \\
        15 & $D_{\ell}$ & $\rm mK^{2}$ & $D_{\ell}$ for the fourteenth $\ell$ bin at $\ell = 254$ \\
        16 & $D_{\ell}$ & $\rm mK^{2}$ & $D_{\ell}$ for the fifteenth $\ell$ bin at $\ell = 308$ \\
        17 & $D_{\ell}$ & $\rm mK^{2}$ & $D_{\ell}$ for the sixteenth $\ell$ bin at $\ell = 360$ \\
        18 & $D_{\ell}$ & $\rm mK^{2}$ & $D_{\ell}$ for the seventeenth $\ell$ bin at $\ell = 422$ \\
        19 & $D_{\ell}$ & $\rm mK^{2}$ & $D_{\ell}$ for the eighteenth $\ell$ bin at $\ell = 491$ \\
        20 & $D_{\ell}$ & $\rm mK^{2}$ & $D_{\ell}$ for the nineteenth $\ell$ bin at $\ell = 565$ \\
        21 & $D_{\ell}$ & $\rm mK^{2}$ & $D_{\ell}$ for the twentieth $\ell$ bin at $\ell = 654$ \\
        22 & $D_{\ell}$ & $\rm mK^{2}$ & $D_{\ell}$ for the twenty-first $\ell$ bin at $\ell = 762$ \\
        23 & $D_{\ell}$ & $\rm mK^{2}$ & $D_{\ell}$ for the twenty-second $\ell$ bin at $\ell = 900$ \\
        24 & $D_{\ell}$ & $\rm mK^{2}$ & $D_{\ell}$ for the twenty-third $\ell$ bin at $\ell = 1055$ \\
        25 & $D_{\ell}$ & $\rm mK^{2}$ & $D_{\ell}$ for the twenty-fourth $\ell$ bin at $\ell = 1254$ \\
        26 & $D_{\ell}$ & $\rm mK^{2}$ & $D_{\ell}$ for the twenty-fifth $\ell$ bin at $\ell = 1428$ \\
		\hline
	\end{tabular}
\end{table*}

\section{Validation}
\label{app:valid}

\begin{figure}
\includegraphics[width=\columnwidth]{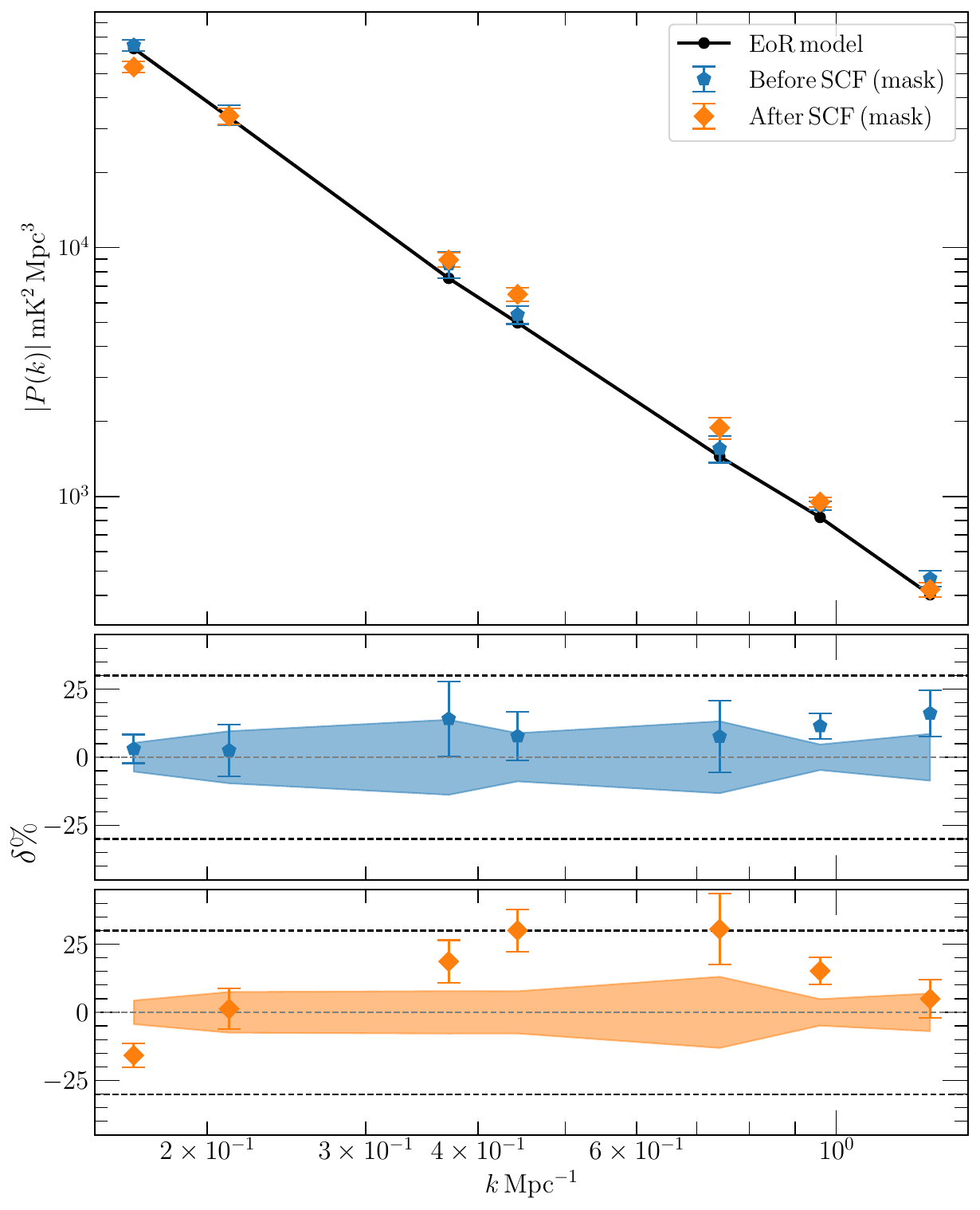}
\caption{The top panel shows the input model EoR 21-cm PS, along with the recovered PS before (after) SCF with $3\sigma$ error bars. 
The fractional deviation of the estimated PS with the input model for the two cases, before and after applying SCF, is shown in the middle and bottom panels, respectively. The shaded regions represent $3\sigma$ uncertainties.}
\label{fig:21cm_validation}
\end{figure}

In this section, we present simulations to validate the entire methodology employed in this work. We generate, in exactly the same way as described in Appendix~A of \citetalias{Elahi2025}, 20 independent realizations of the visibility data at $z=8$ for the EoR 21-cm signal model of \cite{Mondal2017}. We then analyze these in the same way as the actual data. To be specific, we first obtain the gridded visibilities, apply SCF, and use the filtered component to estimate the MAPS $C_\ell(\Delta\nu)$. We then use this to estimate $P(\kpp, \kpar)$. We apply the same masks on $P(\kpp, \kpar)$ as applied on the actual data (Table~\ref{table:mask}), and estimate $P(k)$.

Figure~\ref{fig:21cm_validation} shows the comparison of the estimated $P(k)$ with the input EoR model. The top panel shows the input $P^{m}(k)$ used for the simulation for the $k$ range $0.165{\rm\, Mpc^{-1}}$ to $1.271 {\rm\, Mpc^{-1}}$. The blue pentagons and the orange diamonds show the estimated 21-cm PS with $3\sigma$ error bars before and after applying SCF, respectively. The middle (bottom) panel shows the fractional deviation $(\delta)$ between the model and the estimated $P(k)$ before (after) applying SCF. The shaded regions in the middle (bottom) panel show the $\pm 3\sigma$ uncertainties around zero, derived from 20 random realisations of the signal. The middle panel shows that the estimated values closely match the input $P^{\rm m}(k)$ within $\pm 3\sigma$ uncertainty before applying the SCF in the $k$ range $0.165{\rm\, Mpc^{-1}}$ to $0.742{\rm\, Mpc^{-1}}$. However, for the last two $k$ bins, our estimator overestimates $P(k)$ by $11.4\%$ and $16.1\%$ respectively. In the bottom panel, after applying SCF, we underestimate $P(k)$ by $15.8\%$ at the lowest $k$ bin. However, we see that the estimated $P(k)$ is in reasonably good agreement with the expected $3\sigma$ error bars except in the range $0.371 {\rm\, Mpc^{-1}}$ to $0.959 {\rm\, Mpc^{-1}}$ where the maximum $\delta$ attained is $30.5\%$. We note that these deviations may be due to a low number of $k$ modes contributing to those particular $k$ bins. Furthermore, such uncertainties are acceptable compared with those arising from noise estimation, beam modelling, and calibration. We note that our results are two to three orders of magnitude larger than the predicted 21-cm signal at these redshifts. We also validated using the simulation for another binning scheme mentioned in Section \ref{sec:Incoherent-averaging} (not shown in Figure~\ref{fig:21cm_validation}). In this case, we find that $P(k)$ is underestimated by $18.07\%$ and $0.58\%$ in the lowest two $k$ bins after SCF. We note that like in the previous case, here also, the deviations are larger for the $k$ range $0.408 {\rm \, Mpc^{-1}}$ to $0.813 {\rm \, Mpc^{-1}}$ and goes upto $36.33\%$ at $k = 0.728 {\rm \, Mpc^{-1}}$.

\section{Masking}
\label{app:masking}
\vspace{-0.57em}
The measured $\pk$ inside the black dashed line in Figure~\ref{fig:cylps} still exhibits some very faint streaks around specific $\kpar$ values, even after applying SCF. In Figure~\ref{fig:cylps_mask}, we therefore apply a mask in the $(\kpp, \kpar)$ plane to exclude those $\kpar$ ranges, which are indicated by the grey shaded regions. Table~\ref{table:mask} lists the values of $(\kpp, \kpar)$ corresponding to the unmasked regions used for estimating the spherically averaged PS $P(k)$. We find that for the three highest $\kpp$ values, modes with $\kpar < 0.72\,\mathrm{Mpc}^{-1}$ are masked. This is primarily due to increased foreground leakage at high $\kpp$. However, for the two lowest $\kpp$ values, we are able to include modes down to $\kpar = 0.135\,\mathrm{Mpc}^{-1}$. Our most stringent upper limits are derived from these low-$k$ modes.


\begin{table*}
\centering
\renewcommand{\arraystretch}{1.2}
\caption{The unmasked $\kpar$ ranges corresponding to different $\kpp$ values as shown in Figure~\ref{fig:cylps_mask} and Figure~\ref{fig:CylXsp}.}
\begin{tabular}{c c c c c c}
\hline
$\kpp\,\rm Mpc^{-1}$ & \multicolumn{5}{c}{$\kpar \,\rm Mpc^{-1}$} \\

\hline
0.007 & 0.135--0.228 & 0.360--0.499 & 0.720--0.797 & 0.921--1.095 & 1.171--1.399 \\
0.015 & 0.135--0.228 & 0.360--0.499 & 0.720--0.797 & 0.921--1.095 & 1.171--1.399 \\
0.022 & -- & 0.360--0.499 & 0.720--0.797 & 0.921--1.095 & 1.171--1.399 \\
0.029 & -- & -- & 0.720--0.797 & 0.921--1.095 & 1.171--1.399 \\
0.038 & -- & -- & 0.720--0.797 & 0.921--1.095 & 1.171--1.399 \\
0.045 & -- & -- & 0.720--0.797 & 0.921--1.095 & 1.171--1.399 \\
\hline
\end{tabular}

\label{table:mask}
\end{table*}

\section{Power Spectrum for All PCs}
\label{app:ps_summary}
\begin{table*}
\centering
\arrayrulecolor{black}
\setlength{\arrayrulewidth}{1pt}
\renewcommand{\arraystretch}{1.8}
\large
\caption{The value of $\Delta^2(k)$, predicted errors $\sigma(k)$, and the $2\sigma$ upper limits $\Delta_{\rm UL}^{2}(k)$, all in units of $\rm mK^{2}$, for 2 PCs at $\alpha = 0.0^{\circ}, 11.0^{\circ}$, showing only 4 of the 7 $k$-bins. The best $2\sigma$ upper limit from a single PC is shown in boldface. The entire table, covering all 163 PCs and all 7 $k$-bins, is available in PDF format in the Supplementary Material, along with the data values provided in machine-readable form.}
\label{tab:PS-summary-all}
\resizebox{\textwidth}{!}{%
\begin{tabular}{c|ccc|ccc|ccc|ccc}
\hline
$\alpha$ 
& \multicolumn{3}{c|}{$k_{1} = 0.161\,{\rm Mpc^{-1}}$} 
& \multicolumn{3}{c|}{$k_{2} = 0.212\,{\rm Mpc^{-1}}\,\ldots$}  
& \multicolumn{3}{c|}{$\ldots\,k_{6} = 0.959\,{\rm Mpc^{-1}}$} 
& \multicolumn{3}{c}{$k_{7} = 1.271\,{\rm Mpc^{-1}}$} \\
\hline

& $\Delta^2(k)$ & $\sigma(k)$ & $\Delta_{\rm UL}^{2}(k)$
& $\Delta^2(k)$ & $\sigma(k)$ & $\Delta_{\rm UL}^{2}(k)$
& $\Delta^2(k)$ & $\sigma(k)$ & $\Delta_{\rm UL}^{2}(k)$
& $\Delta^2(k)$ & $\sigma(k)$ & $\Delta_{\rm UL}^{2}(k)$ \\

 
& $\rm mK^2$ & $\rm mK^2$ & $\rm mK^2$
& $\rm mK^2$ & $\rm mK^2$ & $\rm mK^2$
& $\rm mK^2$ & $\rm mK^2$ & $\rm mK^2$
& $\rm mK^2$ & $\rm mK^2$ & $\rm mK^2$ \\

\hline

$0.0^{\circ}$ 
& $(164.60)^2$ & $(151.26)^2$ & $(269.91)^2$
& $-(544.25)^2$ & $(293.26)^2$ & $(414.74)^2$
& $(1929.73)^2$ & $(1698.34)^2$ & $(3081.00)^2$
& $(5059.95)^2$ & $(2563.36)^2$ & $(6224.53)^2$ \\

$11.0^{\circ}$ 
& $-(165.39)^2$ & $(122.42)^2$ & $\mathbf{(173.13)^2}$
& $-(205.69)^2$ & $(245.10)^2$ & $(346.62)^2$
& $(1797.20)^2$ & $(1383.99)^2$ & $(2657.21)^2$
& $(238.92)^2$ & $(2081.15)^2$ & $(2952.88)^2$ \\

\hline
\end{tabular}%
}

\label{tab:ps_summary_alpha}
\end{table*}

In this section, we present the detailed PS results for all PCs. As a representative summary, Table~\ref{tab:ps_summary_alpha} lists the results for two selected PCs at $\alpha = 0.0^{\circ}$ and $11.0^{\circ}$. This table reports the values of $\Delta^{2}(k)$, the corresponding u   ncertainties $\sigma(k)$, and the associated $2\sigma$ upper limits $\Delta_{\rm UL}^{2}(k)$ for four of the seven $k$-bins.
We have included a full table in the Supplementary Material consisting of all 163 PCs and all seven $k$-bins in PDF format similar to Table \ref{tab:PS-summary-all}. A machine-readable format is also provided in the Supplementary Material, whose description of each column is given in Table~\ref{tab:PS-summary-def}.



\begin{table*}
	\centering
	\caption{Definition of all columns in the supplementary table for all 163 PCs. Column number, label, units, and a short explanation are provided. The first row, commented with $\#$, contains the values of the $k$ bins.}
    

	\label{tab:PS-summary-def}

	\renewcommand{\arraystretch}{2.0} 
	
	\begin{tabular}{cccc} 
		\hline
		Column number & Column label & Units & Explanation \\
		\hline
		1 & $\alpha$ & degrees ($^{\circ}$) & Right Ascension of PC \\
        2 & $\Delta^2(k)$ & $\rm mK^{2}$ & $\Delta^2(k)$ for the first $k$ bin at $k = 0.161\, \rm{Mpc}^{-1}$\\
        3 & $\sigma(k)$ & $\rm mK^{2}$ & $\sigma(k)$ for the first $k$ bin at $k = 0.161\, \rm{Mpc}^{-1}$\\
        4 & $\Delta^{2}_{\rm UL}(k)$ & $\rm mK^{2}$ & $\Delta^{2}_{\rm UL}(k)$ for the first $k$ bin at $k = 0.161\, \rm{Mpc}^{-1}$\\
        5 & $\Delta^2(k)$ & $\rm mK^{2}$ & $\Delta^2(k)$ for the second $k$ bin at $k = 0.212\, \rm{Mpc}^{-1}$\\
        6 & $\sigma(k)$ & $\rm mK^{2}$ & $\sigma(k)$ for the second $k$ bin at $k = 0.212\, \rm{Mpc}^{-1}$\\
        7 & $\Delta^{2}_{\rm UL}(k)$ & $\rm mK^{2}$ & $\Delta^{2}_{\rm UL}(k)$ for the second $k$ bin at $k = 0.212\, \rm{Mpc}^{-1}$\\
        8 & $\Delta^2(k)$ & $\rm mK^{2}$ & $\Delta^2(k)$ for the third $k$ bin at $k = 0.371\, \rm{Mpc}^{-1}$\\
        9 & $\sigma(k)$ & $\rm mK^{2}$ & $\sigma(k)$ for the third $k$ bin at $k = 0.371\, \rm{Mpc}^{-1}$\\
        10 & $\Delta^{2}_{\rm UL}(k)$ & $\rm mK^{2}$ & $\Delta^{2}_{\rm UL}(k)$ for the third $k$ bin at $k = 0.371\, \rm{Mpc}^{-1}$\\
        11 & $\Delta^2(k)$ & $\rm mK^{2}$ & $\Delta^2(k)$ for the fourth $k$ bin at $k = 0.443\, \rm{Mpc}^{-1}$\\
        12 & $\sigma(k)$ & $\rm mK^{2}$ & $\sigma(k)$ for the fourth $k$ bin at $k = 0.443\, \rm{Mpc}^{-1}$\\
        13 & $\Delta^{2}_{\rm UL}(k)$ & $\rm mK^{2}$ & $\Delta^{2}_{\rm UL}(k)$ for the fourth $k$ bin at $k = 0.443\, \rm{Mpc}^{-1}$\\
        14 & $\Delta^2(k)$ & $\rm mK^{2}$ & $\Delta^2(k)$ for the fifth $k$ bin at $k = 0.741\, \rm{Mpc}^{-1}$\\
        15 & $\sigma(k)$ & $\rm mK^{2}$ & $\sigma(k)$ for the fifth $k$ bin at $k = 0.741\, \rm{Mpc}^{-1}$\\
        16 & $\Delta^{2}_{\rm UL}(k)$ & $\rm mK^{2}$ & $\Delta^{2}_{\rm UL}(k)$ for the fifth $k$ bin at $k = 0.741\, \rm{Mpc}^{-1}$\\ 
        17 & $\Delta^2(k)$ & $\rm mK^{2}$ & $\Delta^2(k)$ for the sixth $k$ bin at $k = 0.959\, \rm{Mpc}^{-1}$\\
        18 & $\sigma(k)$ & $\rm mK^{2}$ & $\sigma(k)$ for the sixth $k$ bin at $k = 0.959\, \rm{Mpc}^{-1}$\\
        19 & $\Delta^{2}_{\rm UL}(k)$ & $\rm mK^{2}$ & $\Delta^{2}_{\rm UL}(k)$ for the sixth $k$ bin at $k = 0.959\, \rm{Mpc}^{-1}$\\      
        20 & $\Delta^2(k)$ & $\rm mK^{2}$ & $\Delta^2(k)$ for the seventh $k$ bin at $k = 1.271\, \rm{Mpc}^{-1}$\\
        21 & $\sigma(k)$ & $\rm mK^{2}$ & $\sigma(k)$ for the seventh $k$ bin at $k = 1.271\, \rm{Mpc}^{-1}$\\
        22 & $\Delta^{2}_{\rm UL}(k)$ & $\rm mK^{2}$ & $\Delta^{2}_{\rm UL}(k)$ for the seventh $k$ bin at $k = 1.271\, \rm{Mpc}^{-1}$\\         
		\hline
	\end{tabular}
\end{table*}

\section{PC selection}
\label{appdx:PC_selection_caseI}

\begin{figure*}
\centering
\includegraphics[width=\textwidth]{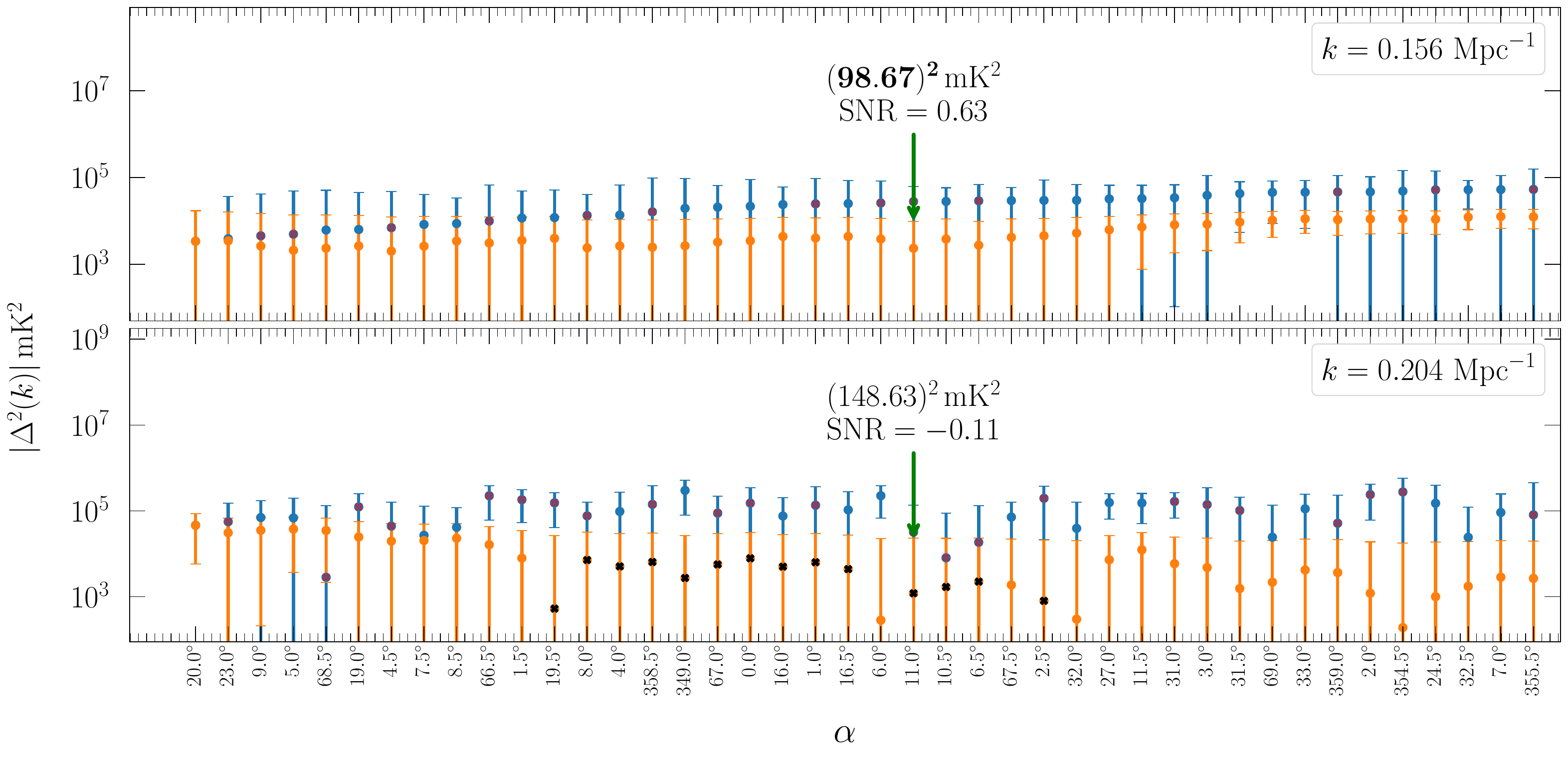}
\caption{The blue circles represent the values of $|\Delta^{2}(k)|$ arranged in ascending order based on the lowest $k$-bin for Case~I. The orange circles show the cumulatively averaged $|\Delta^{2}(k)|$ together with their associated $2\sigma$ uncertainties as successive PCs are added. The green arrow indicates the value of $\Delta^{2}_{\rm UL}(k)$ obtained after including 23 PCs, along with the corresponding signal-to-noise ratios (SNRs). The top and bottom panels correspond to the two lowest $k$-bins indicated in each panel.}
\label{fig:ullowk}
\end{figure*}

\begin{table*}
\centering
\setlength{\tabcolsep}{4pt}  
\renewcommand{\arraystretch}{2.0}
 \caption{Table shows the list of $\alpha$ of the selected PCs used in Cases I and II. The $\alpha$ range $358.5^{\circ}$ to $20.0^{\circ}$ is common in both cases and is marked using boldface along with a sole PC at $\alpha = 66.5^{\circ}$. In both cases, the PCs are arranged in  ascending order of $|\Delta^2(k)|$.}    
\label{tab:RA_cases}
\begin{tabular}{c c c c c c c c c}
\hline
Case & \multicolumn{8}{c}{$\alpha$} \\
\hline
I  
& $349.0^{\circ}$ 
& $\mathbf{358.5^{\circ}}$ 
& $0.0^{\circ}$ 
& $1.0^{\circ}$ 
& $1.5^{\circ}$ 
& $4.0^{\circ}$ 
& $\mathbf{4.5^{\circ}}$ 
& $5.0^{\circ}$ \\

& $6.0^{\circ}$
& $7.5^{\circ}$ 
& $\mathbf{8.0^{\circ}}$ 
& $8.5^{\circ}$ 
& $9.0^{\circ}$ 
& $\mathbf{11.0^{\circ}}$ 
& $16.0^{\circ}$ 
& $16.5^{\circ}$ \\
& $19.0^{\circ}$ 
& $\mathbf{19.5^{\circ}}$ 

& $\mathbf{20.0^{\circ}}$ 
& $23.0^{\circ}$ 
& $\mathbf{66.5^{\circ}}$ 
& $67.0^{\circ}$ 
& $68.5^{\circ}$ \\

\hline
II 
& $\mathbf{358.5^{\circ}}$ 
& $359.0^{\circ}$ 
& $\mathbf{4.5^{\circ}}$ 
& $\mathbf{8.0^{\circ}}$ 
& $10.0^{\circ}$ 
& $\mathbf{11.0^{\circ}}$ 
& $\mathbf{19.5^{\circ}}$ 
& $\mathbf{20.0^{\circ}}$ \\

& $30.5^{\circ}$ 
& $31.0^{\circ}$ 

& $34.5^{\circ}$ 
& $59.0^{\circ}$ 
& $59.5^{\circ}$ 
& $\mathbf{66.5^{\circ}}$ \\

\hline
\end{tabular}
\end{table*}

In this section, we describe the procedure used to select the optimal set of PCs for both Case~I and Case~II, as outlined in Section~\ref{sec:Incoherent-averaging}. For Case~I, we first sort all 163 PCs in ascending order based on the absolute values of $\Delta^{2}(k)$ for the lowest $k$-bin $(k = 0.156\,{\rm Mpc^{-1}})$, shown by the blue circles in the top panel of Figure~\ref{fig:ullowk}. The corresponding $2\sigma$ uncertainties for each measurement are indicated by vertical error bars. We then perform a cumulative averaging of the $\pk$ values for those sorted PCs using inverse-variance weighting, while applying the same binning scheme described in Section~\ref{sec:Incoherent-averaging}. The orange circles in Figure~\ref{fig:ullowk} show the cumulatively averaged $|\Delta^{2}(k)|$ and their associated $2\sigma$ uncertainties as successive PCs are added.

We find that the minimum value of the $2\sigma$ upper limit, $\Delta^{2}_{\rm UL}(k)$, is obtained after including 23 PCs, with a value of $(98.67)^2\,{\rm mK^{2}}$, which is marked by the green vertical arrow in the figure. Beyond 23 PCs, $\Delta^{2}_{\rm UL}(k)$ increases again, suggesting that foreground contamination becomes dominant beyond those 23 PCs at this $k = 0.156\,{\rm Mpc^{-1}}$ mode. For clarity, we restrict the abscissa range displayed in this panel. In the bottom panel of Figure~\ref{fig:ullowk}, we show $|\Delta^{2}(k)|$ (blue circles) for the second $k$-bin $(k = 0.204\,{\rm Mpc^{-1}})$ as a function of the same sorted PC index. As expected, the $|\Delta^{2}(k)|$ values are not in strictly ascending order in this case, and therefore the cumulatively averaged $|\Delta^{2}(k)|$ does not decrease monotonically as in the top panel.

For Case~II, we select the upper region in the $(\kpp,\kpar)$ plane of Figure~\ref{fig:cylps_mask} and apply the same sorting procedure based on $|\Delta^{2}(k)|$ at $k = 0.406\,{\rm Mpc^{-1}}$. Using this approach, we identify 14 PCs for Case~II. In Table~\ref{tab:RA_cases}, we list the $\alpha$ values of the selected PCs for both cases and highlight the common selections in boldface. We note that the PCs at $\alpha = 11.0^{\circ}$ (corresponding to the second row of Figure~\ref{fig:CylXsp}) and at $\alpha = 19.5^{\circ}$ (shown in Figure~\ref{fig:cylps} and Figure~\ref{fig:cylps_mask}) are selected in both Case~I and Case~II.


\begin{figure*}
\centering
\includegraphics[width=\textwidth]{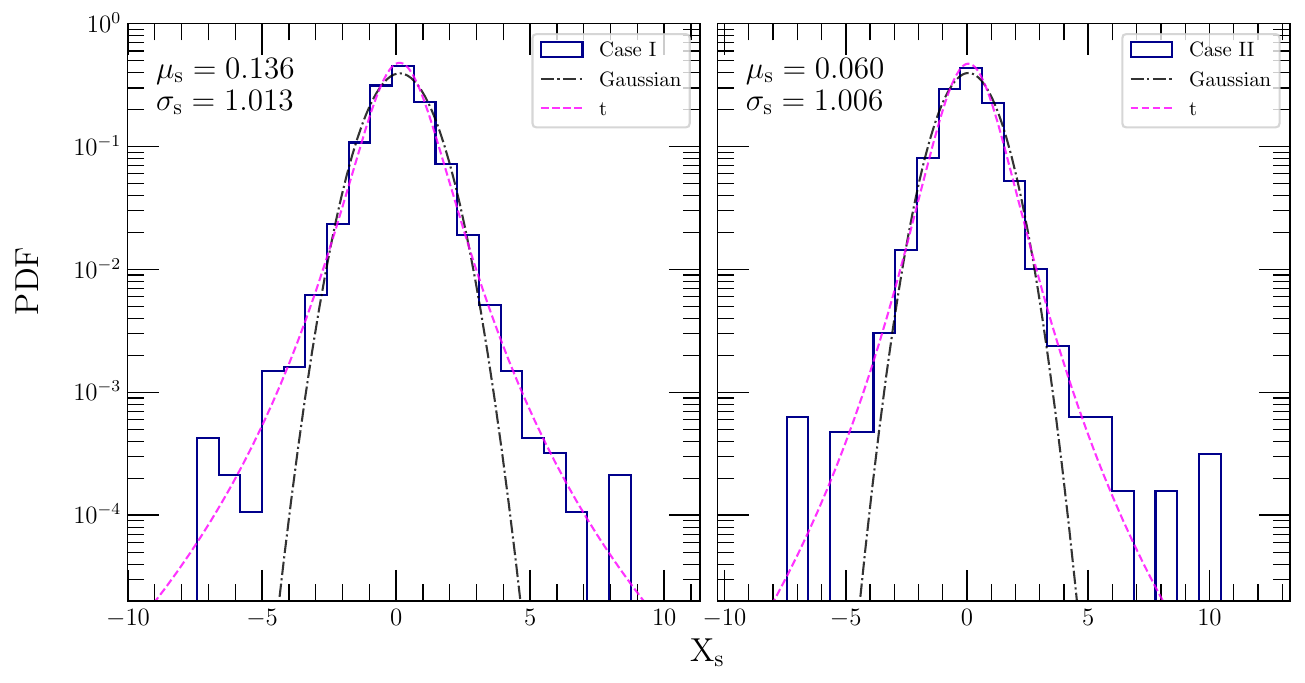}
\caption{The left (right) hand panels show the histograms of the variable $X_{\rm s}$ for Case~I (Case~II). The mean and standard deviation of $X_{\rm s}$ are $\mu_{\rm s}=0.136$ and $\sigma_{\rm s}=1.013$ for Case I, and $\mu_{\rm s}=0.060$ and $\sigma_{\rm s}=1.006$ for Case II, respectively. The best fit t- and Gaussian distributions are shown for both Cases.}
\label{fig:X_comb}
\end{figure*}

Figure~\ref{fig:X_comb} shows the histogram of the variable $X_{\rm s}$, defined in Section~\ref{sub-sec:PS_est}, for Case I (left panel) and II (right panel). In both cases, we use all $(k_{\perp},k_{\parallel})$ modes from the unmasked region shown in Figure~\ref{fig:cylps_mask}, considering 23 PCs for Case I and 14 PCs for Case II. As expected, the combined $X_{\rm s}$ distribution is symmetric around zero. The mean and standard deviation of $X_{\rm s}$ are $\mu_{\rm s}=0.136$ and $\sigma_{\rm s}=1.013$ for Case I, and $\mu_{\rm s}=0.060$ and $\sigma_{\rm s}=1.006$ for Case II, respectively. In both cases, $\sigma_{\rm s}\approx 1$ and $\mu_{\rm s}\approx 0$, consistent with noise-dominated data. The best-fitting Student’s $t$ and Gaussian probability distribution functions (PDFs) are overplotted using magenta dashed and black dot–dashed curves, respectively. We find that the central part of the distribution is well described by a Gaussian, whereas the tails are better captured by a $t$ distribution. The presence of a small number of large positive and negative outliers in both cases indicates that a minor fraction of residual foregrounds and instrumental systematics likely remain, even after applying SCF and incoherent averaging procedures.

\bsp	
\label{lastpage}
\end{document}